\documentclass[journal]{IEEEtran}

\usepackage{balance}
\usepackage{cite}
\usepackage{subcaption}
\usepackage{float}
\usepackage{caption}
\usepackage{color}
\usepackage{array}
\usepackage[cmex10]{amsmath}
\usepackage{amsthm,amsfonts,amssymb,amscd, amsbsy}
\usepackage{mathtools}
\usepackage{enumitem}
\usepackage{dsfont}
\usepackage{stackrel}
\usepackage{mathtools}
\usepackage{bbm}
\usepackage[dvipsnames,table,xcdraw]{xcolor}
\usepackage{tikz}
\usepackage{algorithm}
\usepackage{algpseudocode}
\usepackage{svg} 
\usetikzlibrary{arrows,automata}
\newcommand{\norm}[1]{\left\lVert#1\right\rVert}

\usepackage{multirow}
\usepackage{nicefrac}
\usepackage{placeins}

\DeclareMathSymbol{\mlq}{\mathord}{operators}{``}
\DeclareMathSymbol{\mrq}{\mathord}{operators}{`'}
\DeclareMathOperator*{\argmax}{arg\,max}
\DeclareMathOperator*{\argmin}{arg\,min}

\DeclareMathOperator{\sign}{sign}

\newcommand\NB[1][0.3]{N\kern-#1em\textcolor{red}{B}}

\begin{document}

\title{Reliable Extraction of Semantic Information and Rate of Innovation Estimation for Graph Signals}


\author{Mert~Kalfa,
        Sad{\i}k~Ya\u{g}{\i}z~Yetim,
        Arda~Atalik,
        Mehmetcan~G\"{o}k,
        Yiqun Ge,
        Rong Li,
        Wen Tong,
        Tolga~M.~Duman,
        and~Orhan~Ar{\i}kan% <-this % stops a space
        }


\markboth{}%
{Kalfa \MakeLowercase{\textit{et al.}}: Reliable Extraction of Semantic Information and Rate of Innovation Estimation for Graph Signals}



\maketitle

\begin{abstract}
Semantic signal processing and communications are poised to play a central part in developing the next generation of sensor devices and networks. A crucial component of a semantic system is the extraction of semantic signals from the raw input signals, which has become increasingly tractable with the recent advances in machine learning (ML) and artificial intelligence (AI) techniques. The accurate extraction of semantic signals using the aforementioned ML and AI methods, and the detection of semantic innovation for scheduling transmission and/or storage events are critical tasks for reliable semantic signal processing and communications. In this work, we propose a reliable semantic information extraction framework based on our previous work on semantic signal representations in a hierarchical graph-based structure. The proposed framework includes a time integration method to increase fidelity of ML outputs in a class-aware manner, a graph-edit-distance based metric to detect innovation events at the graph-level and filter out sporadic errors, and a Hidden Markov Model (HMM) to produce smooth and reliable graph signals. The proposed methods within the framework are demonstrated individually and collectively through simulations and case studies based on real-world computer vision examples.
\end{abstract}


% \fi
\IEEEpeerreviewmaketitle

\section{Introduction}
\IEEEPARstart{R}{ecent} advances in machine learning techniques are resulting in a paradigm shift in signal processing, where semantically-rich information about any underlying signal is becoming increasingly available. Extraction of semantic information from videos~\cite{rehman2014features}, sounds~\cite{cakir2017convolutional, mesaros2021sound}, financial time-series data~\cite{cheng2022financial}, and many more signal modalities enables processing, storage, and communication at a semantic level. As a result, the next generation of signal processing and communication systems are expected to include \textit{semantically-aware} agents that can optimize the processing and transmission of information over the inherent semantic meaning~\cite{letaief2019roadmap, Strinati2020, akyildiz20206g, dang2020should}.

In~\cite{kalfa2021}, we introduced a goal-oriented signal processing framework, where any input signal can be mapped to an application-specific graph-based semantic language, in conjunction with internal or external goals that are also based on the same language. Compared to the natural-language-based applications of the semantic processing techniques, the application-specific nature of the proposed semantic language in~\cite{kalfa2021} reduces the computational requirements of the agents at the site of the sensors, and provides inherent privacy and secrecy by mapping raw data into graph-signals. 

In the goal-oriented semantic signal processing framework~\cite{kalfa2021} a \textit{semantic extractor}, which is typically a machine learning algorithm, transforms the input signal into graph signals with a hierarchical structure that also include embedded numerical attributes. Regardless of the adopted semantic signal processing framework, this transformation process in any semantics-enabled system will produce noisy semantic signals due to the signal-to-noise-ratio (SNR) of the input signal (at the technical level), insufficient training, model imperfections, and knowledge-base limitations. For an object classifier implemented on a video signal, the low input SNR can be due to the distance of the objects or due to the improper lighting conditions, whereas model errors can be due to lack of training or simply the limitations of the model itself (i.e., number of layers, structure of the neural network, number of neurons, etc.). The knowledge-base for the framework in~\cite{kalfa2021} can be defined as the range of outcomes of the pre-defined language structure and the logical likelihoods of certain outcomes (see Section~\ref{sec:SSPprimer} for details), where a limited definition can lead to missed detections or wrong classifications. Note that the semantic noise discussed here is not the semantic channel noise~\cite{shi2021new}, nor is it an adversarial agent distorting the input of the algorithm~\cite{hu2022robust} but rather a source noise introduced when the semantic signal is first generated. An advantage in handling the source noise is that we can have access to the statistical characteristics of the semantic extractor, as well as logical characteristics of the semantic information, and exploit these to improve the fidelity of the results. 

Goal-oriented filtering of the semantic information helps focus the remaining signal processing on those graph signals that are of interest. For both storage and transmission purposes, it is important to detect innovation events on the extracted semantic signals~\cite{kalfa2021}. The proposed semantic signals in~\cite{kalfa2021} have a hierarchical structure with graphs and numerical attributes; therefore, innovation events can happen at different levels of the semantic signal. An example of an innovation event at the graph signal level can mean a graph pattern of interest (that is predefined by the internally or externally defined goals) has emerged (or receded) in the semantic description of the raw signal. Once a graph pattern of interest is identified and is being tracked, the numerical attributes such as position and subfeature vectors can be tracked across time to detect innovation events at the attribute level (again, based on the interests defined by the goals). To give a more practical example, consider a computer vision application that is used to provide security for a pedestrian-only street. In this scenario, the external goals can be the detection of objects in the classes of \textit{vehicles} and \textit{suitcases}. Therefore, innovations at the graph level can be identified as events where these classes of interest come into our out of the semantic graph description of the scene. Additionally, once these classes are identified, their accurate positions, as well as some of their features (a large suitcase is more attention-worthy than a small one, etc.) can be tracked across time to be stored or transmitted once they exceed a certain threshold.

Quantification of the innovation events enables efficient and accurate tracking of the semantic signals, filtering out sporadic erroneous detections, and scheduling the storage or transmission events, depending on the application. Coupled with the inherent compression provided by a semantic language~\cite{kalfa2021,xie2021deep}, a scheduler based on semantic innovations can reduce the overall transmission traffic dramatically. 

In this paper, with the objective of improving the fidelity and quantifying the innovation of semantic information, we propose and demonstrate a semantic extraction framework that can be implemented on graph signals with embedded numerical attributes. Specifically, we introduce an integration technique to improve semantic fidelity in a class-aware manner, a modified graph edit distance (GED) metric and a Hidden Markov Model (HMM) to quantify, track, and smooth the semantic signals at the graph level, and a subspace tracking algorithm to quantify and track the semantic information at the numerical attribute level. The proposed framework can be implemented in the next generation of communication networks and sensor devices to enable semantic signal processing and semantic communications. The individual methods can be used either separately or collectively for filtering erroneous detections, fusion of multiple sensors/data-streams, and transmission scheduling for semantic communication applications. The methods presented in this paper are rigorously explained and demonstrated using simulations and computer vision examples on acquired video signals.

The rest of this paper is organized as follows. Section~\ref{sec:RelatedWorks} provides a short literature review on the semantic signal processing, and semantic graph signal extraction. Section~\ref{sec:SSPprimer} briefly re-introduces the semantic signal processing framework of~\cite{kalfa2021}. Section~\ref{sec:SSPFramework} presents the proposed semantic extraction framework and its main building blocks. The details of each block in the proposed framework are given in Sections~\ref{sec:TimeInt}--\ref{sec:GED}. Section~\ref{sec:CaseStudies} demonstrates the performance of the proposed methods via simulations and real-world computer vision case studies. A short discussion on the rate of innovation of semantic signals and the corresponding data throughput is given in Section~\ref{sec:innovation}. Concluding remarks and future research directions are given in Section~\ref{sec:Conclusion}.

\section{Related Works}
\label{sec:RelatedWorks}
The initial work on semantic information is almost as old as Shannon's seminal work~\cite{shannon1948}, with Weaver introducing the semantic communication paradigm~\cite{weaver1953}, then Bar-Hillel and Carnap proposing a Semantic Information Theory (SIT) built on a semantic language based on propositional logic. In the last decade, the quantum leap in the abundance of semantically rich data enabled a resurgence of research on semantic signal processing and communications. To give a clear overview of the state-of-the-art on this emerging but highly active field, we review the literature in the following two subsections, namely on extraction of semantic information and on semantic communications.

\subsection{Extraction of Semantic Information}
Semantic information exists in many signal modalities such as a textual description of an image, a knowledge graph derived from a paragraph, and even in correlation functions of random processes. We refer to semantic transformation or semantic extraction as the mapping from an input modality to a target semantic modality where the target semantic modality can be anything ranging from vectors, texts, and graphs. Most popular semantic transformation techniques (especially in computer vision applications) include object detection~\cite{krizhevsky2012imagenet,girshick2014rich,redmon2016you,redmon2017yolo9000,tan2020efficientdet,wang2021scaled}, semantic segmentation~\cite{van2021unsupervised,xia2017w,ji2018invariant,shotton2008semantic,shotton2011real,tighe2014scene,hao2020brief,yu2015multi,badrinarayanan2017segnet,poudel2019fast,kirillov2019panoptic,lateef2019survey,garcia2018survey}, and captioning~\cite{donahue2015long,karpathy2015deep, mao2014explain,chen2015mind,xu2015show,johnson2016densecap,yang2017dense,kim2021dense,krishna2017visual,krause2017hierarchical,krishna2017dense,xu2019joint,hossain2019comprehensive,aafaq2019video}. 

A powerful form of semantic transformation is to convert signals into graphs that represent a scene and encode the relationships presented in the signal. Scene graphs are proposed in~\cite{johnson2015image} describe image features and object relationships in an explicit and structured way for image retrieval. Some recent papers~\cite{li2017scene, xu2017scene, li2017vip} also consider joint optimization of object detection and relationship recognition parts. Specifically, Factorizable-Net is proposed in~\cite{li2018factorizable} where a Region Proposal Network (RPN) is used to extract object proposals and proposed objects are paired to obtain a fully-connected initial coarse graph. In~\cite{yang2018graph}, Graph-RCNN is introduced. Graph-RCNN uses relation-proposal-network (RePN) to prune the connections in the initial graph and an Attentional Graph Convolutional Network~\cite{velivckovic2017graph} refines the features on the graph. On the other hand, VCTree model~\cite{tang2019learning} constructs a dynamic tree from a scoring matrix where visual context is encoded into the tree structure. Furthermore, some recent works~\cite{li2017scene, xu2017scene} use recurrent architectures for graph inference. Particularly, in~\cite{xu2017scene} a feature refining module consisting of edge and node Gated Recurrent Units (GRU), and in~\cite{zellers2018neural}, stacked bi-directional Long Short-Term Memory models (bi-LSTMs) are used.

Scene graphs can also be extracted from video signals. In~\cite{wang2020storytelling}, each frame in the video is converted into a scene graph as an intermediate semantic representation. Then, using frame and cross-frame level relationships of intermediate scene graphs, a story of the video is generated. Joint parsing of cross-view videos is introduced in~\cite{qi2018scene} where scene-centric and view-centric graphs are hierarchically generated. For more details, we refer the readers to survey papers~\cite{chang2021scene, agarwal2020visual}.

\subsection{Semantic Communications}
Many outlook papers on next-generation communication networks envisioned that some form of semantic communications will pave the way for next generation wireless communications~\cite{guler2018semantic, qin2021semantic, Strinati2020, shi2021new,shi2021semantic, uysal2021semantic, kountouris2021semantics, lan2021semantic}. Recently more practical approaches for viable semantic communications are presented in the literature. In~\cite{xie2021deep}, a deep learning enabled semantic communication architecture called DeepSC is proposed for text transmission. Unlike traditional approaches where the objective is to minimize bit errors, this work focused on minimizing semantic errors while maximizing the system capacity with a mutual information estimation model for semantic channel coding. In~\cite{xie2020lite}, the authors extend their previous work to affordable internet-of-things (IoT) applications where a lite and distributed version of~\cite{xie2021deep} called the L-DeepSC is proposed. In~\cite{wang2021}, the authors propose a semantic communications framework where natural language texts are modeled as knowledge graphs that are transmitted and converted back into text at the destination. The conversion into and from knowledge graphs offer an intuitive and explainable semantic extraction compared to~\cite{xie2020lite,xie2021deep}.

Transmission of speech data for semantic communications is considered in~\cite{weng2021semantic_v2}. The authors proposed a deep learning enabled system called DeepSC-S which extracts essential semantic information from speech audio signals by using squeeze-and-excitation (SE) networks~\cite{hu2018squeeze} and applies attention based weighting to emphasize the essential information where source and channel encoder/decoder is jointly designed to mitigate channel distortion. The proposed architecture is shown to outperform traditional approaches in almost any SNR regime.

In~\cite{yang2022semantic}, a semantic communication architecture (SC-AIT) is introduced where both effectiveness, semantic and technical level of communication is inter-connected via a neural network supported module. The proposed architecture is realized on a real-world test-bed for image classification task to detect surface defections. The authors showed that the proposed architecture outperforms traditional approaches in any SNR regime while having lower latency and higher compression ratio. 

Real-time semantic communications is first studied in~\cite{yoo2022real} where the authors developed a prototype for wireless image transmission based on the field-programmable gate array (FPGA), Vision Transformer (ViT)~\cite{dosovitskiy2020image} and a denoising auto-encoder. The implemented prototype has been shown to be superior to traditional 256-quadrant amplitude modulation (256-QAM) in the low-SNR regime on a measure of structural similarity index (SSIM). In~\cite{tung2022deepjscc} constellation constrained version of DeepJSCC~\cite{bourtsoulatze2019deep} (DeepJSCC-Q) is proposed for wireless image reconstruction. DeepJSCC-Q utilizes a differentiable soft quantization layer to map latent semantic vectors to transmitted symbols such that each quantization level corresponds to a learnable constellation point. It is illustrated that DeepJSCC-Q achieves comparable performances with respect to its unconstrained counterpart and outperforms the traditional methods which rely on separate source and channel coding scheme. 

Deep learning aided end-to-end JSCC for wireless video transmission scheme (DeepWive) is introduced in~\cite{tung2021deepwive} where the proposed architecture performs video compression, channel coding and modulation with a single neural network, resulting in direct mapping from video signals to channel symbols whereas the trained semantic decoder predicts the residuals without distortion feedback. 

In~\cite{tung2021}, effectiveness of semantic communications is studied within a \textit{joint learning and communication framework}. The authors adopted multi-agent reinforcement learning (MARL) approach to develop a systematic structure for collaborative agents participating in treasure hunts. The problem is formulated as a multi-agent partially observed Markov decision process (MA-POMDP). Agents are intended to learn policies to effectively exchange messages over a shared noisy wireless channel for improved coordination and collaboration while taking long-term rewarding actions. As a result, the proposed joint learning and communication framework achieves higher performance than treating action-taking and communication aspects of MARL separately.

In the following section, a brief introduction to our work on semantic information and signal processing~\cite{kalfa2021} will be presented. 

\section{Primer on the Goal-Oriented Semantic Signal Processing Framework}
\label{sec:SSPprimer}
Recently a goal-oriented semantic signal processing framework has been proposed to extract and process semantic information generated at sensor nodes~\cite{kalfa2021}. In this framework, the semantic information is organized using a flexible graph-based language in a structured and hierarchical way. The highly organized and hierarchical approach to construct the semantic language shifts the processing load to the initial semantic information extraction, which in turn leads to ease of processing, filtering, and reduce the overall storage within a device or transfer of information between devices within a network.

The proposed structure is made up of bipartite graphs that include the identified signal \textit{components}, and \textit{predicates} that show the state or relationship of the detected components. Each node in the graphs may also include layers of numerical attributes with different levels of complexity. Although the aforementioned structure of the proposed language is fixed, the particular definitions of the components, predicates, and attributes are application-specific. Therefore, the structures are only as complex as the specific requirements of the implementation at the sensor/agent level and the specific situation at the sensor site. In comparison to the natural-language-based semantic signal processing applications~\cite{Bao2011}, the proposed structure for the semantic information offers a much more efficient and unambiguous representation while inherently providing privacy and secrecy through relatively abstract graph signals. Moreover, the additional computational cost of introducing graph signal processing algorithms is balanced by the inherent low-order graph signals generated by the proposed framework as they are generated within an application-specific and limited dictionary. 

\begin{figure*}[ht]
\centering
  \includegraphics[trim=2cm 0 0 0, clip, width=0.95\linewidth]{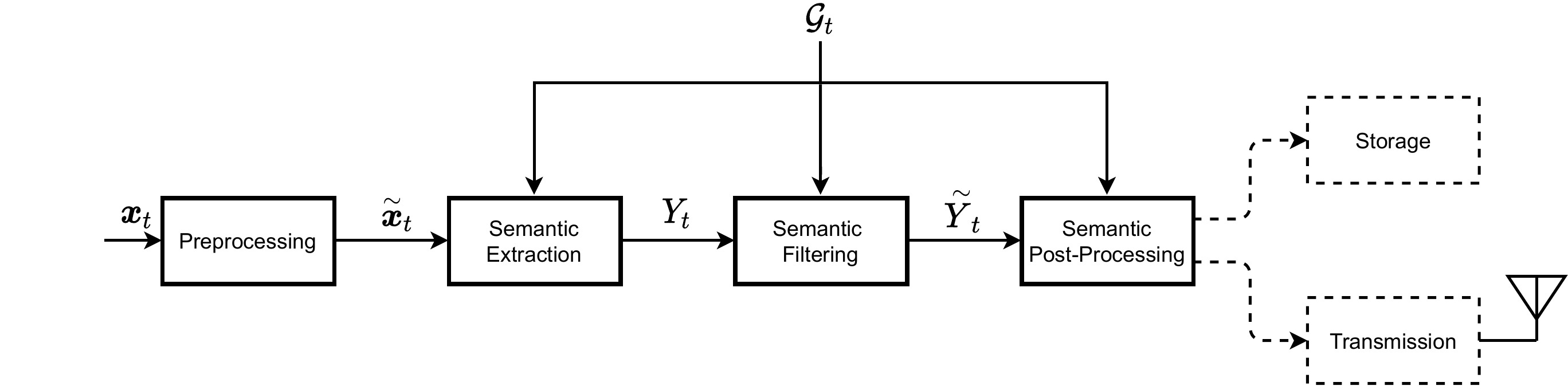}
  \caption{The proposed goal-oriented semantic signal processing framework~\cite{kalfa2021}. Raw signal input is denoted as $x_t$, whereas the external goals are denoted with $\mathcal{G}_t$.}
  \label{fig:SSPframework}
\end{figure*}

The proposed structure of the goal-oriented semantic signal processing framework at a sensor node is shown in Fig.~\ref{fig:SSPframework}. Briefly, a raw input signal generated by the sensor is preprocessed (e.g., through up/downsampling or Fourier transform) before \textit{semantic extraction}, where the sensor output is mapped to the target semantic output. Semantic filtering and post-processing blocks are implemented in a goal-oriented manner to reduce the range of semantic outputs to the specific goals and perform additional processing (e.g., source coding), respectively. Finally, the resulting goal-filtered and compressed semantic output is either transmitted or stored. The most critical component in the entire diagram shown in Fig.~\ref{fig:SSPframework} is the semantic extraction block, since accurate and reliable mapping of the raw signals into the semantic language affects the performance of the remaining blocks that follow in the semantic signal processing chain. Before we present the methods and algorithms to improve the accuracy and reliability of the semantic extractor, we first review the semantic language structure of~\cite{kalfa2021}.

\begin{figure}[ht]
	\centering
	\includegraphics[width=0.9\linewidth]{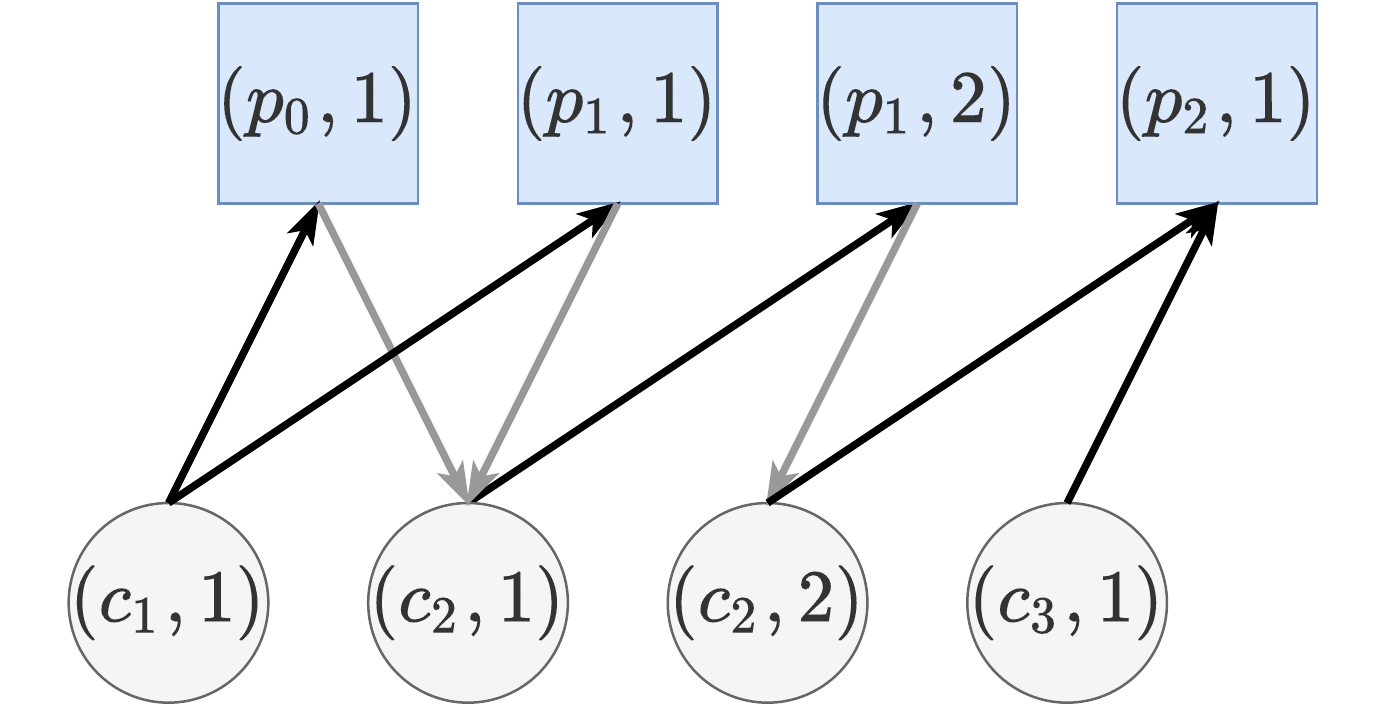}
	\caption{An example of an atomic bipartite graph structure ($D_{t,i}$) with signal components and predicates at the instance level. The two-tuple representation denotes the component or predicate class, and the unique instance identifier.}
	\label{fig:graph_example_Dt}
\end{figure}

As detailed in~\cite{kalfa2021}, the semantic description of a signal starts with the definition of component and predicate classes of interest as
\begin{align}
C &= \{c_1, c_2, \ldots, c_{N_c}\}, \label{eq:CP_C}\\
P &= \{p_0,p_1, p_2, \ldots, p_{N_p-1}\},
\label{eq:CP_P}
\end{align}
where $N_c$ and $N_p$ are the numbers of component and predicate classes in the graph language, respectively. The semantic multi-graph description generated by the detected instances of $C$ and $P$ is defined as 
\begin{equation}
\mathcal{D}_t=\{D_{t,1}, D_{t,2}, \ldots, D_{t,N}\},
\label{eq:Dt}
\end{equation}
where each $D_{t,i}$ is an atomic bipartite graph that includes a connected set of detected component and predicate instances, as illustrated in Fig.~\ref{fig:graph_example_Dt}. Note that the instance graph includes multiple instances of the same component or predicates; hence, each node in Fig.~\ref{fig:graph_example_Dt} is identified with a two-tuple, i.e., the component or predicate, and a unique ID number.

Each component and predicate in an atomic graph $D_{t,i}$ has a corresponding attribute set denoted by
\begin{equation}
\Theta_{t,i}(n_j,k) = \{\boldsymbol{\theta}_{t,i}^{(1)}(n_j,k), \boldsymbol{\theta}^{(2)}_{t,i}(n_j,k), \ldots, \boldsymbol{\theta}^{(L_{n_j})}_{t,i}(n_j,k) \},
\label{eq:theta_ti}
\end{equation}
where $(n_j,k)$ is the corresponding component or predicate node in the graph. Note that for each type of node $n_j$, we define $L_{n_j}$-levels of attributes. This enables an organized way of storing different attributes, preferably in increasing order of complexity for ease of goal-based filtering. We note that in the computer vision case study given in~\cite{kalfa2021}, there are three levels of attributes: the scalar position and velocity in the lowest-level attribute set $\boldsymbol{\theta}_{t,i}^{(1)}$, subfeature vectors in $\boldsymbol{\theta}_{t,i}^{(2)}$, and the full bounding-box images of the detected components in $\boldsymbol{\theta}_{t,i}^{(3)}$. Nonetheless, the definition given in~\eqref{eq:theta_ti} is purposefully general, and can be modified depending on the specific application.

Recent advances in machine learning and artificial intelligence have enabled the extraction of semantic information in the proposed graph structure of~\eqref{eq:CP_C}--\eqref{eq:theta_ti}. In practice, machine learning and artificial intelligence based semantic extractors do not have perfect accuracy or sensitivity; hence, the extracted semantic information does not have perfect correspondence with the objective reality. Moreover, the evolution of the generated semantic output must be tracked accurately to filter out any erroneous detection and identify the points of innovation so that the transmission (or storage) events can be scheduled efficiently. An advantage of using learning-based algorithms is that statistical properties of the algorithms (e.g., confusion matrices) can be modeled through extensive training statistics, which can be used to improve the fidelity of the semantic output by exploiting the natural temporal continuity of the objects in the sensed area. Additionally, logical probabilities of the possible semantic output patterns can be exploited to improve reliability in a Bayesian sense, as well. For example, in a computer vision security application involving a car dealer showroom for a certain brand, identifying cars with other brands are much less likely logically, and thus, semantically. Therefore, the semantic extractor can take this into account during the temporal evolution of its semantic output to determine whether any new detections are the result of a true innovation or a sporadic erroneous event. 

In the next section, with the objective of addressing the problems stated above, we introduce a semantic extraction framework that is based on the semantic structure detailed in this section. We note that the proposed methods are not limited to the structure described here, and they can also be used for different graph-based signals with embedded scalar/vector attributes.

\section{The Proposed Semantic Extraction Framework}
\label{sec:SSPFramework}

In the semantic signal processing framework of~\cite{kalfa2021} summarized in Section~\ref{sec:SSPprimer} and illustrated in Fig.~\ref{fig:SSPframework}, the semantic extractor is assumed to be generating reliable and smooth semantic output signals. On the other hand, the extracted semantic information includes semantic noise that is introduced by input noise, algorithm limitations, etc. In this section, we propose several methods that can be used individually or collectively to exploit prior information about the semantic extractor and environmental characteristics to improve the fidelity of the extracted semantic information. Moreover, the proposed techniques enable asynchronous updates of the semantic signal across time, where the updates can be designed to occur only when there is a significant semantic innovation in the signal. 

\begin{figure}[t]
	\centering
	\includegraphics[trim = 1cm 0 0 0, clip, width=0.95\linewidth]{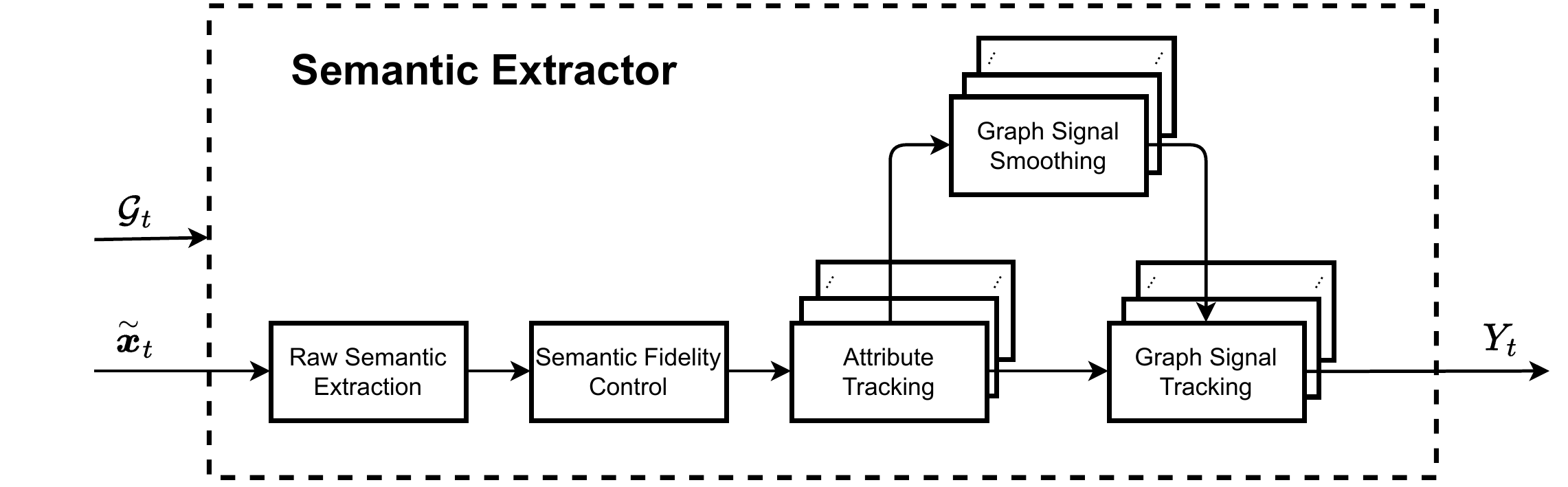}
	\caption{Proposed block diagram for a typical graph-based semantic extractor.}
	\label{fig:semantic_extractor}
\end{figure}

The proposed conceptual block diagram for a semantic extractor is given in Fig.~\ref{fig:semantic_extractor}. Note that the \textit{Raw Semantic Extraction} block typically involves a learning-based algorithm such as a scene-graph generator for image/video signals~\cite{yang2018graph}, segmented and classified objects in LIDAR images~\cite{che2019}, sound event detection (SED) for acoustic signals~\cite{mesaros2021sound}, etc. The following blocks in Fig.~\ref{fig:semantic_extractor} exploit the known characteristics of the raw extractor, such as its confusion matrix; and the statistics of the environment, such as the occurrence probabilities of the possible output signals. Throughout this paper, we assume either these parameters are known a priori or estimated through measurements. 

The proposed \textit{Semantic Fidelity Control} block takes the raw output of the initial extraction and uses time integration techniques to improve the detection characteristics according to internally set or externally requested reliability parameters. The proposed fidelity control method is explained in Section~\ref{sec:TimeInt}.

Note that after this point in the semantic extraction framework shown in Fig.~\ref{fig:semantic_extractor},  tracking and smoothing blocks work in parallel processing banks, with each instance working on an atomic graph as defined in~\eqref{eq:Dt}. This parallel approach is the result of intentional separation of the inter-connected atomic graphs, and reduces the computational complexity of the following blocks. Also note that a list of goals denoted as $\mathcal{G}_t$ is an input of the semantic extraction framework. Either internally or externally defined, the goals can be used to filter the total semantic description of the signal into a goal-oriented subset, which can also greatly reduce the processing load of the following blocks. Since the goal-filtering can be performed anywhere along the processing chain, it is left as a general input in Fig.~\ref{fig:semantic_extractor}.

In a typical application of semantic signal processing or communications, the innovation in the semantic signals will not only come from the graph structure, but also from the low-level numerical attributes embedded in the graphs (see \eqref{eq:theta_ti} in Section~\ref{sec:SSPprimer}). The \textit{Attribute Tracking} is an application-specific block, where these low-level numerical attributes can be tracked across time to detect innovations in the semantic signals of interest. For vector attributes (e.g., subfeature vectors or state vectors), we propose and demonstrate a subspace tracking algorithm in Section~\ref{sec:AttrTrack}.

The \textit{Graph Signal Smoothing} block is proposed as an optional pre-filtering step to smooth out sporadic erroneous detections and reduce the throughput for the following blocks. Hence, this step needs to have a relatively low computational cost compared to the final filtering/tracking block. Therefore, a simple GED algorithm with a modified cost definition is proposed for this block and explained in Section~\ref{sec:GED}.

The final block in Fig.~\ref{fig:semantic_extractor}, the \textit{Graph Signal Tracking}, uses the pre-filtered graph signal and attribute updates in conjunction with the estimated environmental probabilities to produce a reliable, accurate semantic description of the input signal. For this block, we propose an HMM over the graph signals in Section~\ref{sec:HMM} and use state estimation algorithms to optimize the output signal. As a result, the output of this block is a reliable, accurate, and slowly-varying signal (compared to the raw input) that can indicate significant innovations in the graph-signal or its underlying attributes. 

Starting with the next section, we present each sub-block and the potential methods and algorithms that can be incorporated within, in detail. Note that the proposed algorithms and techniques presented in the following sections can be used collectively, as shown in Fig.~\ref{fig:semantic_extractor}, or individually depending on the computational capabilities and requirements of the signal processing hardware.

\section{Semantic Fidelity Control with Time Integration}
\label{sec:TimeInt}
Time integration of signals for improved detection and estimation performance are well documented in the signal processing literature~\cite{dillard1980performance,li2015coherent, li2016clean}. These techniques can also be employed in the semantic signal processing framework in the initial \textit{preprocessing} block, shown in Fig.~\ref{fig:SSPframework}, in front of the \textit{semantic extractor}. With the introduction of a semantic language and a semantic extractor, a semantically-aware way of improving the detection and estimation performance can be achieved by employing time integration techniques at the output of the semantic extractor. 

The \textit{raw semantic extractor} shown in Fig.~\ref{fig:semantic_extractor} typically includes a machine learning algorithm that detects the signal components and their inter-relationships. After the training and testing of these algorithms, we assume that we have access to the confusion matrix that provides statistics on the characteristics of the trained algorithm~\cite{visa2011confusion}. The confusion matrix of a raw extractor is in the following form:
\begin{equation}
    CM(\tau) = \left[ \begin{matrix} n_{1,1} & n_{1,2} & \cdots & n_{1,K} \\ n_{2,1} & n_{2,2} & \cdots & n_{2,K} \\ \vdots & \vdots & \ddots & \vdots \\ n_{K,1} & n_{K,2} & \cdots & n_{K,K} 
    \end{matrix}
    \right],
    \label{eq:CM}
\end{equation}
where $\tau$ is the score threshold used for detection, $n_{i,j}$ corresponds to the number of times a pattern $j$ is detected given an actual pattern of $i$, with $K$ being the total number of output patterns. The output patterns in this definition can be simple components/predicates or more complex graph outputs as shown in Fig.~\ref{fig:graph_example_Dt}, i.e., the range of possible semantic outputs that the extractor can generate. Note that, the total number of possible components and predicates ($N_c$ and $N_p$) are limited, since they are defined for a specific application with specific goals. Moreover, in a given time instant, the connected atomic semantic graph outputs will have even fewer components and predicates ($\eta_c$ and $\eta_p$, respectively), which will reduce the computational complexity of processing the graphs even further. 

The confusion matrix can be used to infer other statistical metrics about the output patterns. Let $N_i \geq \sum_j n_{i,j}$ be the total number of samples where the pattern $i$ exists, and $N = \sum N_i$ be the total number of samples. With~\eqref{eq:CM} and $N_i$, other pertinent confusion metrics for each pattern can be written as 
\begin{align}
    \text{True Positive Rate:  } &TPR_i = \frac{n_{i,i}}{N_i}, \label{eq:TPR} \\
    \text{False Positive Rate:  } &FPR_i = \frac{\sum\limits_{j \neq i} n_{j,i}}{N-N_i}, \label{eq:FPR} \\
    \text{False Negative Rate:  } &FNR_i = \frac{N_i-n_{i,i}}{N_i}, \label{eq:FNR} \\
    \text{True Negative Rate:  } &TNR_i = \frac{\sum\limits_{j \neq i}\sum\limits_{k \neq i} n_{j,k}}{N-N_i}. \label{eq:TNR}
\end{align}
Note that \eqref{eq:TPR}--\eqref{eq:TNR} also depend on the detection threshold $\tau$. By changing $\tau$, different detection characteristics for a given pattern can be obtained to generate the Receiver Operating Characteristic (ROC) curve. The ROC for the semantic extractor can also be constructed for different environmental factors such as time of day, weather, season, etc., depending on the application and the desired accuracy. 

To control the fidelity of the extractor output with a semantically-aware approach, we propose temporal integration of the output scores of a classifier that identifies the signal components to improve the ROC. Assuming underlying Gaussian processes, the output score distribution ($S_i$) for each pattern under binary hypothesis testing ($\mathcal{H}_0$: the pattern does not exist and $\mathcal{H}_1$: the pattern exists) can be modeled as
\begin{equation}
    P(O_i) \sim \begin{cases}
        \mathcal{N}(\mu_{i,0}, \, \sigma_{i,0}^2), & \text{under } \mathcal{H}_0 \\
        \mathcal{N}(\mu_{i,1}, \, \sigma_{i,1}^2), & \text{under } \mathcal{H}_1 \\
    \end{cases}
    \label{eq:h0_pO}
\end{equation}
where $\mu$ and $\sigma$ are the mean and standard deviation for the score distributions in each case. Assuming a jointly Gaussian distribution, the time-average of $T_i$ consecutive samples yields
\begin{equation}
    P(O_i^{T_i}) = P\left( \frac{1}{T_i} \sum \limits_{n=1}^{T_i} O_i(t_n) \right) \sim \begin{cases}
        \mathcal{N}(\mu_{i,0}, \, \hat{\sigma}_{i,0}^2), & \!\!\!\!\text{under } \mathcal{H}_0 \\
        \mathcal{N}(\mu_{i,1}, \, \hat{\sigma}_{i,1}^2), & \!\!\!\!\text{under } \mathcal{H}_1 \\
    \end{cases}
    \label{eq:h0_pO_Ti}
\end{equation}
where $O_i(t_n)$ is the output score at time $t_n$. As expected, the mean values do not change and the standard deviations are bounded as 
\begin{equation}
    \sigma_i/T_i \leq \hat{\sigma}_i \leq \sigma_i,
    \label{eq:sigmas}
\end{equation}
for both hypotheses. The lower and upper bound for the standard deviations are achieved when the consecutive samples are independent or perfectly correlated, respectively. To illustrate the correlation between consecutive samples, consider the semantic extraction of a video signal. If the scene is perfectly stationary, i.e., the same still images are recorded across consecutive time instants, the score distributions will be identical, rendering the time integration ineffective. However, even small changes in the video image can lead to different score output realizations of the same underlying distribution, hence improving the detection performance. 

Since the standard deviations for the distributions under both hypotheses are reduced, it is straightforward to show that the ROC metrics given in \eqref{eq:TPR}--\eqref{eq:TNR} can be improved as
\begin{align}
    TPR_{i,T_i} \geq TPR_i, \label{eq:TPR2} \\
    FPR_{i,T_i} \leq FPR_i, \label{eq:FPR2} \\
    FNR_{i,T_i} \leq FNR_i, \label{eq:FNR2} \\
    TNR_{i,T_i} \geq TNR_i, \label{eq:TNR2}
\end{align}
with subscript $T_i$ denoting the time-integrated metrics. Again, the equalities in \eqref{eq:TPR2}--\eqref{eq:TNR2} hold only when the consecutive samples are perfectly correlated. Also, note that the assumption of Gaussianity is not essential, i.e., similar expressions can be written for other distribution models as well.

\begin{figure}[t]
	\centering
	\includegraphics[trim = 0  3cm 0 0, clip, width=0.8\linewidth]{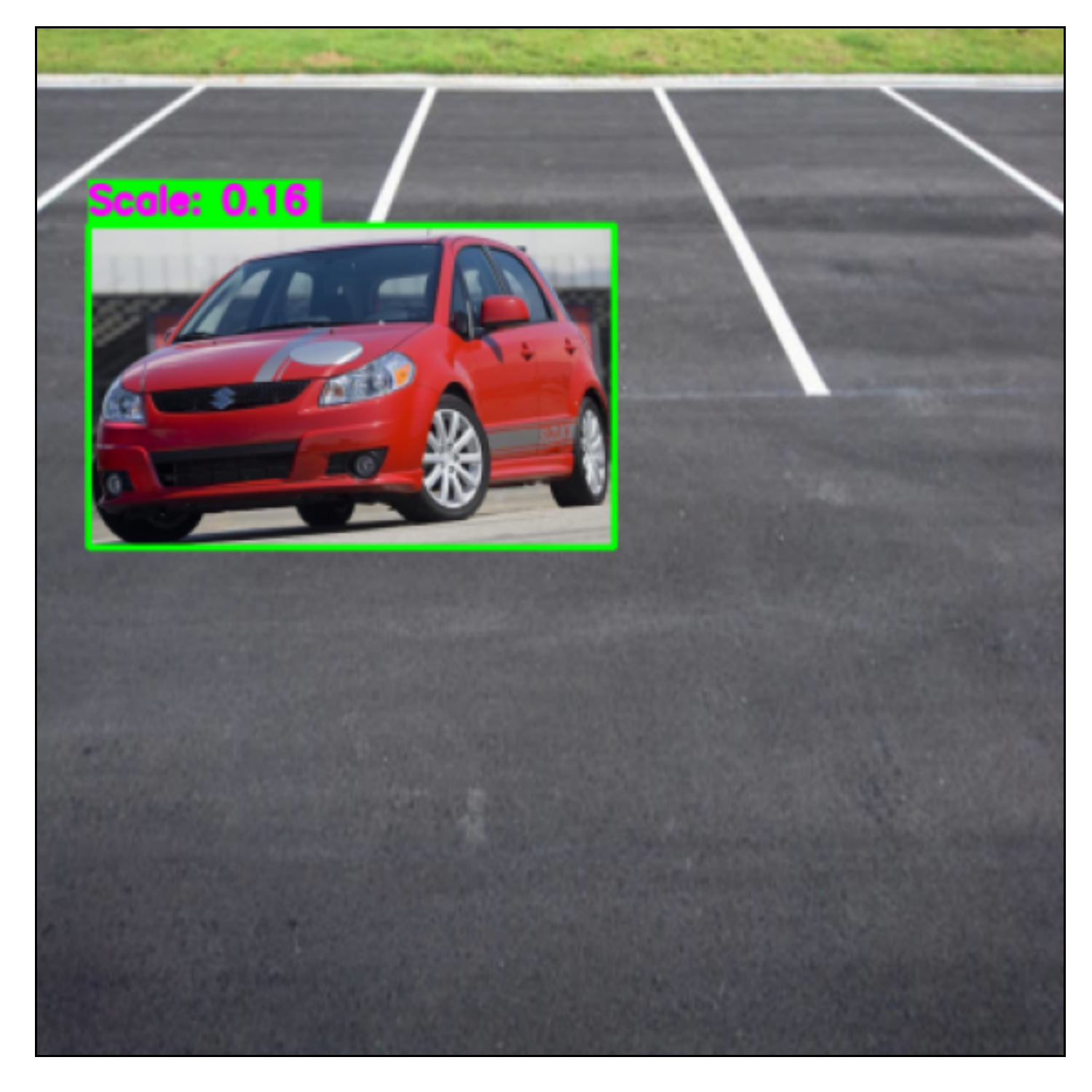}
	\caption{An example car image placed onto a stationary parking lot background. The scale number annotation corresponds to the ratio of the size of the car compared to the full image.}
	\label{fig:carSNR_example}
\end{figure}

\begin{figure*}[t]
	\centering
	\begin{subfigure}[b]{.32\textwidth}
  \centering  \includegraphics[trim={0 0 0 0},clip, width=1\linewidth]{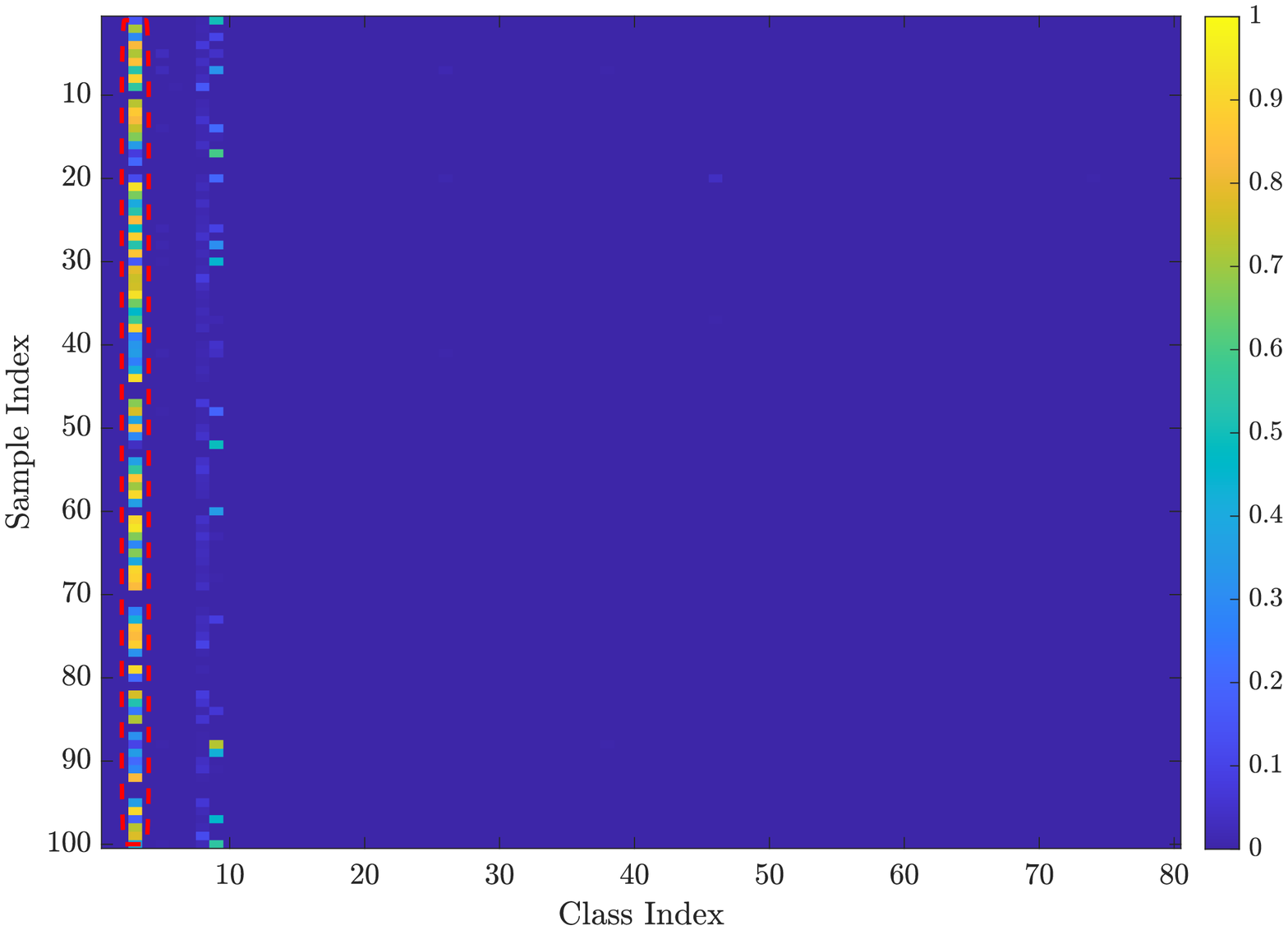}
  \caption{Output scores at Scale: 10\%}
  \label{fig:carSNR_a}
\end{subfigure}%
\begin{subfigure}[b]{.32\textwidth}
  \centering  \includegraphics[trim={0 0 0 0},clip, width=1\linewidth]{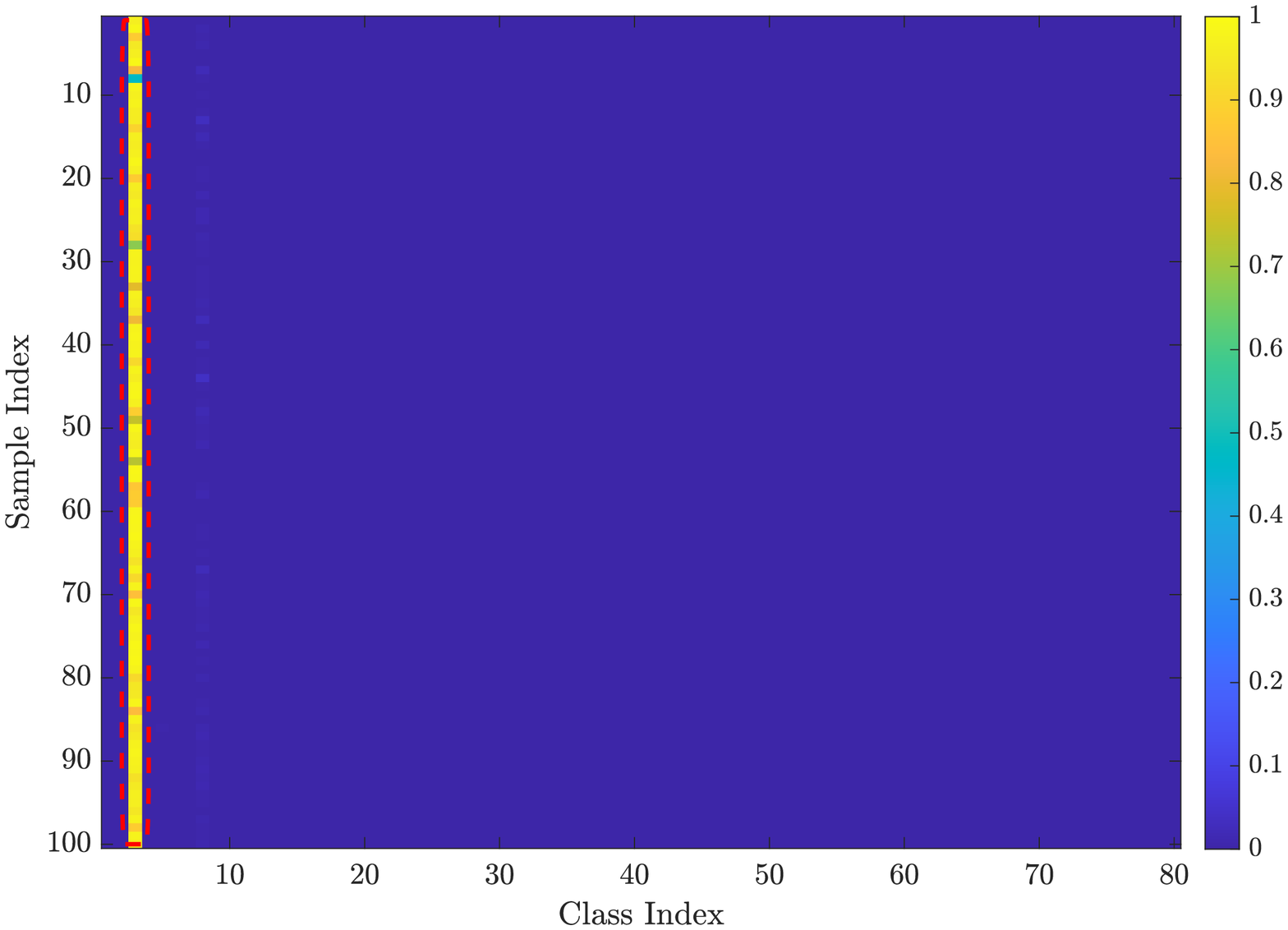}
  \caption{Output scores at Scale: 20\%}
  \label{fig:carSNR_b}
\end{subfigure}%
\begin{subfigure}[b]{.32\textwidth}
  \centering  \includegraphics[trim={0 0 0 0},clip, width=1\linewidth]{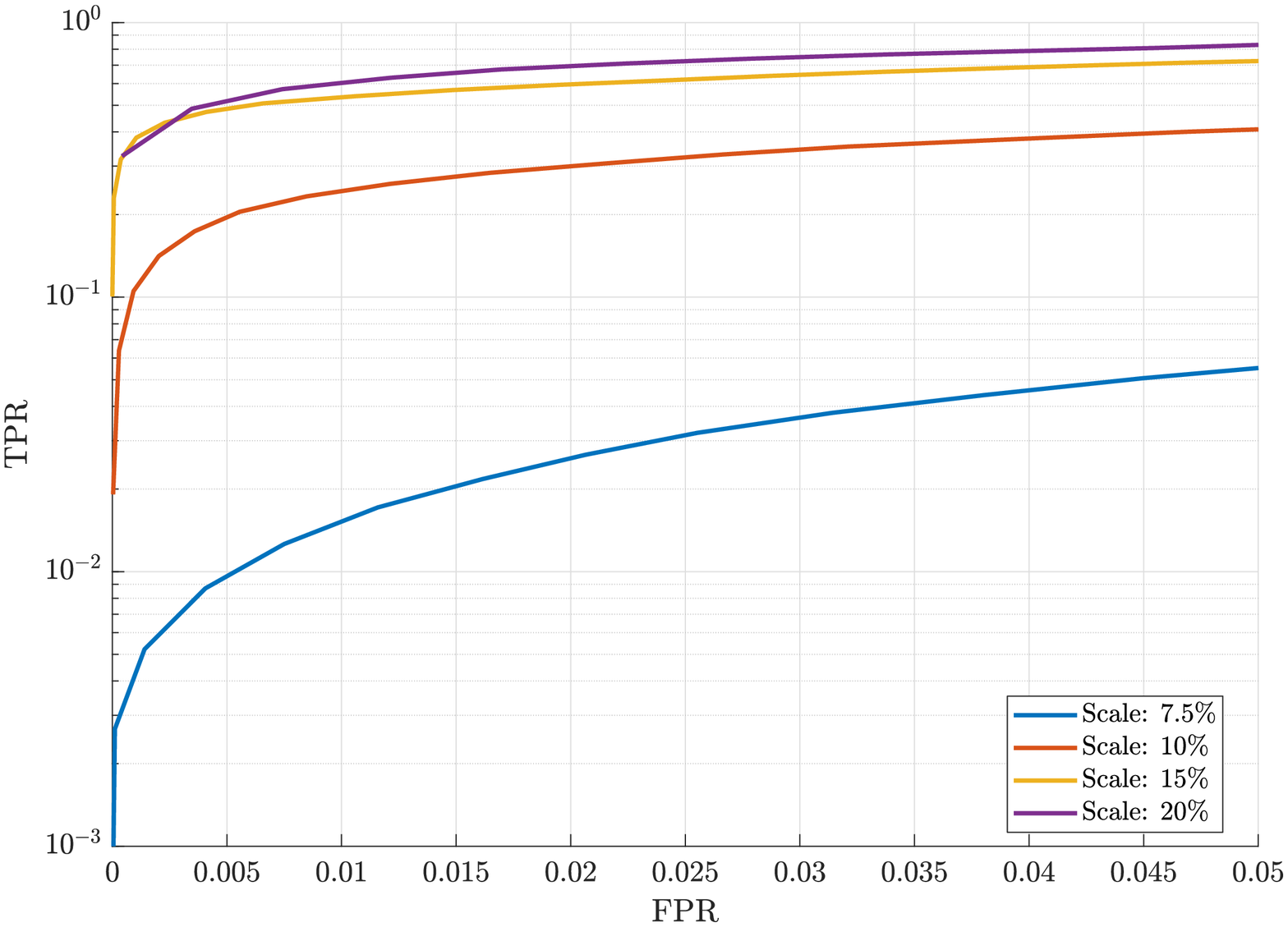}
  \caption{Average ROC curves}
  \label{fig:carSNR_c}
\end{subfigure}%
\caption{Output scores and the ROC curves generated by YOLOv4~\cite{wang2021scaled}. Note that the correct class detection for the \textit{car} class is highlighted in (a) and (b).}
	\label{fig:carSNR_ROC}
\end{figure*}

\begin{figure}[t]
	\centering
	\includegraphics[trim = 0 0 0 0, clip, width=0.95\linewidth]{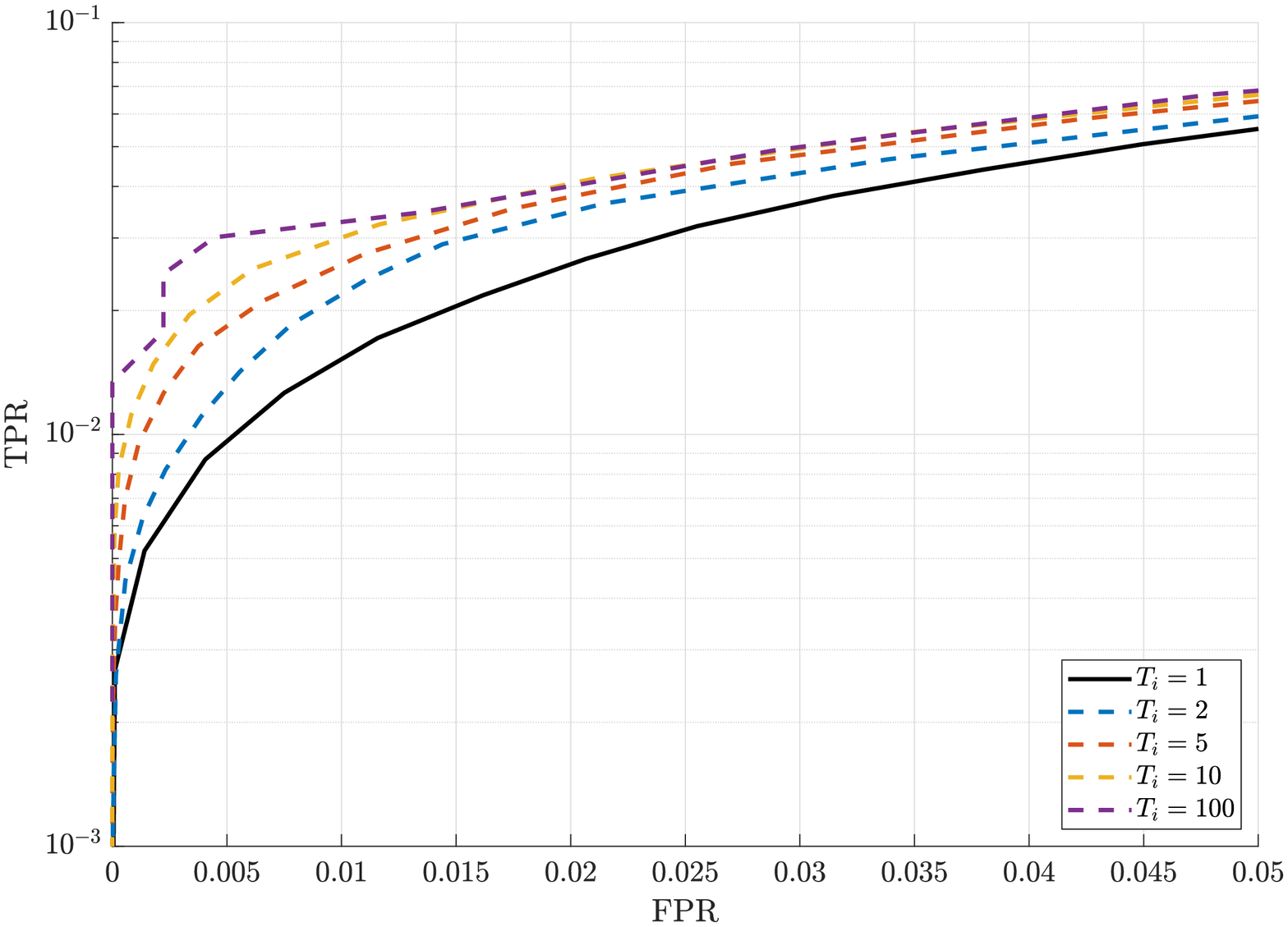}
	\caption{ROC curves for $5\%$ scaled car images at different integration lengths.}
	\label{fig:carSNR_ROC_5pct}
\end{figure}

\begin{figure}[t]
	\centering
	\includegraphics[trim = 0 0 0 0, clip, width=0.95\linewidth]{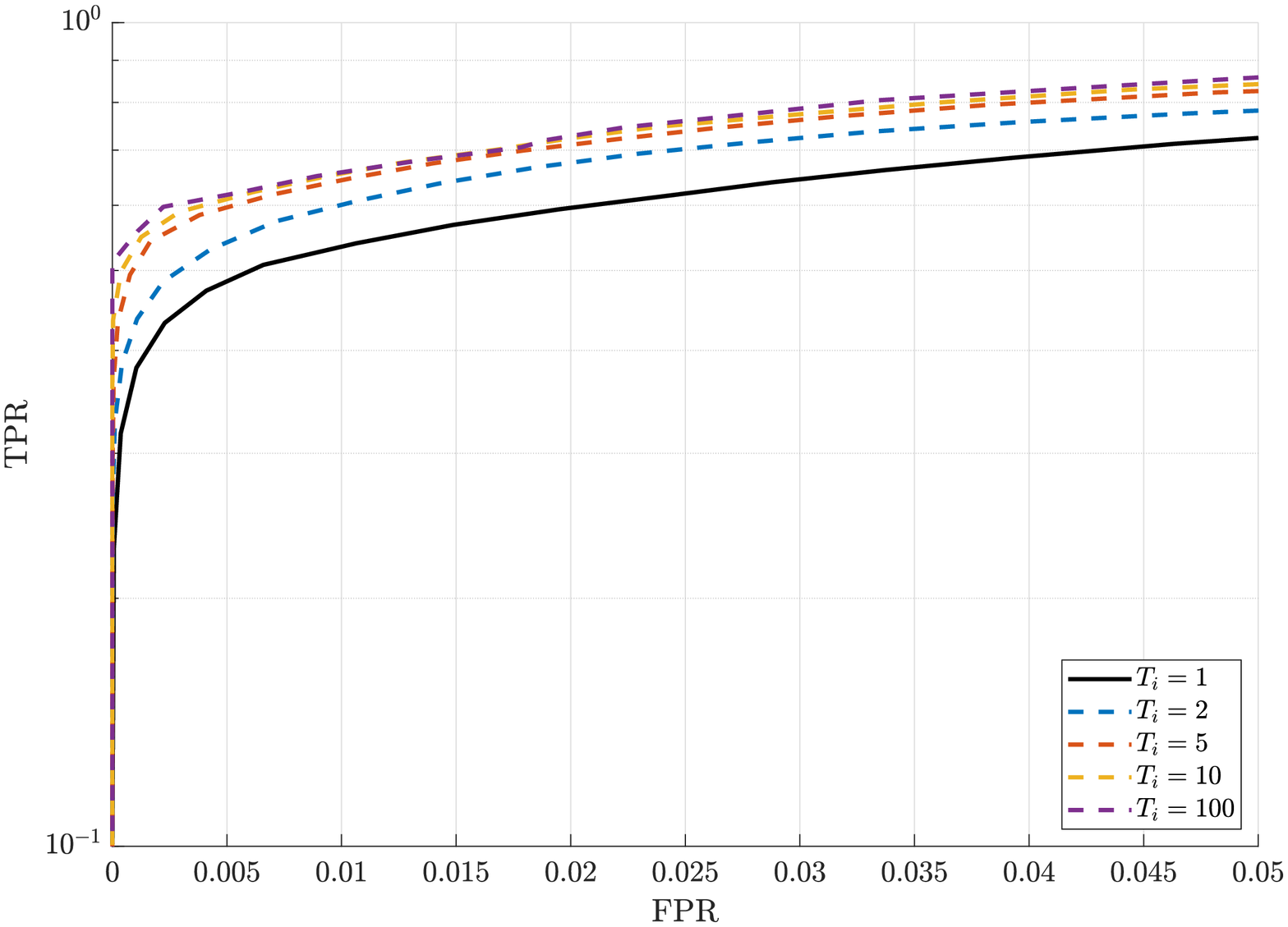}
	\caption{ROC curves for $15\%$ scaled car images at different integration lengths.}
	\label{fig:carSNR_ROC_15pct}
\end{figure}

To illustrate the effects of integrating the output scores across time, we have built a test dataset based on the images from the~\textit{Cars Dataset} given in~\cite{krause2013}, where we placed scaled versions of the car images randomly onto a stationary parking lot background image, as shown in Fig.~\ref{fig:carSNR_example}.
A total of $897$ cars randomly selected from the \textit{Cars Dataset} were placed randomly onto the background shown in Fig.~\ref{fig:carSNR_example} at different size scales, each for a total of $100$ realizations. Then, the YOLOv4 algorithm~\cite{wang2021scaled} is used to generate the output scores for each case, resulting in the output scores and the ROC curves shown in Fig.~\ref{fig:carSNR_ROC}.
As illustrated in Fig.~\ref{fig:carSNR_ROC}, the output scores and the corresponding detection performance deteriorate monotonically with the decreasing size of the object. Using the output scores generated by YOLOv4, we integrate different realizations of the same car objects for various integration window lengths. The resulting improvement in the ROC curves for scales $5\%$ and $15\%$ are illustrated in Figs.~\ref{fig:carSNR_ROC_5pct}~and~\ref{fig:carSNR_ROC_15pct}, respectively, which clearly show the expected monotonic increase in the detection performance as the integration window length increases.

Integration across time also means a delay in producing the desired detections. Hence, depending on the desired $FPR$ and maximum allowable \textit{time-on-target}, the raw score output can be integrated to achieve the maximum possible $TPR$. With a semantically-aware approach, these parameters can be independently defined or estimated for different types of targets (e.g., time-on-target can be a function of the coherence time and the maximum delay) and environmental factors as well. This idea will be illustrated via case studies in Section~\ref{sec:CaseStudies}.

\section{Subspace Tracking of Vector Attributes}
\label{sec:AttrTrack}
The overall goal of the semantic extractor framework shown in Fig.~\ref{fig:semantic_extractor} is to provide reliable and smooth semantic signals while also identifying time indices of significant innovation for scheduling transmission or storage events. The potential fidelity improvement provided by the time integration method only works over short time periods where the semantic information content within the signal does not change (analogous to coherence time in wireless channels) significantly. The following blocks in Fig.~\ref{fig:semantic_extractor}, starting with the \textit{attribute tracking} aim to smooth the semantic signal and identify points of significant innovation over extended periods of time. 

In the raw semantic extraction block shown in Fig.~\ref{fig:semantic_extractor}, the semantic data are organized to include goal-filtered components and their corresponding attributes. In a typical computer vision problem, these attributes can be position, speed for scalar attributes, sub-feature vectors for vector attributes, or even the encoded video signal itself. Tracking these attributes enables identification of strong changes at the attribute level, which in turn can be used to update the transmitted or stored attributes, even when the overarching semantic graph structure stays the same. On the other hand, if the semantic graph signal is erroneously updated (or kept the same), attribute tracking may be used to reconcile consecutive attributes to detect and correct these errors. 

\begin{figure}[!htb]
\centering
  \includegraphics[width=.95\linewidth]{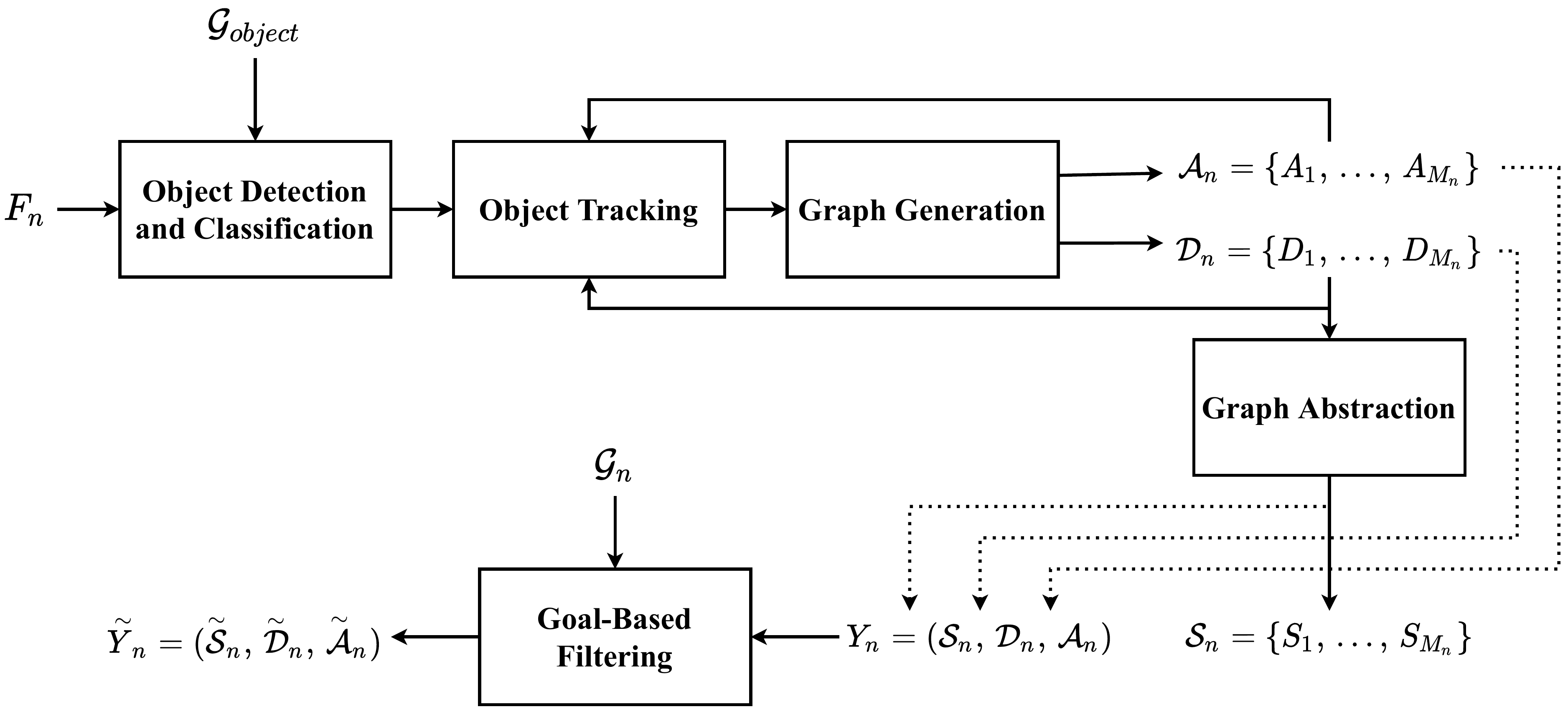}
   \caption{Semantic extraction for real-time video signals as presented in \cite{kalfa2021}.}
   \label{fig:sspframework_casestudy}
\end{figure}

We now present a subspace tracking algorithm that can be used for vector attributes. To illustrate the algorithm more clearly, we use the real-world semantic extraction of video-streams presented in~\cite{kalfa2021} where we define signal components as detected objects in the stream. As illustrated in Fig.~\ref{fig:sspframework_casestudy}, the proposed semantic extractor in~\cite{kalfa2021} utilizes YOLOv4-CSP~\cite{wang2021scaled} model to detect occurring objects in each input frame $F_n$. For temporal extension, we adopt \textit{tracking by detection} paradigm and utilize DeepSORT~\cite{deepsort} to track individual objects in the video stream which utilizes Kalman filter to recursively predict future positions of detected objects. Furthermore, DeepSORT inherits another small convolutional neural network model trained on re-identification
task to extract visual feature vectors of each detected object to be used for association of existing tracks and recent detections. We refer readers to~\cite{deepsort, aharon2022bot, zhang2021bytetrack, du2022strongsort} for details about tracking by detection paradigm. We denote these feature vectors with $\boldsymbol{r}_i \in \mathbb{R}^{128}$, where $||\boldsymbol{r}_i||_{2} = 1$, and utilize them as second-level vector attributes of component nodes in our semantic graph output (see~\cite{kalfa2021} for details). We would like to point out that the feature vector $\boldsymbol{r}_i$ for a particular detected component is a 128-dimensional vector and carries most of the attribute-level \emph{innovation} in the semantic signal processing framework. In Fig.~\ref{fig:sspframework_casestudy}, the object-level goal $\mathcal{G}_{object}$ restricts the components and the predicates that must be detected. The corresponding \textit{multi-graph instance representation} $\mathcal{D}_n = \left \{ D_1, \ldots, D_{M_n} \right \}$ composed of $M_n$ disconnected subgraphs and a corresponding attribute set $\mathcal{A}_n = \left \{ A_1, \, \ldots, \, A_{M_n} \right \}$ extracted from the video signal. The \textit{multi-graph class representation} $\mathcal{S}_n$ is constructed in the Graph Abstraction block, and the complete semantic description $Y_n = (\mathcal{S}_n, \, \mathcal{D}_n, \, \mathcal{A}_n)$ is generated. Finally, if there are external goals $\mathcal{G}_{n}$, the semantic description is filtered to obtain the goal-filtered semantic description.

\begin{figure*}[t]
\centering
\begin{subfigure}{.32\textwidth}
  \centering
  \includegraphics[width=\linewidth]{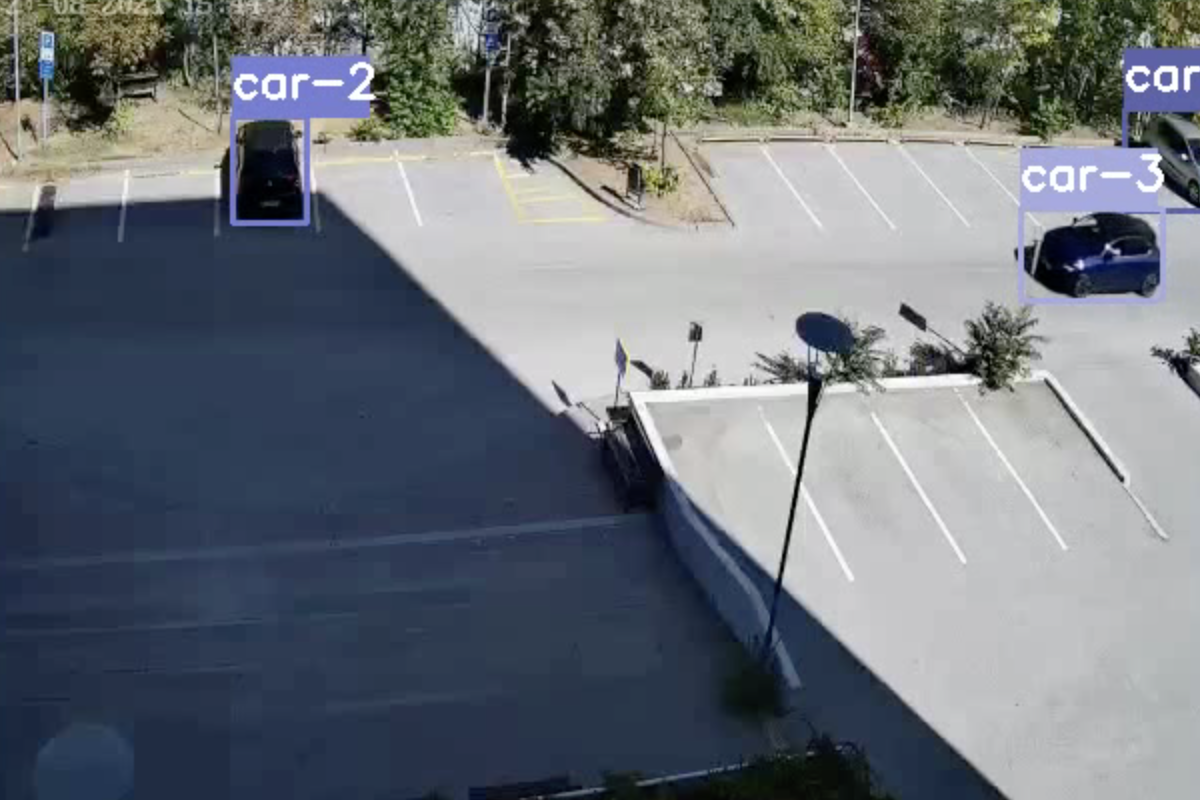}
%   \caption{car1}
  \label{fig:car_1}
\end{subfigure}%
\hspace*{.135cm}
\begin{subfigure}{.32\linewidth}
  \centering
  \includegraphics[width=\linewidth]{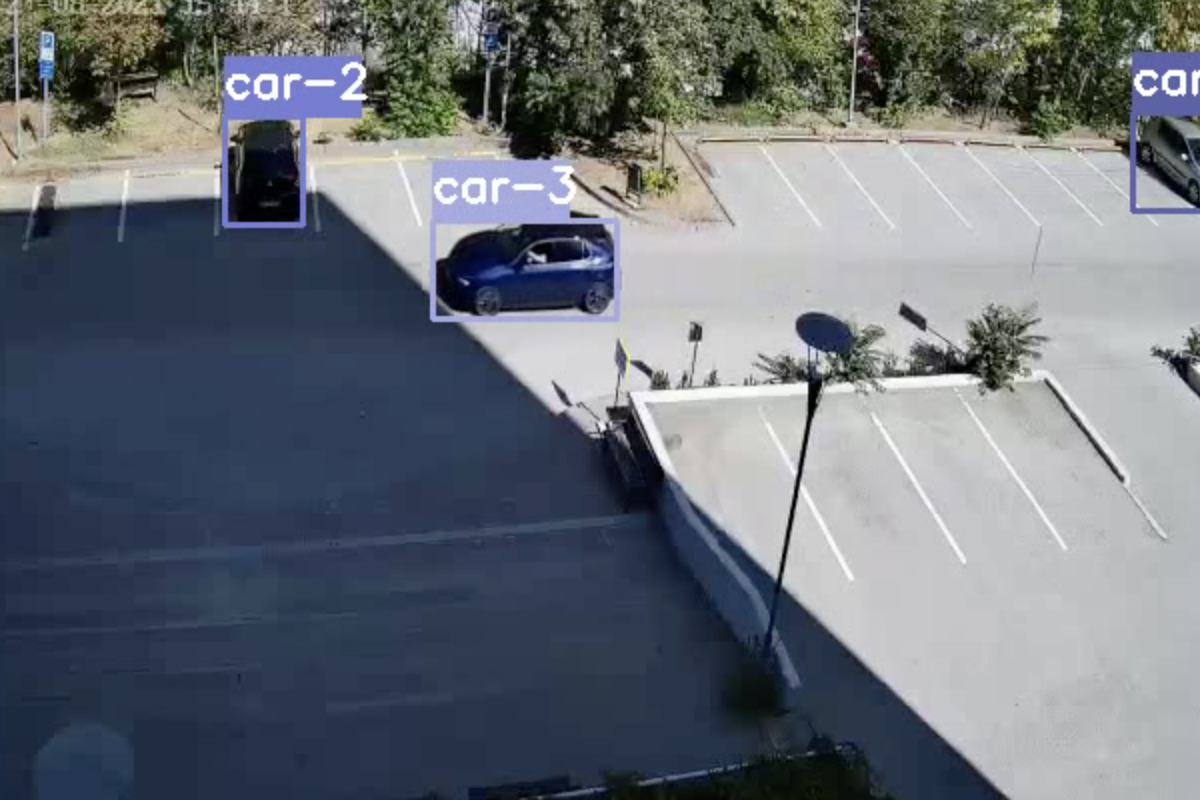}
   %\caption{}
  \label{fig:car_2}
\end{subfigure}%
\hspace*{.135cm}
\begin{subfigure}{.32\linewidth}
  \centering
  \includegraphics[width=\linewidth]{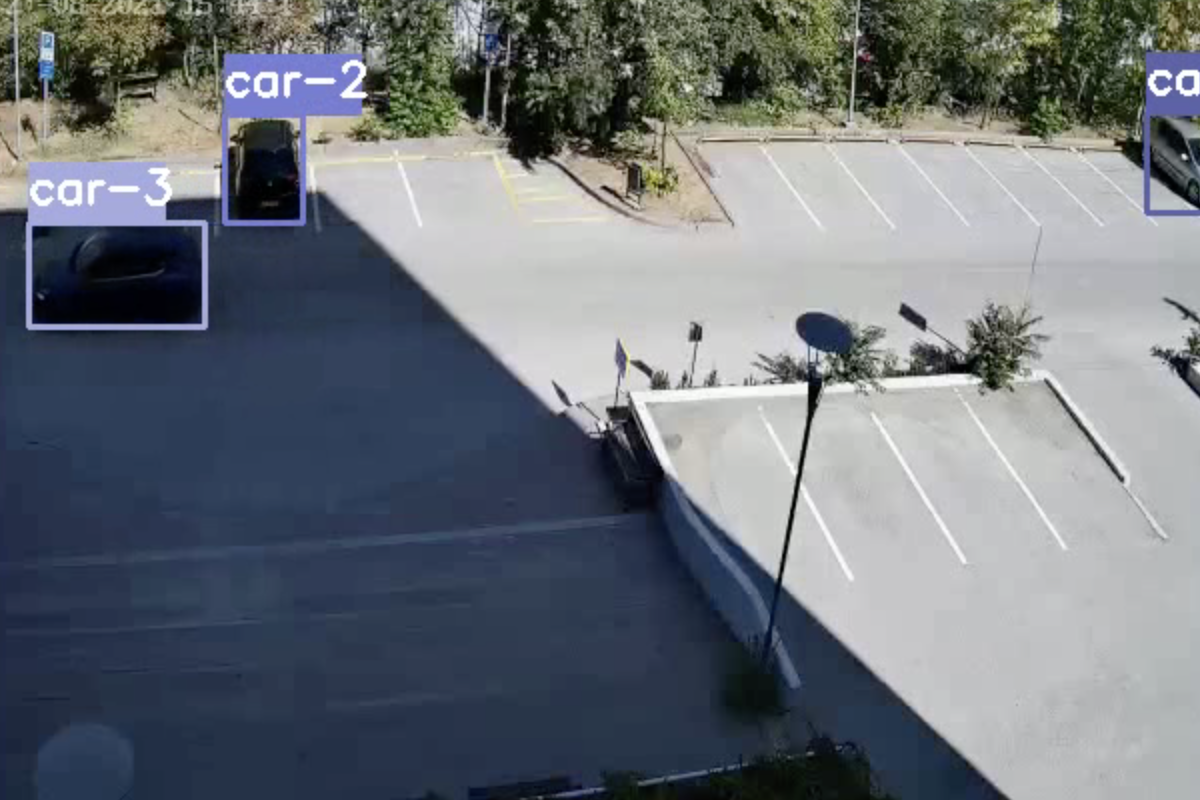}
  %\caption{}
  \label{fig:car_3}
\end{subfigure}%
\caption{A car is passing from a shaded region to a sunny region. The middle and right sub-figures demonstrate the \emph{innovation} event at the attribute level, while the semantic components stay the same.}
\label{fig:car}
\end{figure*}
To process and track attribute vectors efficiently, we use dimensionality reduction techniques such as Robust PCA and subspace tracking \cite{vaswani2018}. To track the evolution of the subspace of feature vectors with respect to time, we use Fast Principal Component Pursuit (PCP) via alternating minimization \cite{rodriguez2013}. 

Defining a feature matrix $D$, with its rows being the subfeature vectors across time, and decomposing the matrix into a compressed basis matrix $L$ and residual matrix $S$, the PCP problem is formulated as 
\begin{equation}
    \argmin_{L,S} \lVert L \rVert_{*}+\lambda \lVert S \rVert_1 \quad\text{subject to}\quad D = L+S \label{originalPCP}
\end{equation}
where $\lVert L \rVert_{*}$ is the nuclear norm of matrix $L$ and $\lVert S \rVert_1$ is the $l_1$-norm of matrix $S$.
A variant of (\ref{originalPCP}) can be constructed as 
\begin{equation}
     \argmin_{L,S} \frac{1}{2}\lVert L+S-D \rVert_{F}+\lambda \lVert S \rVert_1 \quad\text{subject to}\quad \text{rank}(L) = t.
\end{equation}
Finally, we can solve the problem via alternating minimization, namely,
\begin{align}
     L_{k+1} &= \argmin_{L} \lVert L+S_k-D \rVert_{F} \;\text{subject to}\; \text{rank}(L) = t \label{sp1}\\
     S_{k+1} &= \argmin_{S} \lVert L_{k+1}+S-D \rVert_{F}+\lambda \lVert S \rVert_1. \label{sp2}
\end{align}

The only computationally demanding step is \eqref{sp1}, and it can be solved by computing a partial SVD of $D-S_k$ with $t$ components. The solution to \eqref{sp2} is \emph{element-wise shrinkage}, i.e.,  $\sign(D-L_{k+1}) \left(\left|D-L_{k+1}\right|-\lambda\right)^+$, where $(x)^+ = \max\{0,x\}$. Note that $t$ in \eqref{sp1} can be chosen by a heuristic procedure. We increment $t$ and estimate the contribution of the new singular vector. Comparing the contribution with a threshold, the algorithm stops at a particular $t$. In numerical simulations, we observed that $t$ is usually small (does not exceed 3 in our setting).

To utilize PCP to track the evolution of the feature vectors in a practical application, one can solve the alternating minimization problem given in \eqref{sp1} and \eqref{sp2} periodically on a moving window. The size of this window (buffer) depends on the application and the rate of change in the semantic content. Then, a simple heuristic approach can be taken to detect innovations by checking the $l_1$-norm of the sparse component $S$ after convergence and compare it with a predefined threshold.

\begin{figure}[!htb]
    \centering
    \begin{subfigure}[t]{\linewidth}
        \centering
        \includegraphics[trim=0 0 0 0, clip, width = 1\linewidth]{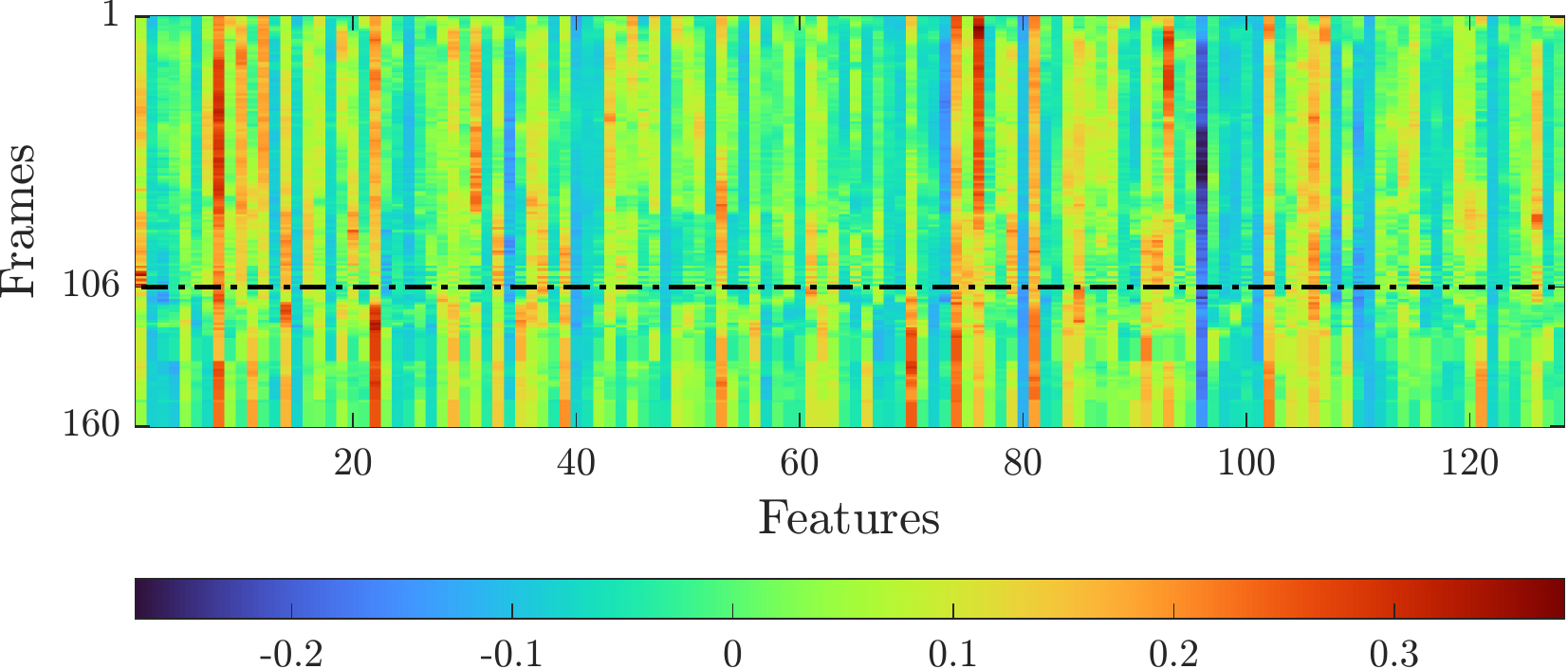}
        \caption{Correlation of subfeature vectors across consecutive frames. The innovation at the attribute level is illustrated by a black dashdotted line.}
        \label{fig:corr}
    \end{subfigure}
    \vskip 1em
    \begin{subfigure}[t]{\linewidth}
        \centering
        \includegraphics[trim=0 0 0 0, clip, width = 1\linewidth]{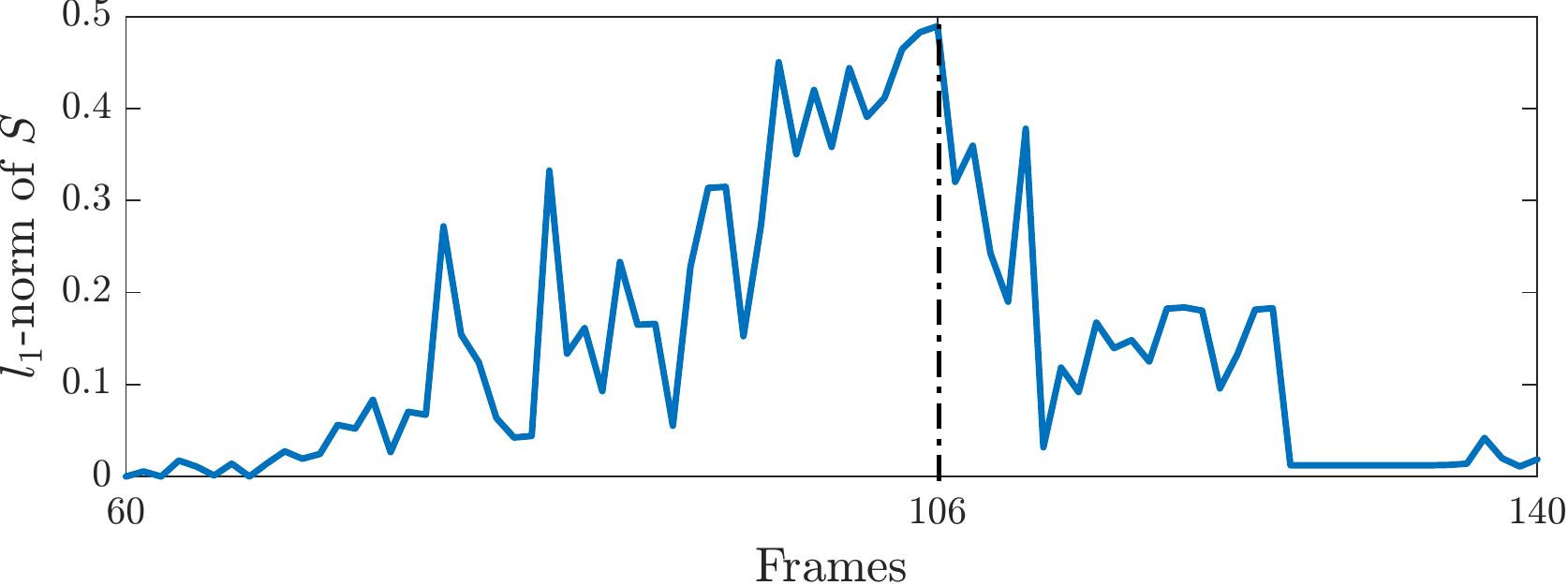}
        \caption{Frames vs. $l_1$-norm of the sparse component $S$. Note that the attribute level innovation is visible around the frame-106 since the $l_1$-norm of the sparse component achieves its maximum.}
        \label{fig:corr2}
    \end{subfigure}
    \caption{Attribute level innovation analysis for the car example given in Fig.~\ref{fig:car}}
\end{figure}

To illustrate the inherent redundancy for consecutive frames, we show the still images from a 10s video that includes a car passing from a shaded region to a sunny region in Fig.~\ref{fig:car}. The correlations of consecutive attribute vectors are given in Fig.~\ref{fig:corr}. Note that in the scenario given in Fig.~\ref{fig:car}, the semantic components and predicates stay the same (i.e., the same objects and relationships are present throughout the clip), while the attributes of the car component (car-3) change due to the image brightness, contrast, etc.
As shown in Fig.~\ref{fig:corr}, an innovation at the attribute level can be identified around frame-106, when the lighting changes for the car-3 component in the overall semantic graph signal. In Fig.~\ref{fig:corr2}, the attribute level innovation has been illustrated by analyzing the $l_1$-norm of the sparse component of the feature matrix $D$. As the norm is maximum at frame 106, we identify the innovation and locate the precise time instance quantitatively.

\section{Graph Signal Tracking using a Hidden Markov Model}
\label{sec:HMM}
After the \textit{semantic fidelity control} and \textit{attribute tracking} blocks, we now have a more reliable signal in which we can detect significant innovations at the attribute level. To provide a smooth output while also identifying significant innovations at the graph level, we introduce a stochastic model for the temporal evolution of the graph signals. More specifically, a Markov Chain can be built by defining distinct component-predicate connections in the graph representation in Section~\ref{sec:SSPprimer} as its distinct states. However, these states are not directly observable, as there may be erroneous detections due to semantic or technical noise. Therefore, we augment the Markov chain with an HMM, where the observed output sequence is generated by the semantic extractor as a function of the actual states of the underlying Markov chain.

The HMM assumption of the graph representation enables estimation of the underlying state sequence through the output sequence generated by the previous blocks. This stochastic model also enables quantifying the semantic information through the entropy rate of the Markov chain, and it can be used to compress the semantic signals. Throughout this section, we rigorously define the HMM and explain how the underlying state sequence can be extracted from the observations. 

\subsection{Markov Chains and the Hidden Markov Model} 

As shown in Fig.~\ref{fig:graph_example_Dt}, a graph signal can have various connections among component and predicate nodes. Each possible graph configuration can be defined as distinct states of the graph. Since the number of component and predicate nodes in the graph is finite (and very small in the proposed language structure of~\cite{kalfa2021} with goal filtering), the number of possible graph states is limited. The state-space of the graph can be defined as \(S = \{S_0,S_1,...,S_{N-1}\}\). With the Markovity assumption, the state transition probability of the graph can be expressed as $P_{S_j,S_i}$ for making a transition from $S_j$ to $S_i$ in a single frame
\begin{align}
     P_{S_j,S_i} = &P(Q_{t+1} = S_i | Q_{t} = S_j, Q_{t-1} = S_{Q_{t-1}}, \ldots \nonumber \\
     &,Q_{0} = S_{Q_{0}}) = P(Q_{t+1} = S_i | Q_{t} = S_j),
\end{align}
where \(Q_t\) represents the state variable at time instance $t$.

The temporal evolution of the graph can be characterized entirely through its state transition matrix (\(\textbf{A}\)) and the initial probability distribution (\(\textbf{p}\)). \(\textbf{A}\) is defined such that its $(j,i)$'th entry is equal to $P_{S_j,S_i}$, which is the probability of moving to state \(S_j\) from state \(S_i\) in a single step. The $i$'th element of (\(\textbf{p}\)), denoted by \(p_i\), which is the probability of being in state \(S_i\) in the beginning.

The HMM of the graph consists of two sequences: state sequence (\(\textbf{Q}\)) and observation sequence (\(\textbf{O}\)). The state sequence (\(\textbf{Q}\))  consists of actual states, and the output sequence (\(\textbf{O}\)) is the observed state sequence. Elements of both sequences are from the state set \(S\). At each time instance, the observed state can be the actual state, or it can be replaced with another element from the state set, leading to an incorrect observation.

\begin{figure}[t]
\centering
\includegraphics[width=0.9\linewidth]{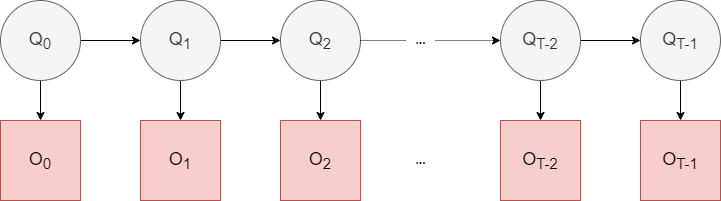}
\caption{Hidden Markov Model: \(Q_t\) is the actual state and \(O_t\) represents the observed state at time t. \(O_t\) depends solely on \(Q_t\). }  
\label{fig:hmm}
\end{figure}

In an HMM, the graph evolves as a Markov Chain according to its state transition matrix (\(\textbf{A}\)) and initial state distribution (\(\textbf{p}\)). As the graph moves between different states (\(Q_t\)) over time, the observed state (\(O_t\)) changes accordingly. \(O_t\) is a random variable which is a function of the actual state (\(Q_t\)) at time $t$. Probability of observing $S_j$ when the underlying state is $S_i$ at time $t$ is denoted by $P_{S_i,O_j}$, i.e.,
\begin{equation}
            P_{S_i,O_j} = P(O_{t} = S_i | Q_{t}= S_j).
\end{equation}

The HMM corresponding to the underlying semantic information can be represented completely through the state transition matrix (\(\textbf{A}\)), initial probability distribution (\(\textbf{p}\)) and the observation matrix (\(\textbf{B}\)). The $(i,j)$'th element of (\(\textbf{B}\)) is equal to $P_{S_i,O_j}$, which is the probability of observing \(S_j\) when the actual state is \(S_i\).

As previously mentioned, a major motivation behind the stochastic model is to solve the smoothing problem by estimating the true nature of the observed process through the noisy output of the semantic extractor. To solve the smoothing problem, we are interested in estimation of the underlying state sequence (\(\textbf{Q}\)) from the observed state sequence (\(\textbf{O}\)) when the model is completely known. This is a well-known problem in HMM framework \cite{hmm_tutorial}; however, the definition of optimality for the solution varies. There are two common approaches: The first one is to find the most likely state sequence producing the observed state sequence (\(\textbf{O}\)), for which the state sequence (\(\textbf{Q}\)) can be estimated using the Viterbi Algorithm \cite{viterbi_algo}. The second approach is to find the state sequence consisting of the most likely states at each step, maximizing the number of correctly estimated states on average. The solution to the second formulation follows from the Forward-Backward Algorithm \cite{fwbw_algo}.

The goal of the proposed framework is to find the most likely state sequence producing the observed state sequence; hence, we focus on the application of the Viterbi Algorithm. Viterbi Algorithm is a Dynamic Programming based recursive algorithm that finds the state sequence maximizing the posterior probability given the output sequence and model parameters $(P(Q|O;\mathbf{A},\mathbf{B},\mathbf{p}))$. 

Some definitions are required to describe the algorithm fully. The underlying state sequence is represented as \(Q = \{Q_0,Q_1,...,Q_{T-1}\}\) where $T$ is the sequence length. The observed state sequence corresponding to (\(Q\)) is denoted as \(O = \{O_0,O_1,...,O_{T-1}\}\). The model parameters: state transition matrix (\(\textbf{A}\)), observation matrix (\(\textbf{B}\)), and initial state distribution (\(\textbf{p}\)) are represented compactly as $\lambda$. At each time step $t$, Viterbi Algorithm considers every state $i$ visited. Then, the probability of observing \(\{O_0,O_1,...,O_{t}\}\) and terminating the path at state $i$ \(\{Q_t = S_i\}\) is maximized over the previous states \(\{Q_0,Q_1,...,Q_{t-1}\}\) \cite{viterbi_algo}, i.e.,
\begin{align}
  \delta_t(i) = \max_{\{Q_0,Q_1,...,Q_{t-1}\}} P(&Q_0,Q_1,...,Q_{t-1}, \nonumber \\
  &Q_t=S_i,O_0,O_1,...,O_t|\lambda).
\end{align}

The heart of the Viterbi Algorithm is the induction step which is used to find \(\delta_{t+1}(j)\) from \(\delta_{t}(i)\) as
\begin{equation}
\label{eqn:8}
  \delta_{t+1}(j) = [\max_{S_i}\delta_t(i) a_{i,j}] b_{j,O_{t+1}},
\end{equation}
where \(\delta_{t}(i)\) denotes the probability of most likely state sequence \(\{Q_0,Q_1,...,Q_{t-1}\}\) ending at \(\ Q_t = S_i\) given the observation sequence \(\{O_0,O_1,...,O_{t-1}\}\) for the first $t$ steps. To find the most likely state sequence, each state for each time instance must keep a pointer \(\psi_t(j)\) to the previous state maximizing~\eqref{eqn:8}. The most likely state sequence is found by backtracking the pointer of the state maximizing \(\delta_{T-1}(i)\). The complete procedure is summarized in Algorithm~\ref{alg:cap}.

\begin{algorithm}
\caption{Viterbi Algorithm}\label{alg:cap}
\begin{algorithmic}
\State $\textbf{Input: } \textbf{A,B,p,O} $ \Comment{Model parameters and output sequence}
\State $\textbf{Output: } \textbf{Q} $ \Comment{Most likely state sequence}
\State $\textbf{Initialization}$
\For{i = 0 upto N-1} 
    \State $\delta_{0}(i) = p_{i}b_{i,O_0}$ 
    \State $\psi_0(i) = -1$
\EndFor
\State $\textbf{Induction}$
\For{t = 1 up to T-1} 
\For{j = 0 up to N-1} 
    \State $ \delta_{t}(j) = \max_{i}[\delta_{t-1}(i)a_{i,j}]b_{i,O_{t}}$ 
    \State $ \psi_t(j) = \argmax_i[\delta_{t-1}(i)a_{i,j}]$
\EndFor
\EndFor
\State $\textbf{Termination}$
\State $ P^* = \max_{i}[\delta_{T-1}(i)] $ \Comment{Probability of most likely state sequence}
\State $ Q_{T-1}^* = \argmax_{i} [\delta_{T-1}(i)]$ 
\State $\textbf{Backtracking}$
\For{t = T-2 down to 0} 
    \State $Q_{t}^* = \psi_{t+1}(q_{t+1}^*)$  \Comment{Backtrack the most likely path} 
\EndFor
\State \Return $ Q^* $  \Comment{Return the most likely state sequence} 
\end{algorithmic}
\end{algorithm}

The Viterbi Algorithm finds the most likely state sequence from the observations in $\mathcal{O}(N^2T)$ steps, where $N$ denotes the number of states and $T$ is the sequence length. Hence, the computation of the algorithm can become intractable for very large number of states. However, in our proposed hierarchical graph representation explained in Section~\ref{sec:SSPprimer}, we separate the semantic graph description of the raw signal into connected atomic graphs that are expected to have a very limited number of states (components and predicates), especially considering that the proposed semantic graph description already starts with a limited component and predicate set for a specific application. Another dramatic reduction over the complexity of a practical implementation is introduced by the goal-oriented nature of the proposed framework, i.e., only a subset of the whole semantic description of a signal is of interest, and is being processed, stored, or transmitted, as discussed in Section~\ref{sec:SSPFramework}.

Even further reductions in the complexity of an HMM implementation is achievable with the use of approximate algorithms such as the M-Algorithm, which is a greedy algorithm that considers only the most likely $M$ states as the predecessor during the induction step. M-Algorithm has a complexity of $\mathcal{O}(MNT)$ where $M$ is strictly less than $N$~\cite{m_algorithm}.

%baum-welch
Aside from the computational complexity, another important aspect of successfully implementing an HMM-based algorithm is the accurate estimation of the model parameters. In this work, we assume that the parameters of HMM such as the state transition probabilities are known a priori. For practical applications, the model parameters can be initialized with available logical prior information (some state descriptions can be logically impossible etc.), then can be estimated from the observations to provide an accurate model for HMM. The Baum-Welch algorithm is a popular method for the model estimation problem in the HMM implementations~\cite{bw_algorithm}. The algorithm starts with an initial guess on the model parameters. It finds the new parameters iteratively by maximizing the observation likelihood for the estimated parameters at the current step. This two-step procedure is repeated until the model parameters converge.

%case for HSMM
Another potential shortcoming of the proposed HMM setting is the restrictions on the waiting time distributions. A Markov model enforces each state waiting time to follow a geometric distribution. In many practical cases, the observed semantic signal might contain states that cannot be modeled accurately with a geometric distribution, leading to a discrepancy between the underlying semantic signal and its Markov model. To avoid such discrepancies, HMM can be replaced with a Hidden Semi-Markov Model (HSMM). Unlike HMM, a state in HSMM might have any distribution for its waiting time, enabling the modeling of a broader range of semantic signals in practice, albeit with a higher computational cost. The observed graph sequence, which follows a HSMM, can be smoothed with a slightly modified version of the Viterbi Algorithm~\cite{hsmm_paper}.

\section{Graph Signal Smoothing using GED}
\label{sec:GED}
In the proposed semantic extraction framework, the \textit{Graph Signal Tracking} block uses an HMM-based estimation algorithm to smooth and track the semantic graph signals. However, due to the computational complexity of the state sequence estimation procedure given in Algorithm~\ref{alg:cap}, a relatively simple algorithm to identify semantic innovation and smooth erroneous detections is introduced in this section as part of the \textit{Graph Signal Smoothing} block in Fig.~\ref{fig:SSPframework}. The algorithm and the smoothing block presented here can either be used as a pre-filtering step before the \textit{Graph Signal Tracking} block to reduce the computational load on the HMM state estimation algorithm, or it can be used as an alternative to the HMM for a low complexity implementation. 

To identify actual or erroneous changes in the graph signals with relative simplicity, we use a graph distance metric for consecutive graphs across time. GED~\cite{alberto1983,gao2010survey} is one such popular metric. For two graphs $g_1$ and $g_2$, the GED is defined as follows:
\begin{equation}
    GED(g_1,g_2)= \min_{(e_1,e_2,\ldots,e_k) \in \mathcal{P}(g_1,g_2)} \sum\limits_{i=1}^{k} c(e_i),
    \label{eq:GED_1}
\end{equation}
where $\mathcal{P}(g_1,g_2)$ is the set of all edit paths that transform $g_1$ into $g_2$, $e_i$'s are the elementary edit operations, and $c(e_i)$ is the cost of performing the edit $e_i$. Elementary edit operations include the insertion, deletion, and substitution of nodes and edges. Note that with the semantic language defined in Section~\ref{sec:SSPprimer}, we have a bipartite graph structure whose edges are always strictly dependent on the predicates that connect different components. Therefore, for the proposed language, we need to define edit costs only for the nodes of the bipartite graphs.

In most applications of the GED, the cost functions are chosen to be constant scalars depending on the type of operation~\cite{gao2010survey,blumenthal2020exact}. In this work, we define the edit costs to be a function of the confusion metrics of the underlying semantic extractor. More specifically, we exploit the known statistical characteristics of the semantic extractor that generates the graphs to obtain the GED as a statistical similarity between two consecutive graphs. As such, we define the cost as
\begin{equation}
    c(e_i)= -\log(P(e_i)),
    \label{eq:GED_2}
\end{equation}
where $P(e_i)$ is the probability of the edit $e_i$ occurring at the output of the semantic extractor. Therefore, the GED becomes an estimator for the probability of obtaining the new graph $g_2$, given the initial graph $g_1$. Note that this probability can be estimated using the confusion matrix and statistical metrics given in~\eqref{eq:CM}--\eqref{eq:TNR}. The relationships between elementary edits and corresponding confusion metrics are given in Table~\ref{tb:GED_CM}. Note that, if the occurrence probabilities of certain graph realizations are available, posterior probabilities can be used as the distance metrics using Bayesian inference. 
\begin{table}[htb]
\caption{GED Cost Definitions}
\label{tb:GED_CM}
\centering
\begin{tabular}{c|c}
\textbf{Elementary Edits}           & \textbf{Confusion Metric} \\ \hline
Node Insertion - $i$                & $FPR_i$                   \\
Node Deletion - $i$                 & $FNR_i$                   \\
Node Substitution $i \rightarrow j$ & $n_{i,j}/N_i$             \\
Node Substitution $j \rightarrow i$ & $n_{j,i}/N_j$            
\end{tabular}
\end{table}

If the confusion metrics for the composite edit paths between $g_1$ and $g_2$ are available, they can be used to estimate the GED cost directly. However, if only elementary edit confusion metrics are available, the GED given in~\eqref{eq:GED_1} becomes a rough estimate since this definition assumes that the individual edits that make up an edit path are statistically independent, as the log-probabilities add up to form the total GED. As the numerical examples and the experimental results show, this assumption is still accurate enough to track the semantic information and identify points of significant innovation. In Section~\ref{sec:CaseStudies}, we present several case studies involving simulations and experimental results that showcase the potential GED as an elegant pre-filtering step to filter semantic noise and provide indications of strong semantic innovation.

\section{Case Studies}
\label{sec:CaseStudies}

In this section, we present case studies for each of the techniques through simulations and experimental results. For showcasing the methods that rely on the semantic extractor's confusion matrix without having to limit ourselves to a potentially limited training set, we have constructed a simulation framework that can generate a semantic extractor model with random characteristics. This extractor is then used to produce noisy semantic signals, given a randomly generated time evolution of a noiseless semantic signal as the ground truth. For experimental demonstrations, we use the semantic extractor given in~\cite{kalfa2021}, which is a YOLOv4-based algorithm that generates noisy semantic signals from video signals. The semantic extractor in~\cite{kalfa2021} uses as its component list the YOLOv4 class labels (see~\eqref{eq:CP_C}). Then, three predicates (\textit{exists, moving together, conjunct}) are defined and detected among the components to generate the full bipartite graph outputs. Each component node in the semantic graph also has an embedded attribute list with $L_{n_j}=3$ according to~\eqref{eq:theta_ti}. The first level attribute contains the bounding box position and velocities, the second level attribute contains vector subfeatures of length $128$, and the third level attribute includes the raw image of the detected bounding box. More details on these definitions can be found in~\cite{kalfa2021}.

Using the extractor described above, we generate noisy semantic signals from several short clips that we recorded. These noisy signals are then post-processed with the proposed methods, and their performance is demonstrated.

\subsection{Semantic Fidelity with Time Integration}
\begin{figure}[tb]
	\centering
	\includegraphics[width=0.95\linewidth]{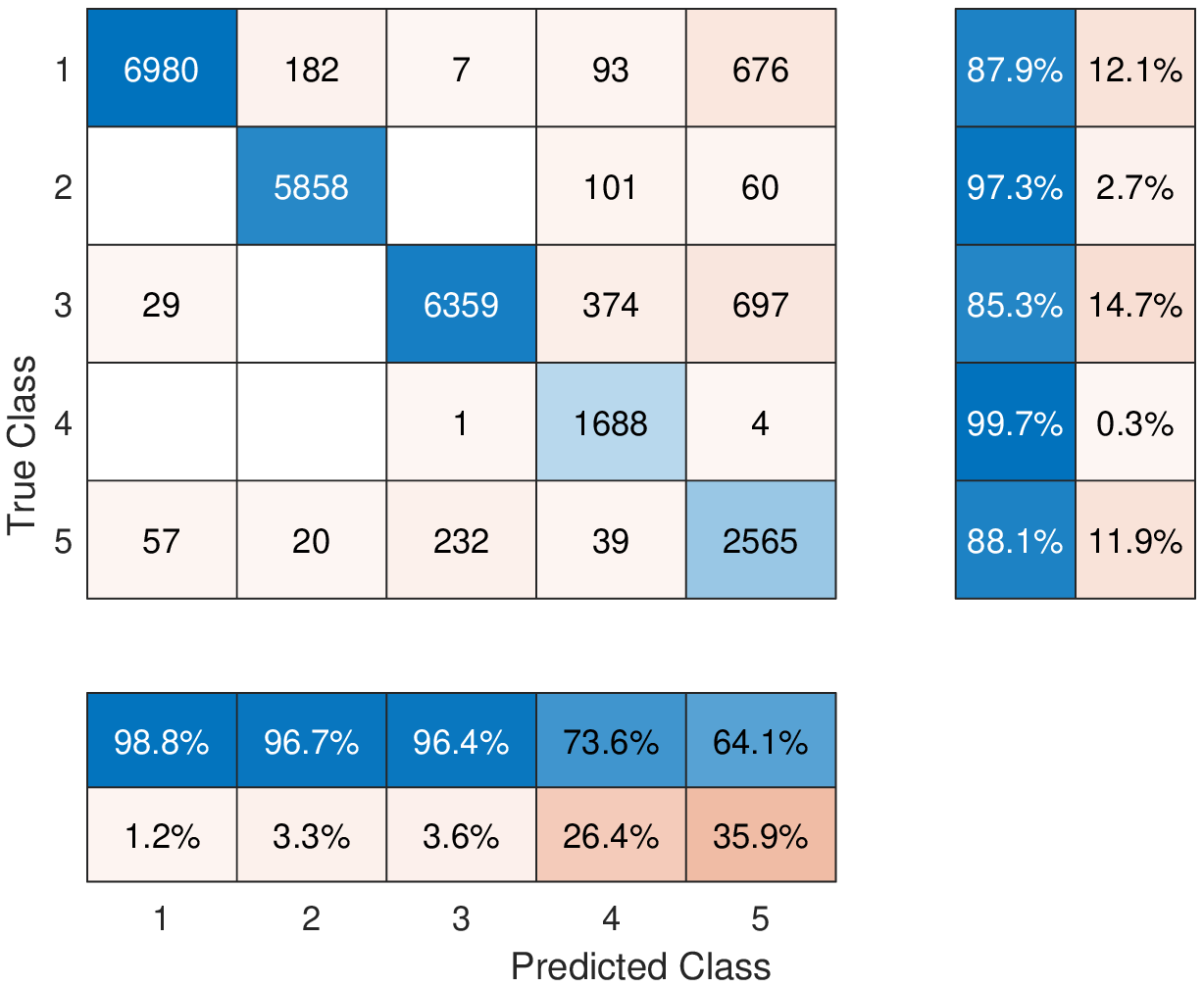}
	\caption{Confusion matrix from a randomly generated semantic extractor with $\tau=0.5$. The additional tables at the right and bottom side denote the accuracy rates of each row and column, respectively.}
	\label{fig:TI_randomextractor}
\end{figure}

\begin{figure}[tb]
	\centering
	\includegraphics[width=0.95\linewidth]{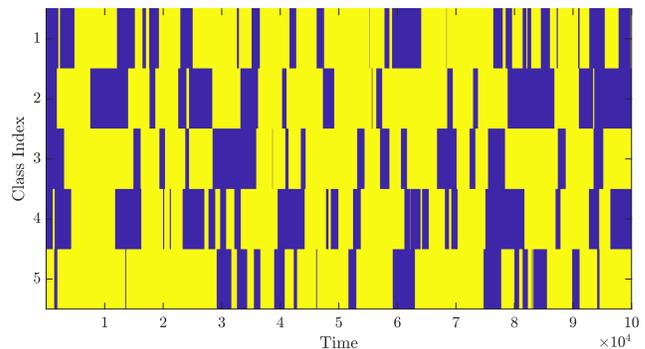}
	\caption{Randomly generated time evolution for the semantic ground truth. Yellow regions denote where the patterns exist.}
	\label{fig:timeevolution}
\end{figure}

\begin{figure}[tb]
	\centering
	\includegraphics[trim=0cm 0 0cm 0cm, clip, width=0.95\linewidth]{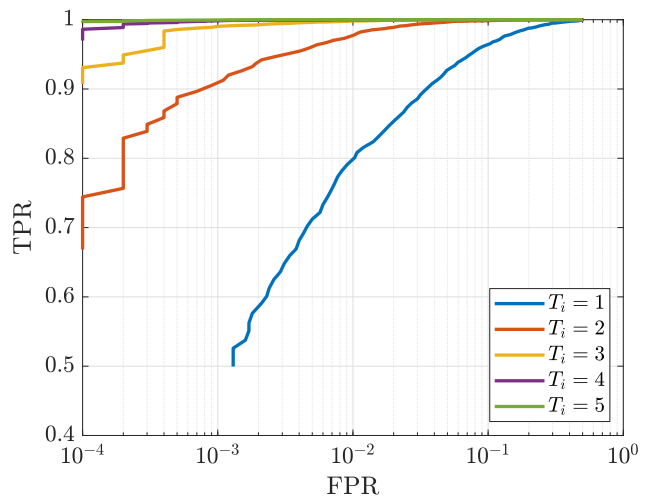}
	\caption{Improvements in the ROC curves of Class-3 using time integration of the output scores at different window lengths.}
	\label{fig:TI_ROC_improvement}
\end{figure}

\begin{figure}[tb]
	\centering
	\includegraphics[trim = 0cm 0 0cm 0, clip, width=0.95\linewidth]{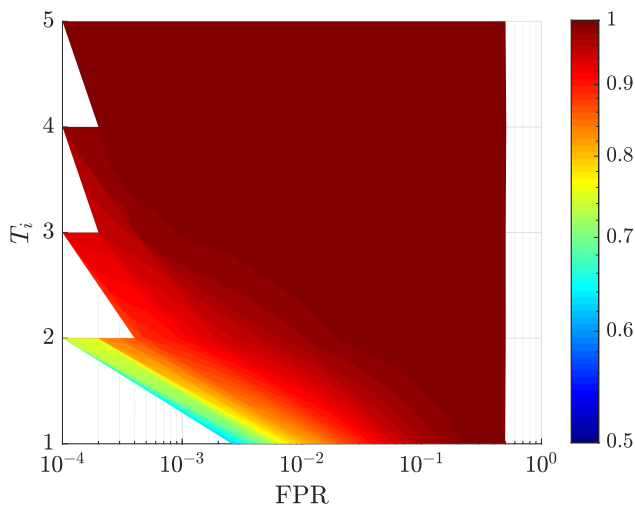}
	\caption{Heatmap of TPR values for different time integration windows and FPR values.}
	\label{fig:TI_heatmap}
\end{figure}

Using the simulation framework described above, we generate a random semantic extractor that can detect five different semantic patterns. The normalized output scores of this random extractor are thresholded at different levels to generate the confusion matrices and the ROC curves for its output patterns. An example confusion matrix with the threshold set at $\tau = 0.5$ is shown in~Fig.~\ref{fig:TI_randomextractor}.
With the semantic extractor characterized in Fig.~\ref{fig:TI_randomextractor}, we generate a random time series of inputs of length $10,000$ as shown in Fig.~\ref{fig:timeevolution} and the corresponding output scores. 
Then, a moving integration window of lengths 1-to-5 is implemented to observe the effects of time integration on the detection performance. In Figs.~\ref{fig:TI_ROC_improvement}~and~\ref{fig:TI_heatmap}, we present the improvements in ROC curves for the detection of class-3 ($C_3$). As seen in both graphs, time integration of the extractor output greatly improves the detection performance in this particular scenario. Note that the integration window lengths, TPR, and FPR values presented here are to give a general idea on how time integration affects the performance of the extractor. In practice, these values depend on the input signal acquisition characteristics and the time evolution of the actual semantic content of the signals. On a case-by-case basis, parameter surveys as shown in Fig.~\ref{fig:TI_heatmap} can be performed to decide on a \textit{time-on-target} parameter to maximize the detection performance in a class-aware fashion.

\subsection{Attribute Subspace Tracking}
\label{subsec:attr}
\begin{figure}[!htb]
    \centering
    \includegraphics[width=\linewidth]{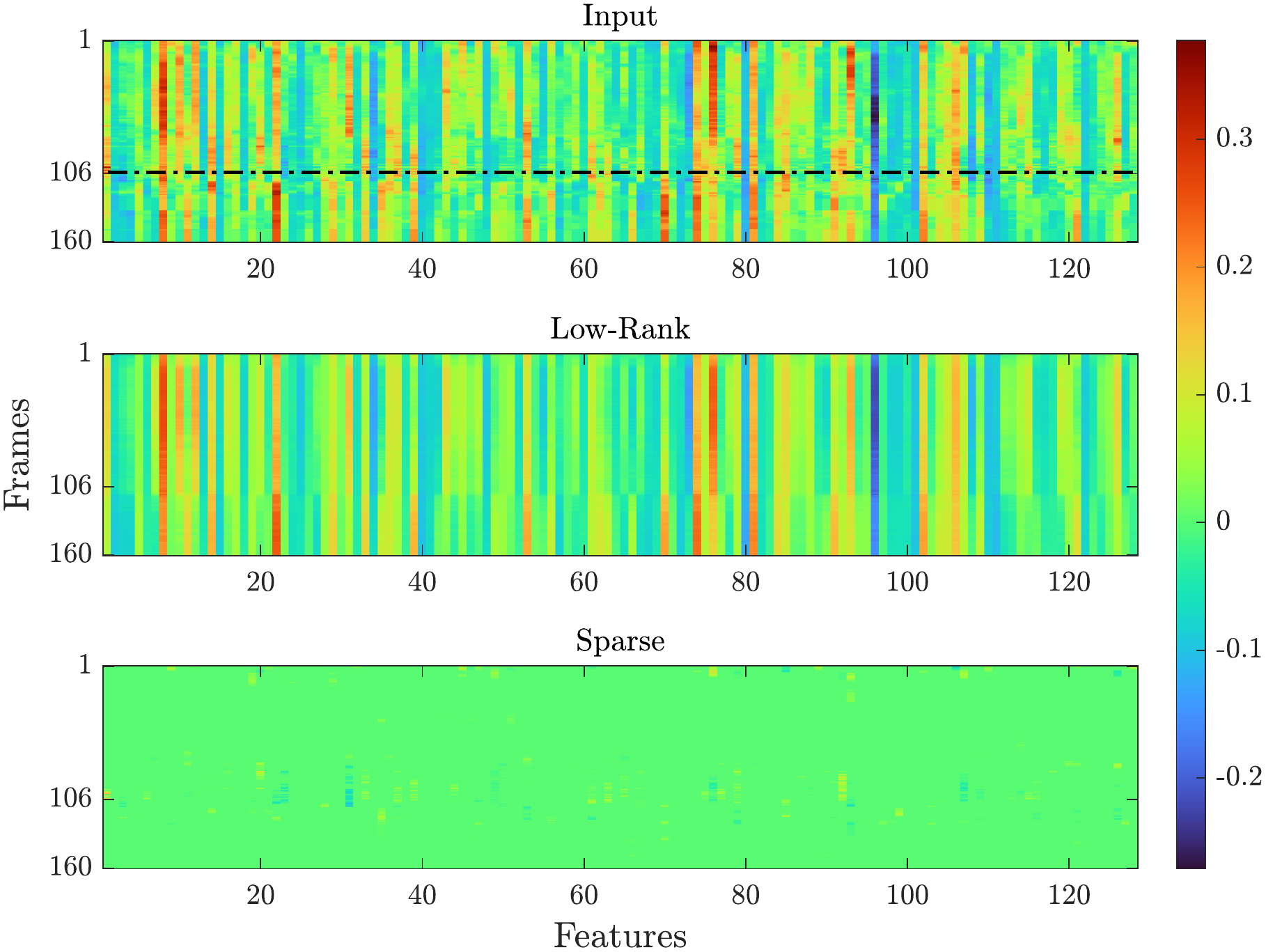}
    \caption{PCP output of the car video}
    \label{fig:pcp_output}
\end{figure}

To illustrate the vector subspace tracking method developed in Section~\ref{sec:AttrTrack}, we use the same recorded video example shown in Fig.~\ref{fig:car}. Running the PCP algorithm given in \eqref{originalPCP}--\eqref{sp2} on the video clip, we obtain the subspace breakdown of vector attributes shown in Fig.~\ref{fig:pcp_output}.
As it is seen in the sparse component plot, we have two innovations, i.e., when the car first appears in the scene and when it passes to the shaded region (around frame 106). By plotting the $l_1$-norm of the sparse component, we can track the innovation versus time. 

\begin{figure}[!htb]
    \centering
    \includegraphics[width = \linewidth]{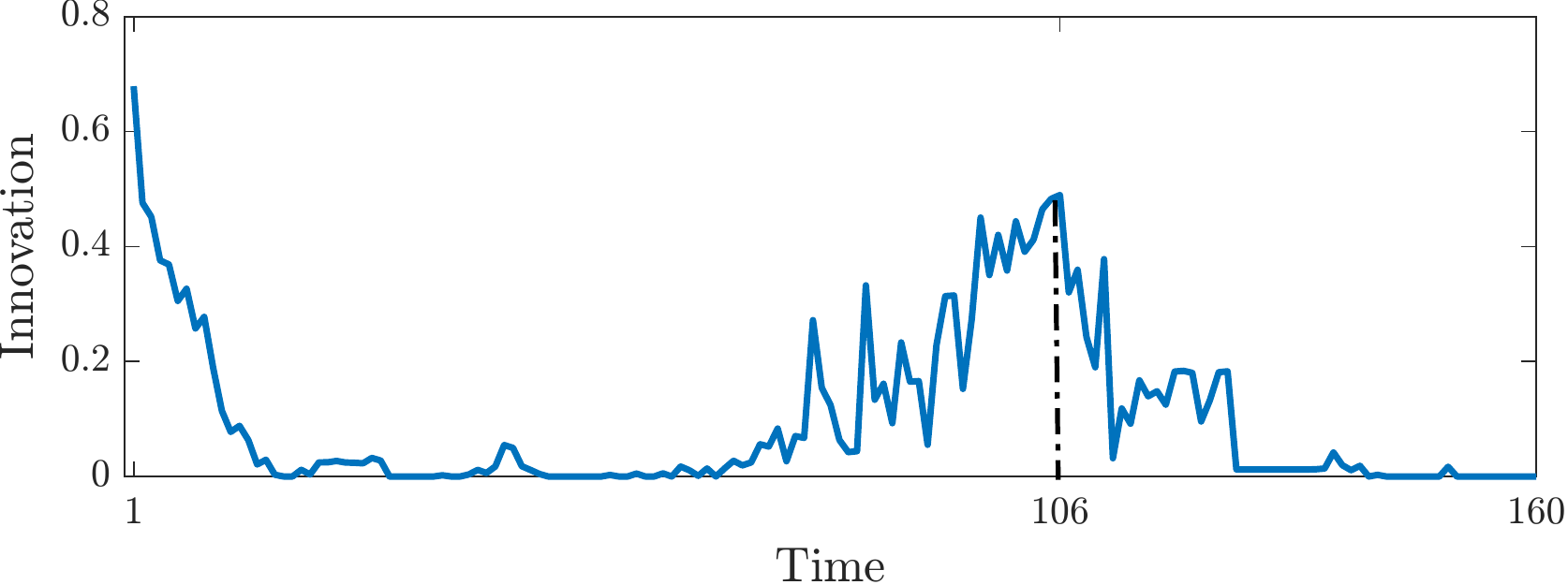}
    \caption{Innovation vs time for the car video}
    \label{fig:innovationvstime}
\end{figure}

\begin{figure}[!htb]
    \centering
    \includegraphics[width = \linewidth]{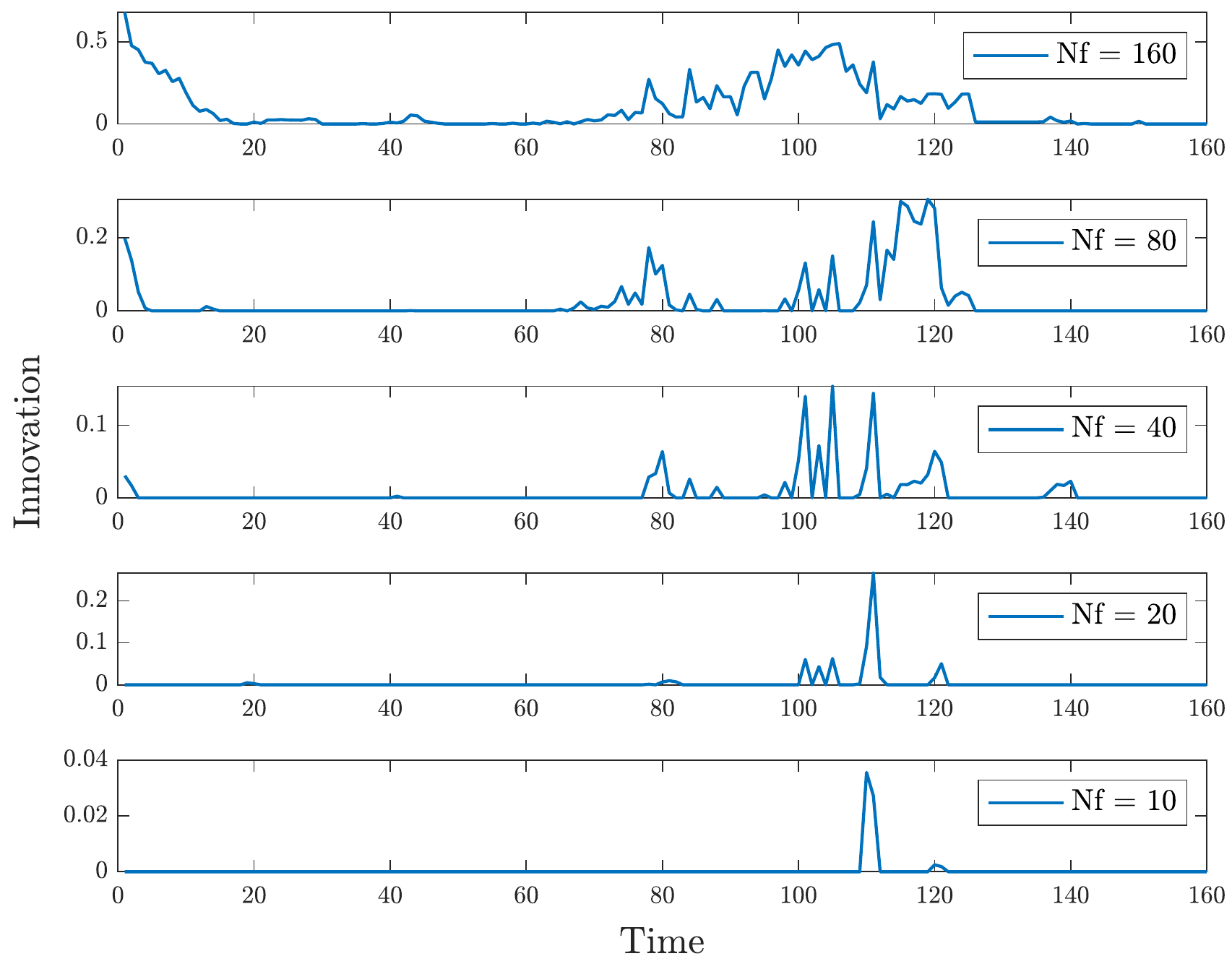}
    \caption{Innovation vs time for various buffer sizes for the car video}
    \label{fig:innovationvstimedifferentbuffers}
\end{figure}

To design an attribute tracking algorithm that can work in real-time, one can use fixed-sized buffers and run the algorithm continuously. Again, we can track the $l_1$-norm of the sparse components to detect an attribute-level innovation in the scene, as shown in Fig.~\ref{fig:innovationvstime}. Moreover, Fig.~\ref{fig:innovationvstimedifferentbuffers} illustrates the effect of buffer size on the innovation detection. As expected, smaller buffer sizes result in highly concentrated, spike-shaped innovations. Furthermore, the innovation at the beginning diminishes as the buffer size gets smaller. 

\begin{figure}[tb]
    \centering
    \includegraphics[width = 0.95\linewidth]{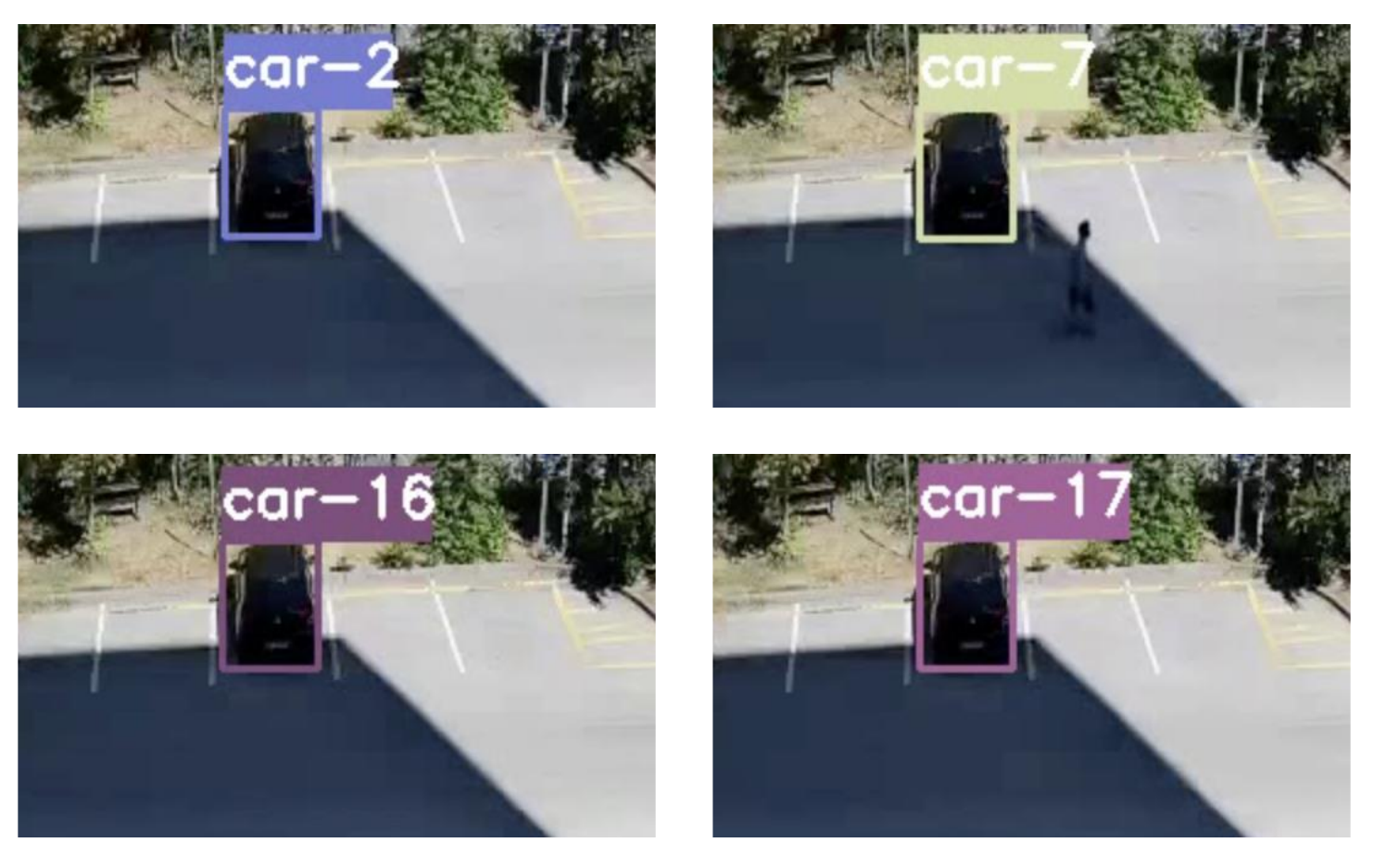}
    \caption{An object is assigned different IDs during the video}
    \label{fig:occlusions}
\end{figure}

\begin{figure}[!ht]
    \centering
    \includegraphics[width = 0.9\linewidth]{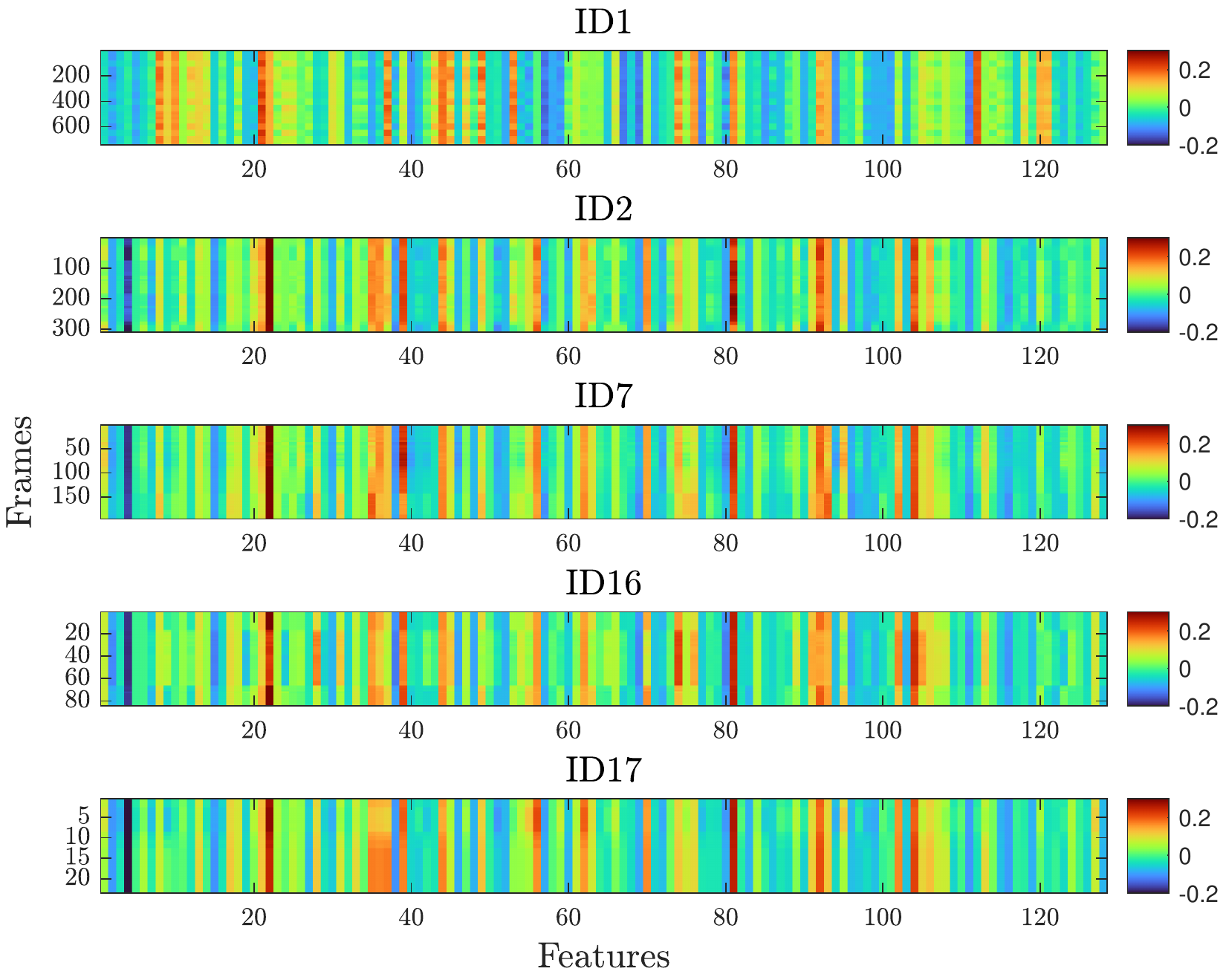}
    \caption{Low-rank features vs time different IDs}
    \label{fig:Objectreconciliation1}
\end{figure}

\begin{figure}[!ht]
    \centering
    \includegraphics[width = 0.9\linewidth]{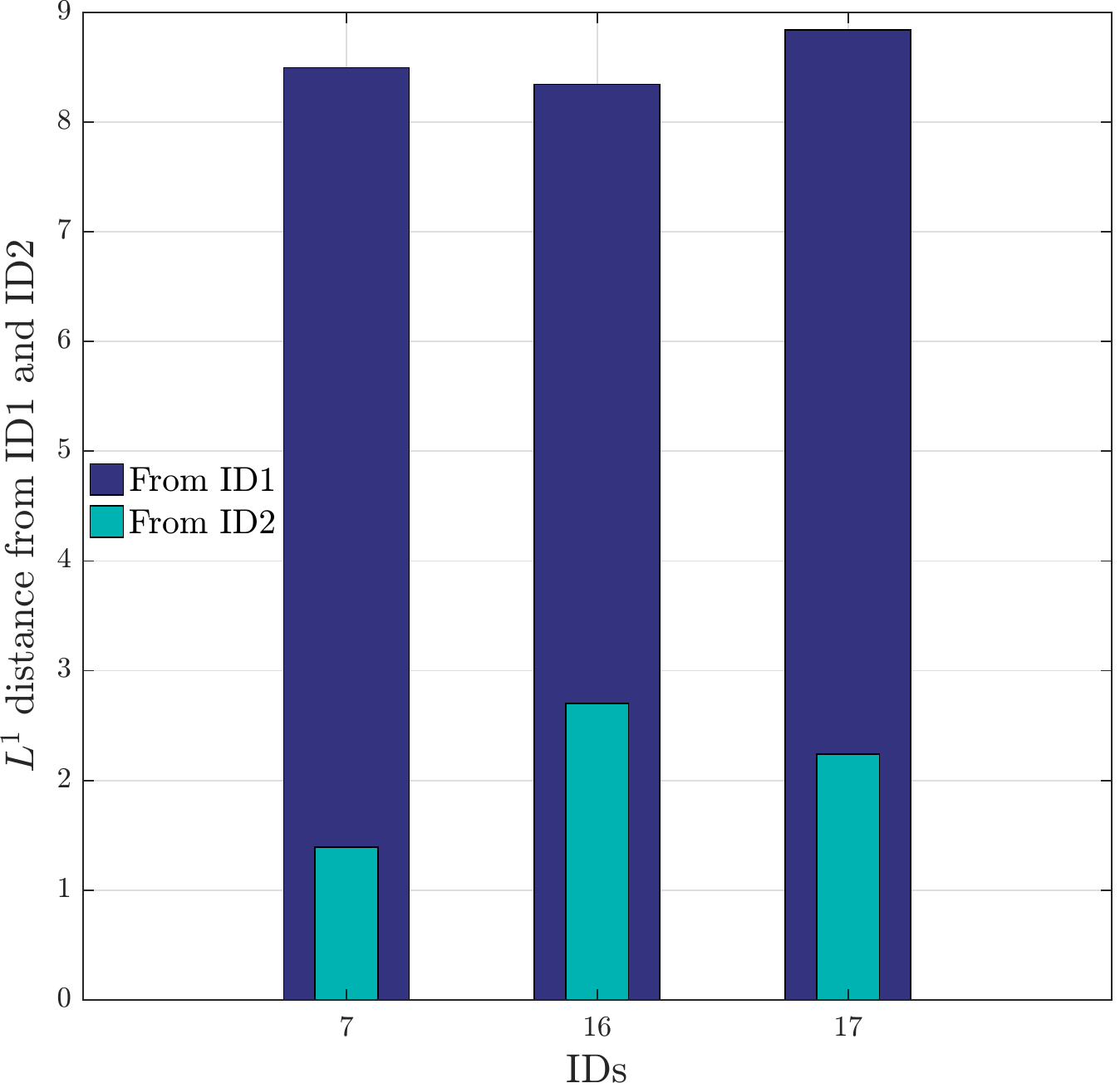}
    \caption{Innovation vs time for various buffer sizes for the car video}
    \label{fig:Objectreconciliation2}
\end{figure}

In certain scenarios where multiple instances of an object are classified with unique identifiers (ID), the same instance of the class can be detected; however, ID numbers can change due to occlusions, buffer sizes, or missed detections. We demonstrate that even in these cases, the proposed method is able to reconcile the different identifications. In Fig~\ref{fig:occlusions}, an object detected with multiple IDs is shown throughout the video. Using the PCP algorithm, we analyze the low rank components of these seemingly separate IDs in Fig~\ref{fig:Objectreconciliation1}. To reconcile these objects, we calculate the mean of low-rank components over time and compare the Manhattan distance between them, e.g., $\norm{L_7-L_1}_1$ vs. $\norm{L_7-L_2}_1$ to reconcile the instance with ID 7. The resulting distance comparison is given in Fig~\ref{fig:Objectreconciliation2}. As illustrated, IDs 7, 16, and 17 are correctly reconciled with the instance ID 2. 

\subsection{Graph Edit Distance}
\begin{figure}[ht]
	\centering
	%trim: left bottom right top
	\includegraphics[trim={0 0 0 0},clip, width=0.95\linewidth]{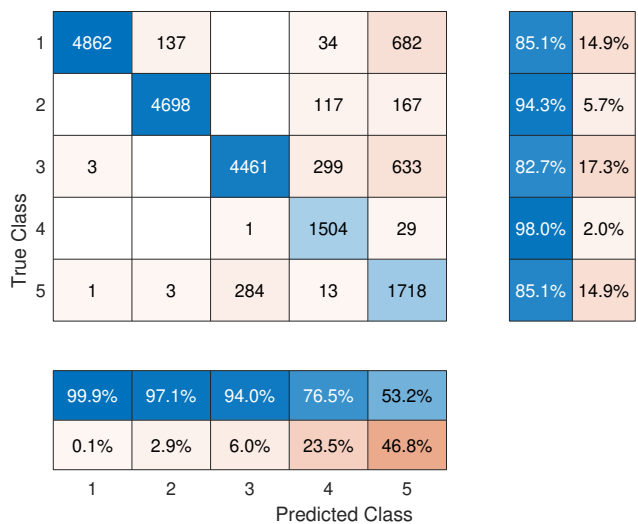}
	\caption{Confusion matrix example for a classifier with five outputs with $\tau = 0.9$. The additional tables at the right and bottom side denote the accuracy rates of each row and column, respectively.}
	\label{fig:conf_matrix_example}
\end{figure}

\begin{figure}[ht]
	\centering
	%trim: left bottom right top
	\includegraphics[trim={0 0 0 0},clip, width=0.95\linewidth]{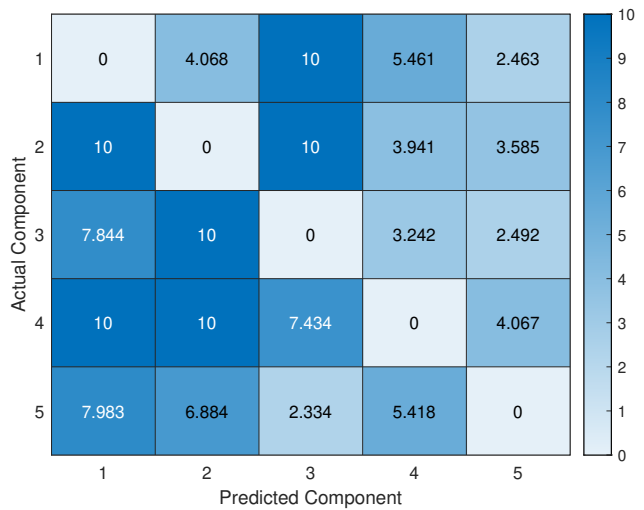}
	\caption{Node substitution costs for the confusion matrix given in Fig.~\ref{fig:conf_matrix_example}.}
	\label{fig:conf_matrix_example_GED}
\end{figure}

\begin{figure}[ht]
	\centering
	%trim: left bottom right top
	\includegraphics[trim={0 0 0 0},clip, width=0.9\linewidth]{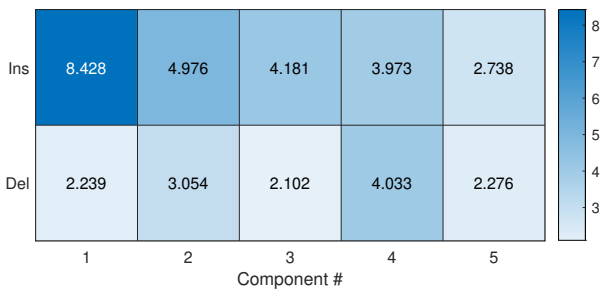}
	\caption{Node insertion/deletion costs for the GED example.}
	\label{fig:conf_matrix_example_GED2}
\end{figure}

To demonstrate the potential use of GED in a semantic extractor, we again take a simple classifier, and its confusion matrix given in Fig.~\ref{fig:conf_matrix_example}.
Note that such a simple classifier will only generate isolated nodes based on its detected classes. Therefore, the edit costs that need to be defined are the node insertion, deletion, and substitution costs, as listed in Table~\ref{tb:GED_CM}. The substitution costs are defined as the probability of confusion between any two classes; and are shown in Fig.~\ref{fig:conf_matrix_example_GED}. Note that some entries show an infinite cost, i.e., meaning some specific patterns were never confused during the training. The insertion and deletion costs are calculated using~\eqref{eq:GED_2} and Table~\ref{tb:GED_CM}, and are shown in Fig.~\ref{fig:conf_matrix_example_GED2}. These cost definitions can be used in~\eqref{eq:GED_1} to compute GED metrics for any output this particular semantic extractor can generate. 
    
\begin{figure}[ht]
	\centering
	\includegraphics[trim= 0 0 0 0, clip, width=0.95\linewidth]{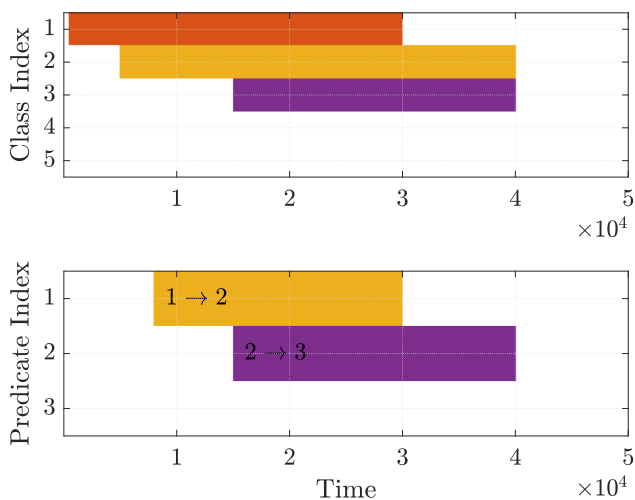}
	\caption{Simulated scenario for the classifier and predicate outputs for $50,000$ frames. The text annotations on the predicates indicate which two classes the predicate connects. }
	\label{fig:GED_groundtruth}
\end{figure}

\begin{figure*}[t]
\centering
\begin{subfigure}[b]{.45\textwidth}
  \centering  \includegraphics[trim={0 0 0 0},clip, width=1\linewidth]{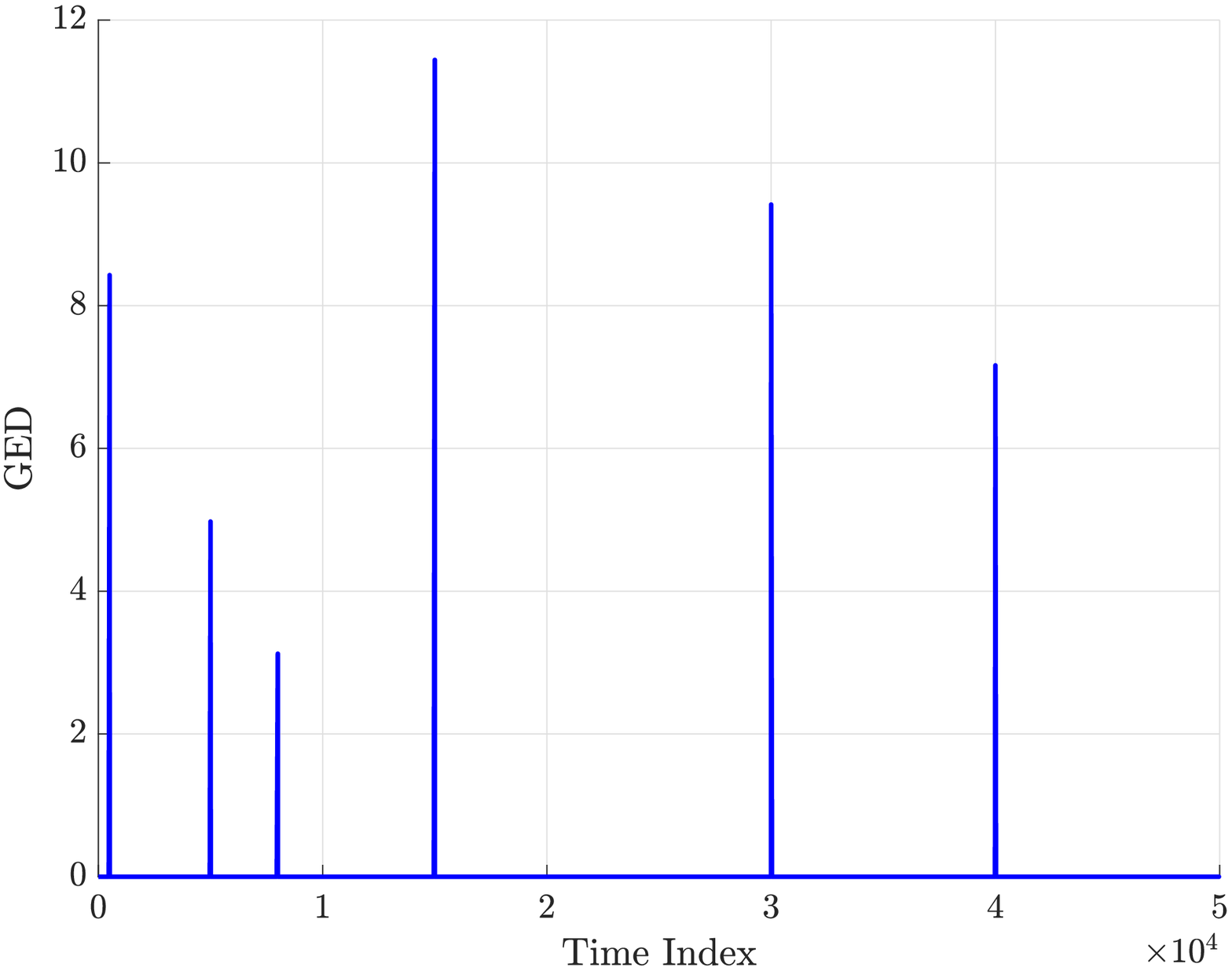}
  \caption{GED ground truth with updated baselines}
  \label{fig:GED_a}
\end{subfigure}%
\begin{subfigure}[b]{.45\textwidth}
  \centering
  \includegraphics[trim={0cm 0 0 0},clip, width=1\linewidth]{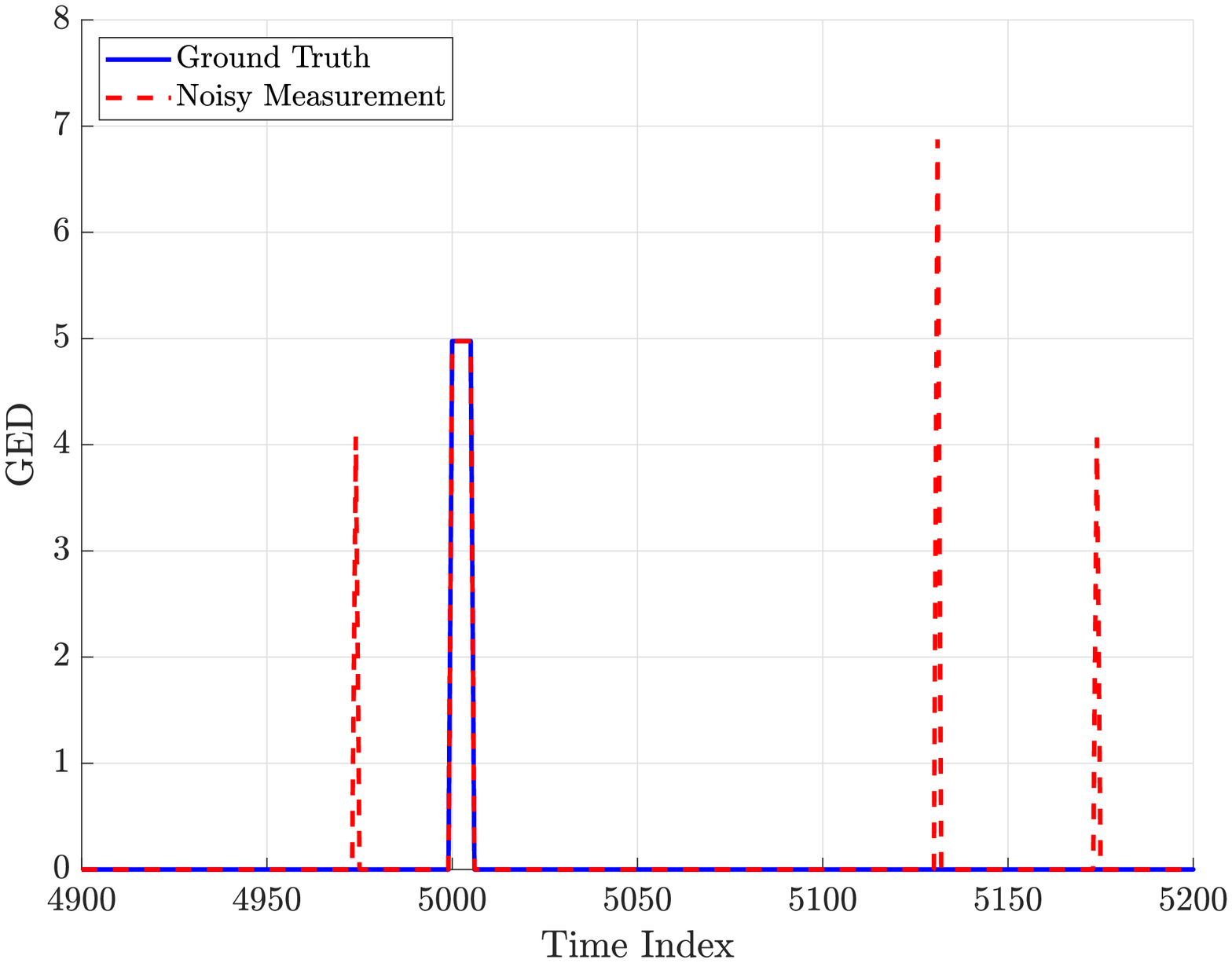}
  \caption{Calculated GED around $t = 5,000$}
  \label{fig:GED_b}
\end{subfigure}
\begin{subfigure}[b]{.45\textwidth}
  \centering
  \includegraphics[trim={0cm 0 0 0},clip, width=1\linewidth]{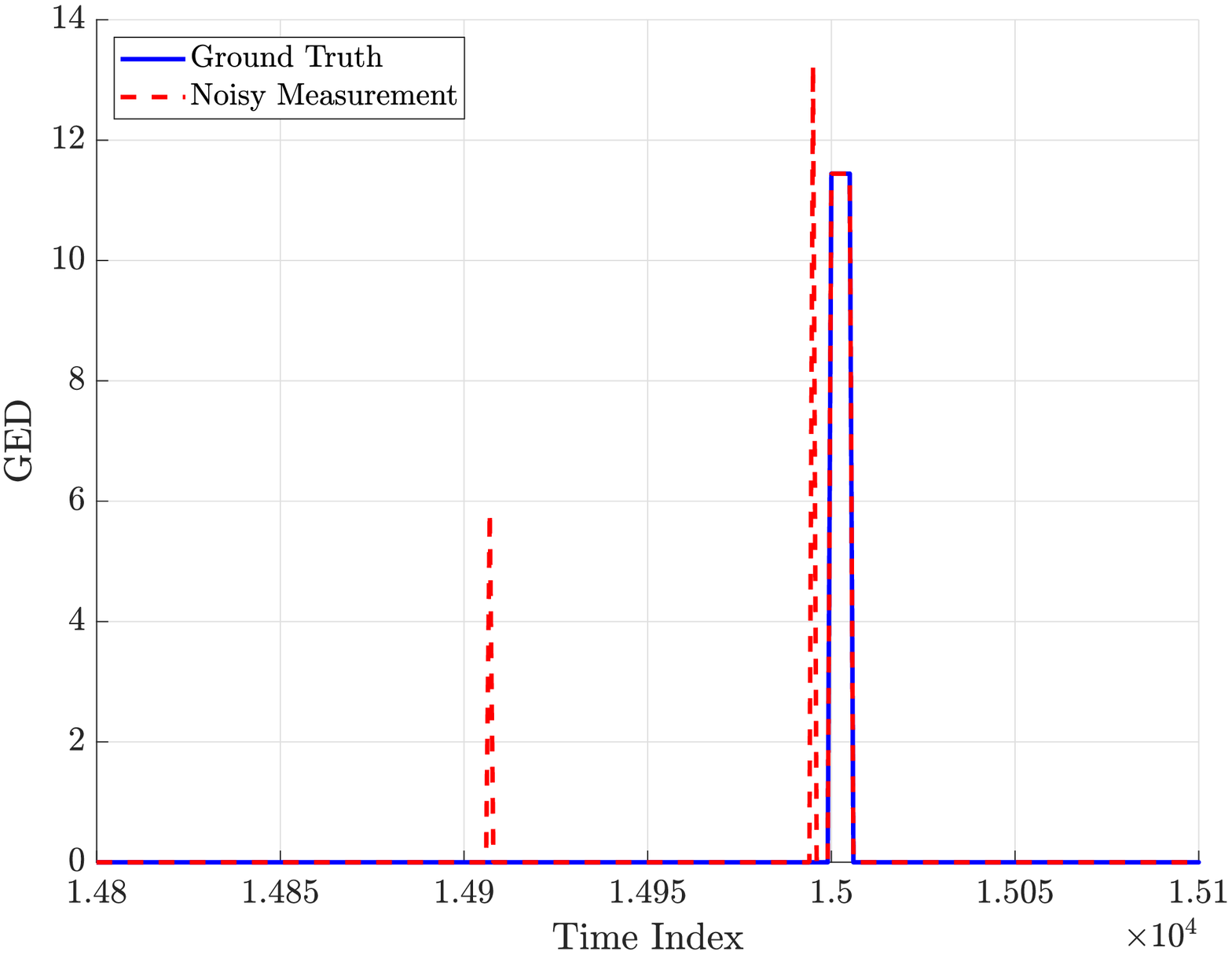}
  \caption{Calculated GED around $t = 15,000$}
  \label{fig:GED_c}
\end{subfigure}
\begin{subfigure}[b]{.45\textwidth}
  \centering
  \includegraphics[trim={0cm 0 0 0},clip, width=1\linewidth]{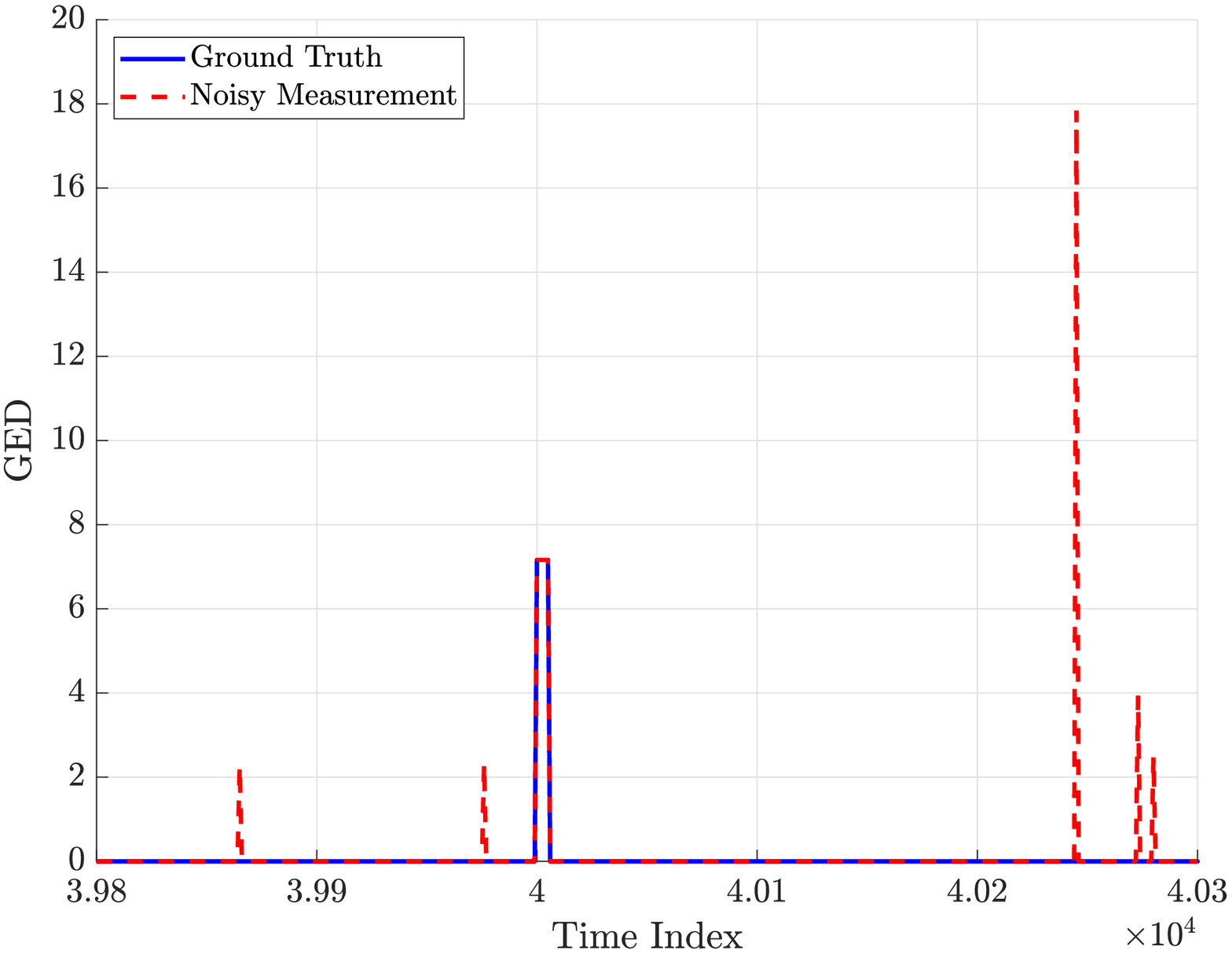}
  \caption{Calculated GED around $t = 40,000$}
  \label{fig:GED_d}
\end{subfigure}
\caption{GED example with updated baselines.}
\label{fig:GED}
\end{figure*}

To further showcase the application of the GED in tracking and filtering of semantic information, we introduce an imperfect predicate detector that can identify three different types of relationships among the components detected by the classifier. The \textit{conditional} confusion matrix and the corresponding GED costs for the predicate detector are calculated but not given here for brevity. Then, we simulate a temporal scenario given in Fig.~\ref{fig:GED_groundtruth}, where some of these five classes and three predicates can be identified (sometimes erroneously). 
The generated scenario is then processed with the example classifier and predicate detector explained in this section. The resulting graph signals $g_t$ at time $t$ are compared to a baseline graph $g_{base}$, which is updated to be the latest detected graph with at least five consecutive detections. The expected and measured GED metrics with this configuration are presented in Fig.~\ref{fig:GED}.
As shown in Fig.~\ref{fig:GED_a}, each of the actual innovation events in the graph signal is identified with a jump in the GED metric, followed by an updated baseline (return to zero) 5 frames after. Measured GEDs around different innovation time indices in Figs.~\ref{fig:GED_b}--\ref{fig:GED_d} show that sometimes there are erroneous detections that give rise to a nonzero GED; however, they can be easily filtered out by the straightforward algorithm explained above.

\begin{figure*}[t]
\centering
\begin{subfigure}[b]{.32\textwidth}
  \centering
  \includegraphics[width=1\linewidth]{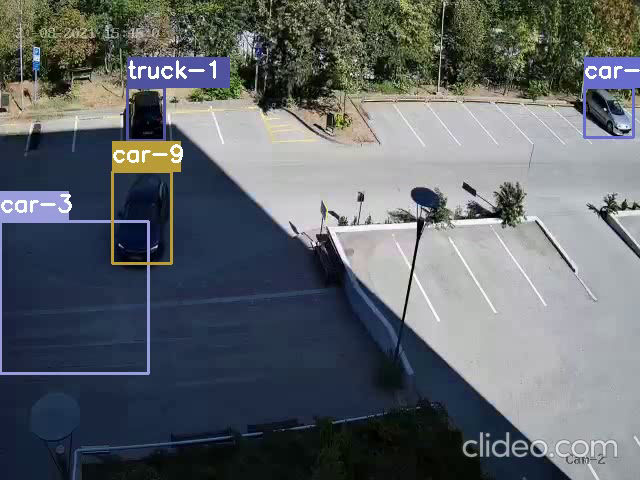}
  \caption{Frame 180}
  \label{fig:GED_video_a}
\end{subfigure}%
\begin{subfigure}[b]{.32\textwidth}
  \centering
  \includegraphics[width=1\linewidth]{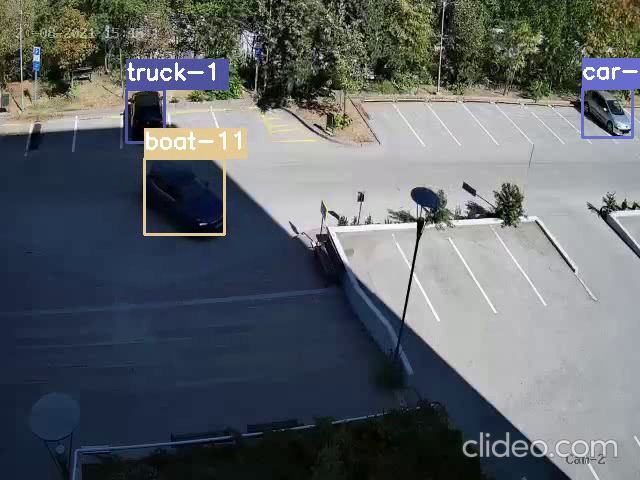}
  \caption{Frame 300}
  \label{fig:GED_video_b}
\end{subfigure}
\begin{subfigure}[b]{.32\textwidth}
  \centering
  \includegraphics[width=1\linewidth]{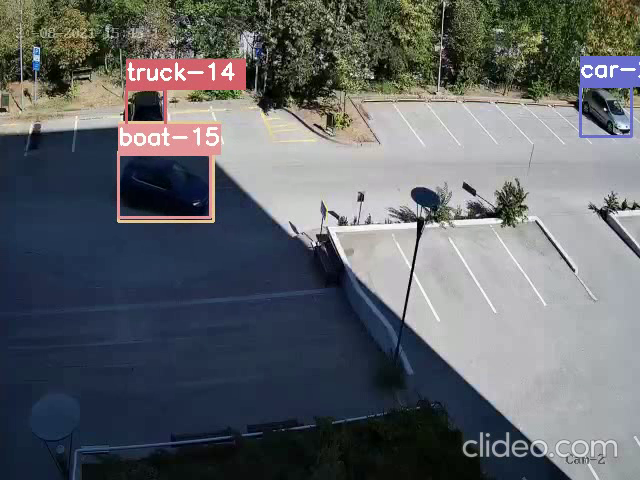}
  \caption{Frame 480}
  \label{fig:GED_video_c}
\end{subfigure}
\caption{Computer vision example with misdetected class outputs.}
\label{fig:GED_video}
\end{figure*}

\begin{figure*}[t]
\centering
\begin{subfigure}[b]{.45\textwidth}
  \centering
  \includegraphics[width=1\linewidth]{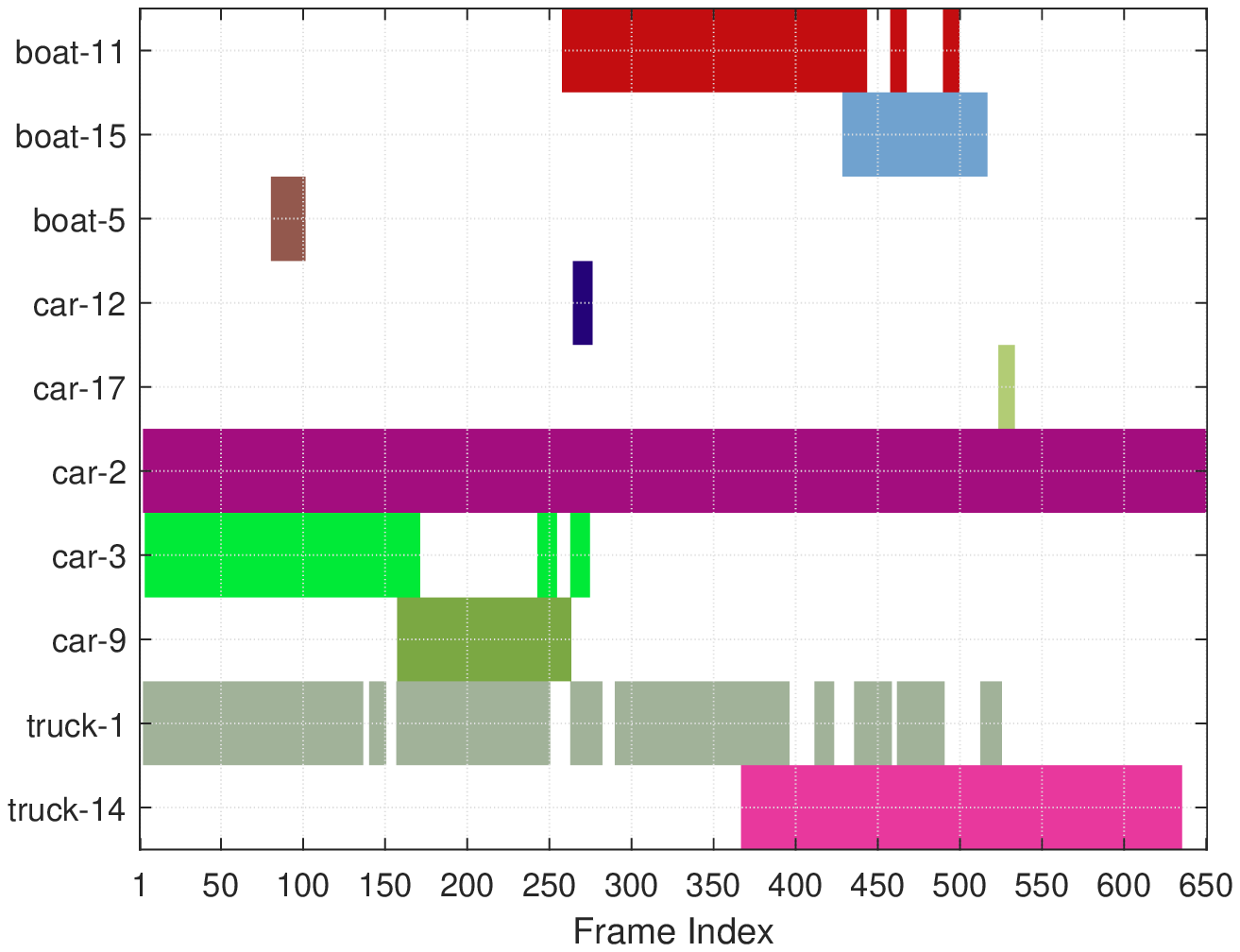}
  \caption{All detections from the semantic extractor.}
  \label{fig:GED_video_det_a}
\end{subfigure}%
\begin{subfigure}[b]{.45\textwidth}
  \centering
  \includegraphics[width=1\linewidth]{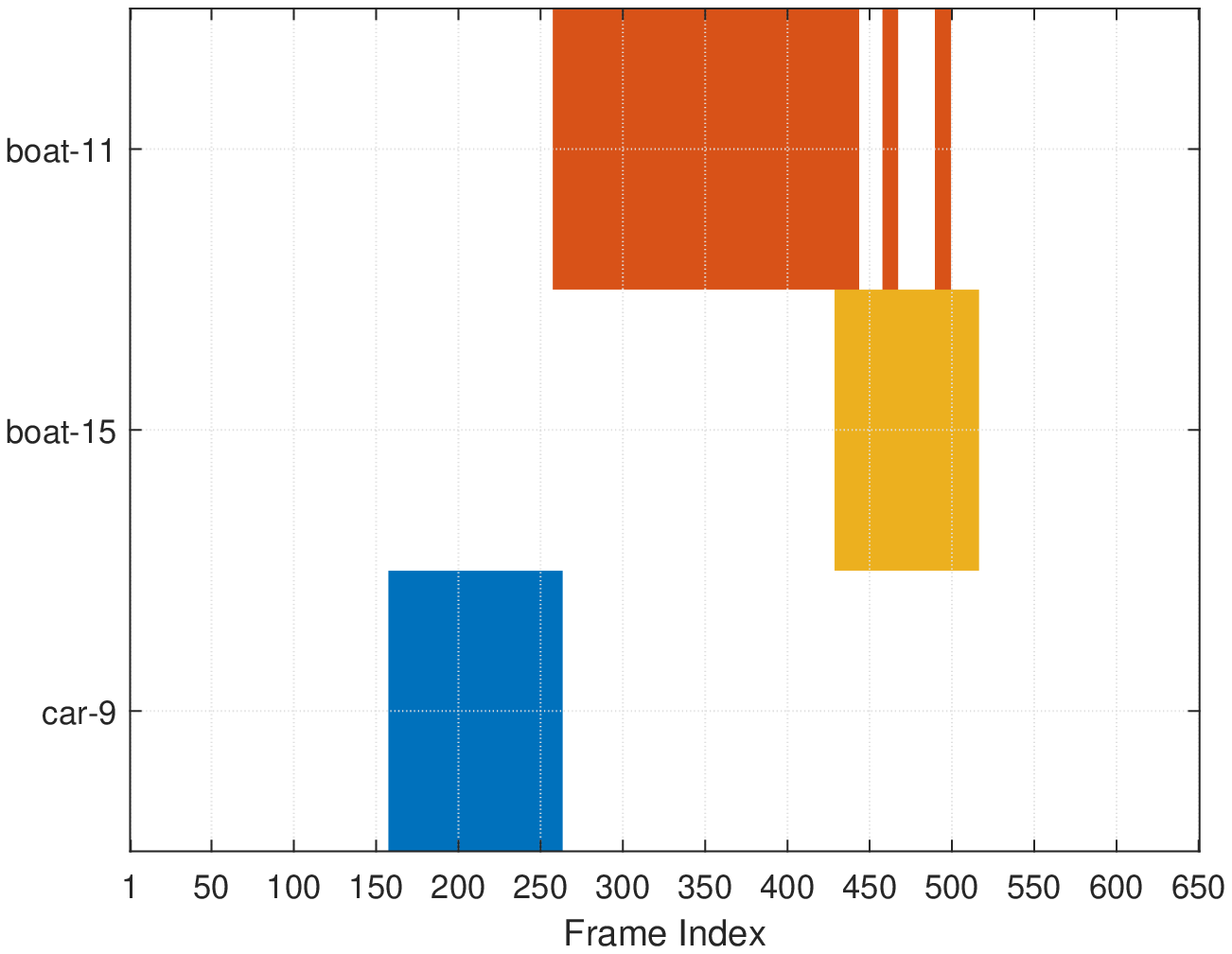}
  \caption{Detections focused on the misdetection of Car-9 instance.}
  \label{fig:GED_video_det_b}
\end{subfigure}
\caption{Semantic extractor output with misdetected class outputs.}
\label{fig:GED_video_detections}
\end{figure*}

We also perform a GED-based pre-filtering on a video signal using the semantic extractor given in~\cite{kalfa2021}. In this example, we have a 650-frame-long video (with 30fps) from a parking lot, where a car is maneuvering near several parked cars. This specific video is selected as its semantic output includes many false alarms, wrong categorizations, and missed detections. In particular, we are interested in showcasing the GED method's ability to reconcile semantic confusions by exploiting the prior knowledge about the statistical similarity of the confused categories. A series of still images from the video example is shown in Fig.~\ref{fig:GED_video}, with the semantic output summarized in Fig.~\ref{fig:GED_video_detections}.

\begin{figure}[tb]
	\centering
	\includegraphics[width=0.95\linewidth]{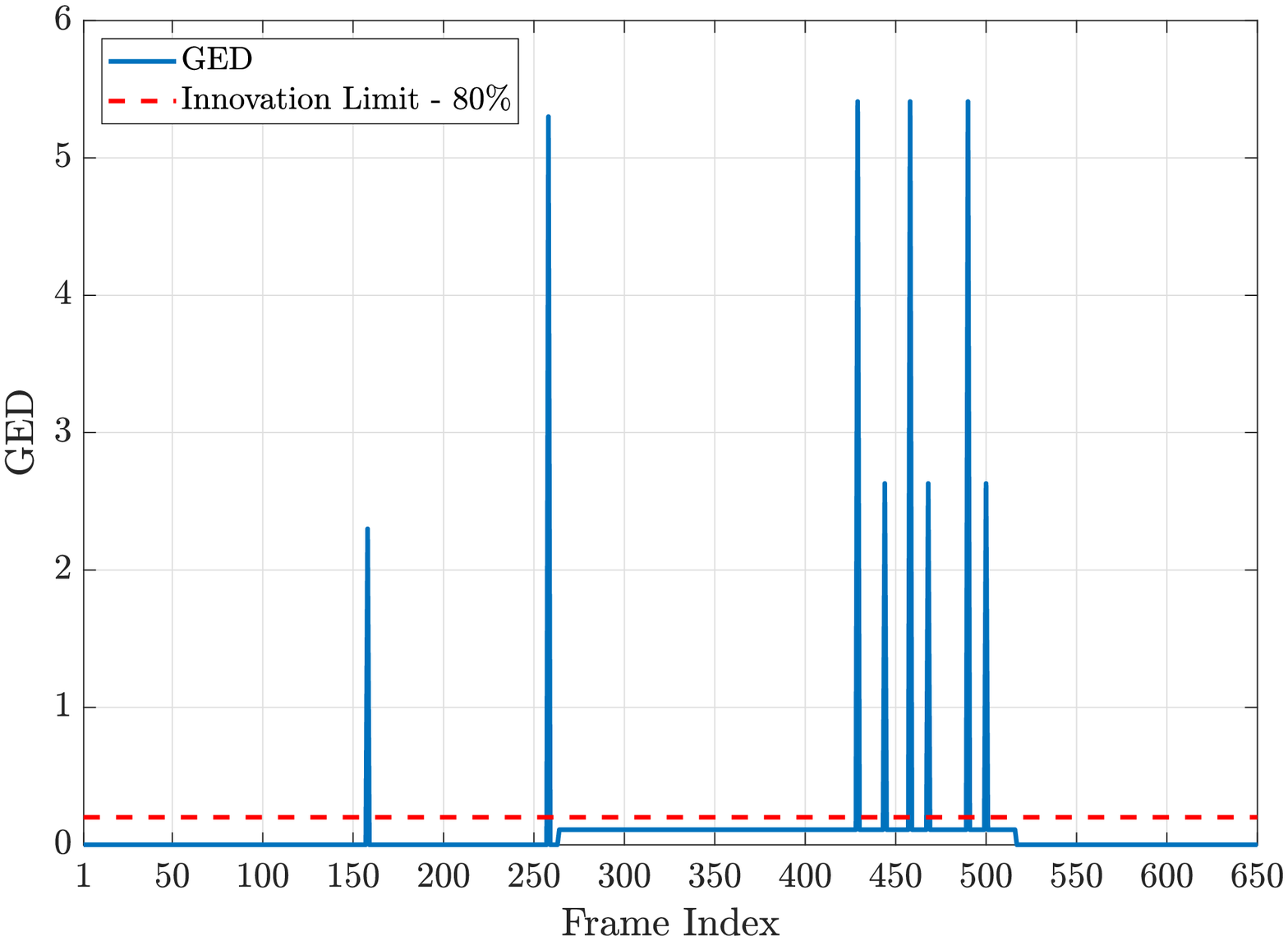}
	\caption{GED across time for the events in Fig.~\ref{fig:GED_video_detections}.}
	\label{fig:GED_vs_time}
\end{figure}

\begin{figure}[tb]
	\centering
	\includegraphics[trim=0cm 0 0cm 0, clip, width=0.95\linewidth]{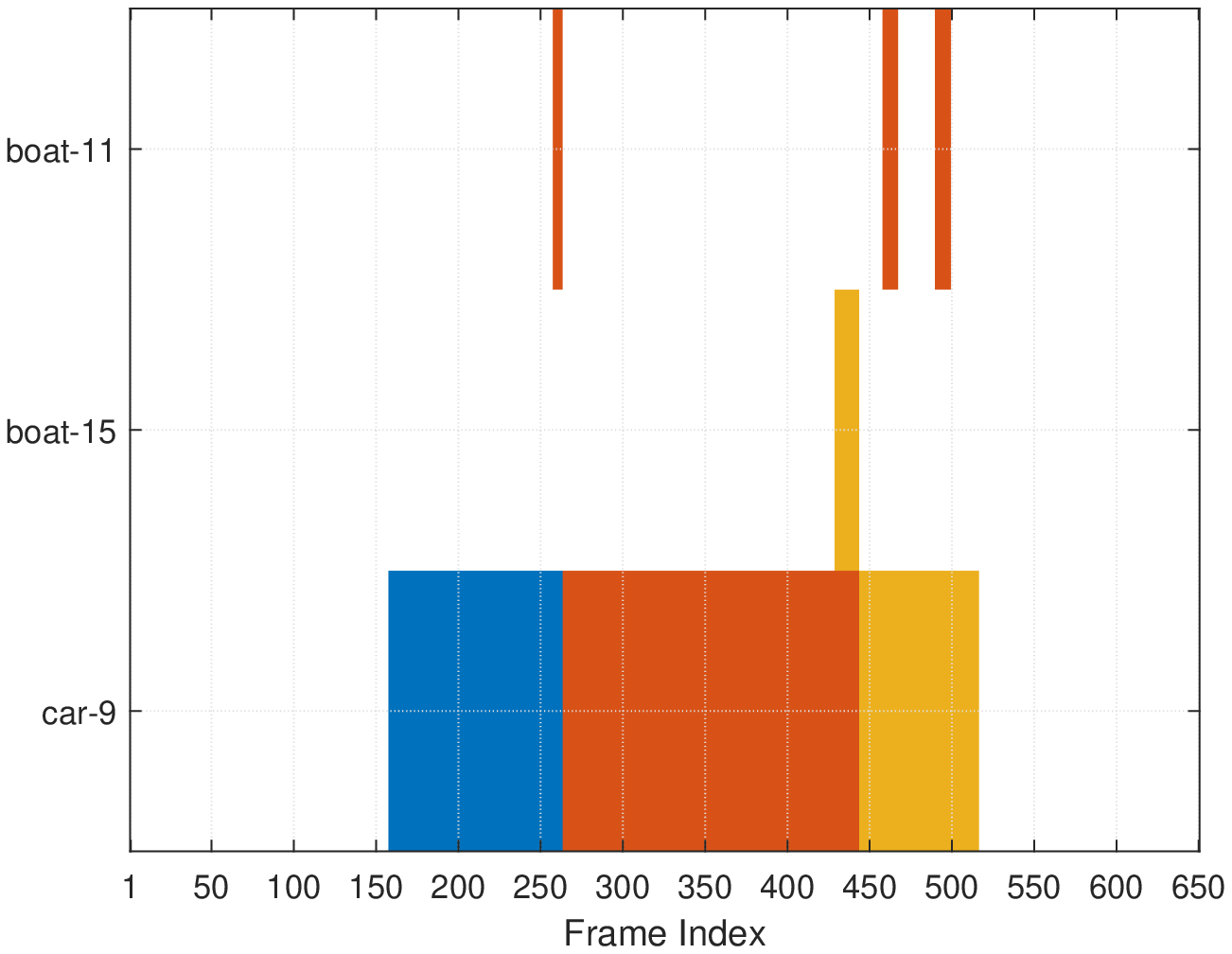}
	\caption{Smoothed semantic evolution after GED filtering.}
	\label{fig:GED_filtered}
\end{figure}

As shown in Figs.~\ref{fig:GED_video} and \ref{fig:GED_video_detections}, the maneuvering car (\textit{Car-9}) is classified correctly as first, but then misclassified as two different boats as the video progresses. Focusing on this part of the video, as shown in Fig.~\ref{fig:GED_video_det_b}, we now demonstrate the use of GED in filtering out the semantic noise in this event. Firstly, we generate similar semantic outputs from a set of videos recorded by the same camera, in a similar time of the day to generate the confusion probabilities between the \textit{Car} and \textit{Boat} classes, as well as their prevalence (estimated rate of occurrence) in this particular setup. Using these statistics, the GED edit costs are calculated and shown in Table~\ref{tb:GED_CM_carboat}.
\begin{table}[tb]
\centering
\caption{Derived statistics for Boat and Car classes. Values on the left are the probabilities and the ones on the right are the corresponding GED costs. }
\label{tb:GED_CM_carboat}
\begin{tabular}{ccc}
 & \textbf{Car (observed)} & \textbf{Boat (observed)} \\ \cline{2-3} 
\multicolumn{1}{c|}{\textbf{Car (actual)}} & 95\% / \textbf{0.02} & 4.5\% / \textbf{3.1} \\
\multicolumn{1}{c|}{\textbf{Prevalence}} & 10\% / \textbf{2.3} & 0.5\% / \textbf{5.3}
\end{tabular}
\end{table}
As shown in the top row of Table~\ref{tb:GED_CM_carboat}, the confusion rate among the \textit{Car} and the \textit{Boat} classes is fairly small. However, taking into account the estimated prevalence of these classes and using Bayes' rule, we can infer the posterior probability of a boat being confused by a car as $P(\text{Car } exists | \text{Boat } observed) = 90\%$ and the corresponding GED cost as $-\log(0.9) \approx 0.11$, which shows the statistical likelihood of a confusion is in fact very high. Using these cost definitions, the time evolution of the GED between consecutive semantic outputs is illustrated in Fig.~\ref{fig:GED_vs_time}. Note that by implementing a simple threshold, we can easily identify significant innovations and filter out the events that are not statistically significant. As shown in Fig.~\ref{fig:GED_vs_time}, by using a statistical significance threshold of $80\%$, or a corresponding GED cost limit of $0.2$, we can reconcile the boat detections with the original correct identification. The GED output and the statistical significance limit are shown in~Fig.~\ref{fig:GED_filtered}. 

As illustrated in Fig.~\ref{fig:GED_filtered}, all misclassifications are correctly filtered except the duplicate detections over the same object. Since the GED method in its proposed form here only focuses on the time evolution of its scalar value, duplicate detections of the same object register as innovations. However, these are easily filtered with the next step in our proposed semantic extraction framework, with the attribute tracking module that can identify these detections to be from the same object. 

\subsection{Hidden Markov Model}
In this part, HMM is applied to the raw semantic extraction of video signals to test its usefulness in practical scenarios. The main question is whether the smoothing algorithm (employing the Viterbi Algorithm) can correct the erroneously specified states when the model parameters ($\textbf{A,B,p}$) are known. It is assumed that the scene starts with an empty graph; that is, there is not any component in the scene at the beginning. This is due to the design of the algorithm, which requires multiple consecutive detections of the same pattern to decide that it exists. 
    
To illustrate the method, two video signals that contain cars and a person are used. For simplicity, a graph configuration with two predicates is employed. The predicates express the existence or absence of components in the graph language. So, the generated graph moves to a different state when a component enters or leaves the scene. The components leave or enter the scene independently of the other ones; hence if a scene contains $N$ components, there are $2^N$ possible states in total. There are two classes of components, and the components can be connected to two predicates (can be in one of the two states). Assuming components of the same class follow the same statistical distribution, state transition matrices for car and person can be expressed as follows:

\begin{align}
A_{car} =& \begin{bmatrix} 
    P^c_{0,0} && P^c_{0,1} (1-P^c_{0,0})\\
    P^c_{1,0} (1-P^c_{1,1}) && P^c_{1,1}\\
    \end{bmatrix},\ \nonumber \\
A_{person} =& \begin{bmatrix} 
    P^p_{0,0} && P^p_{0,1} (1-P^p_{0,0})\\
    P^p_{1,0} (1-P^p_{1,1}) && P^p_{1,1}\\
    \end{bmatrix}.\ \nonumber
\end{align}

The state transition matrix and observation kernel assumed for these videos are as follows:
\begin{align}
A_{car} = \begin{bmatrix} 
    0.75 && 0.25\\
    0.20 && 0.80\\
    \end{bmatrix}\
B_{car} = \begin{bmatrix} 
    0.75 && 0.25\\
    0.40 && 0.60\\
    \end{bmatrix}\ \nonumber
\end{align}

\begin{equation}
A_{person} = \begin{bmatrix} 
    0.70 && 0.30\\
    0.20 && 0.80\\
    \end{bmatrix}\
B_{person} = \begin{bmatrix} 
    0.70 && 0.30\\
    0.40 && 0.60\\
    \end{bmatrix}.\ \nonumber
\end{equation}

\begin{figure*}[tb]
 \centering
\begin{subfigure}[b]{.32\textwidth}
  \centering
  \includegraphics[width=1\linewidth]{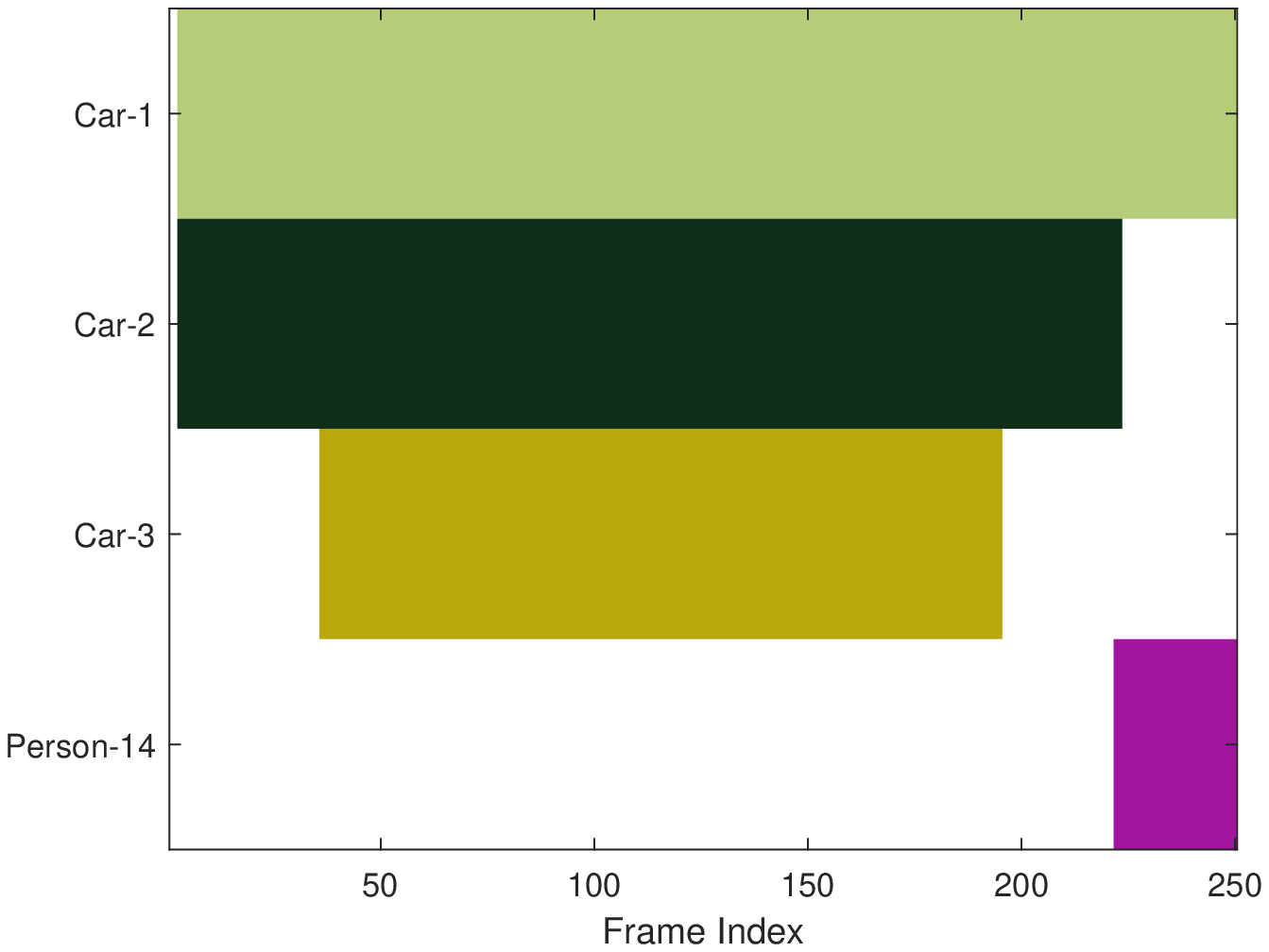}
  \caption{True State Sequence}
%   \label{fig:GED_video_det_a}
\end{subfigure}%
\begin{subfigure}[b]{.32\textwidth}
  \centering
  \includegraphics[width=1\linewidth]{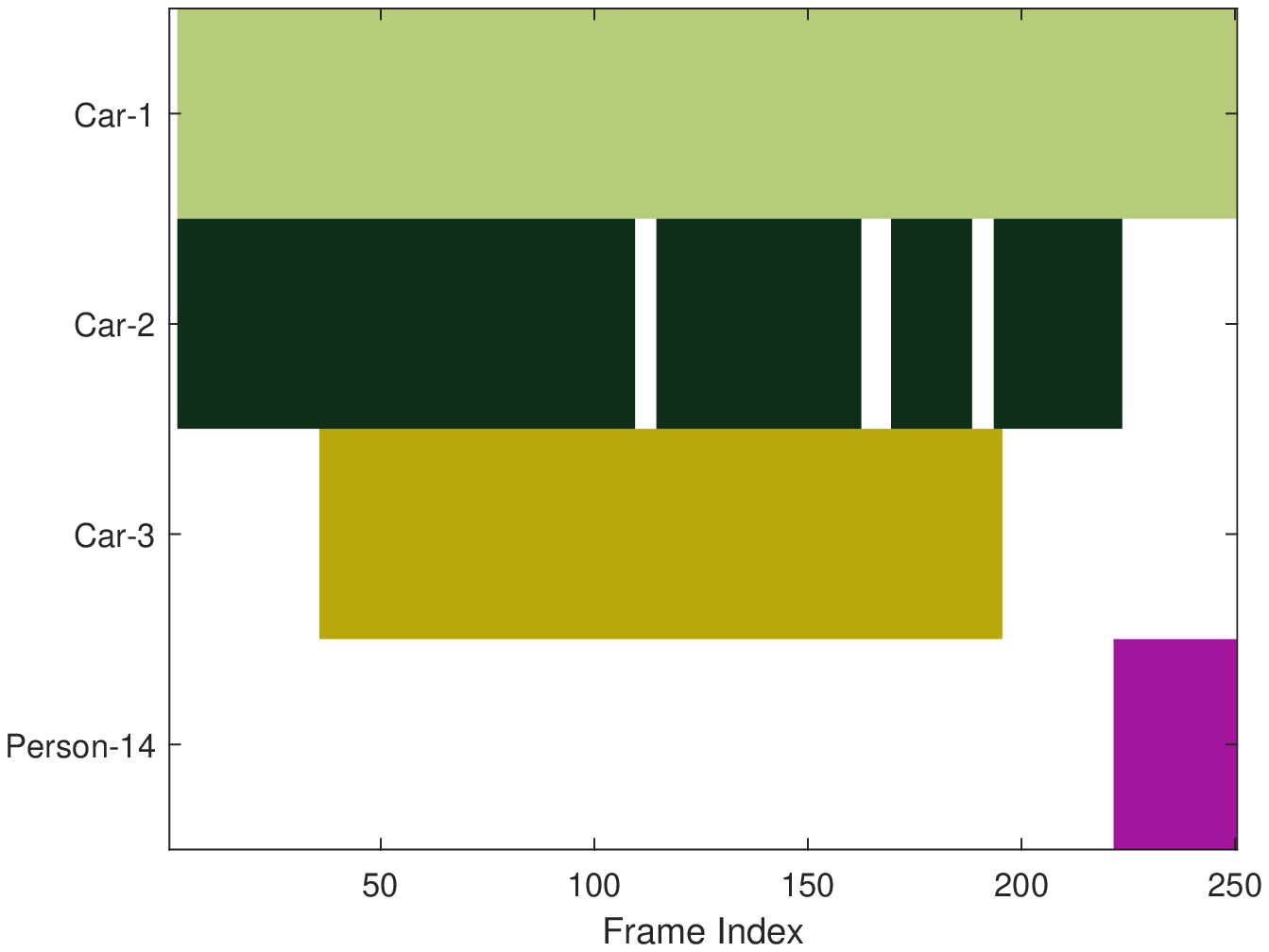}
  \caption{Extracted State Sequence}
%   \label{fig:GED_video_det_a}
\end{subfigure}%
\begin{subfigure}[b]{.32\textwidth}
  \centering
  \includegraphics[width=1\linewidth]{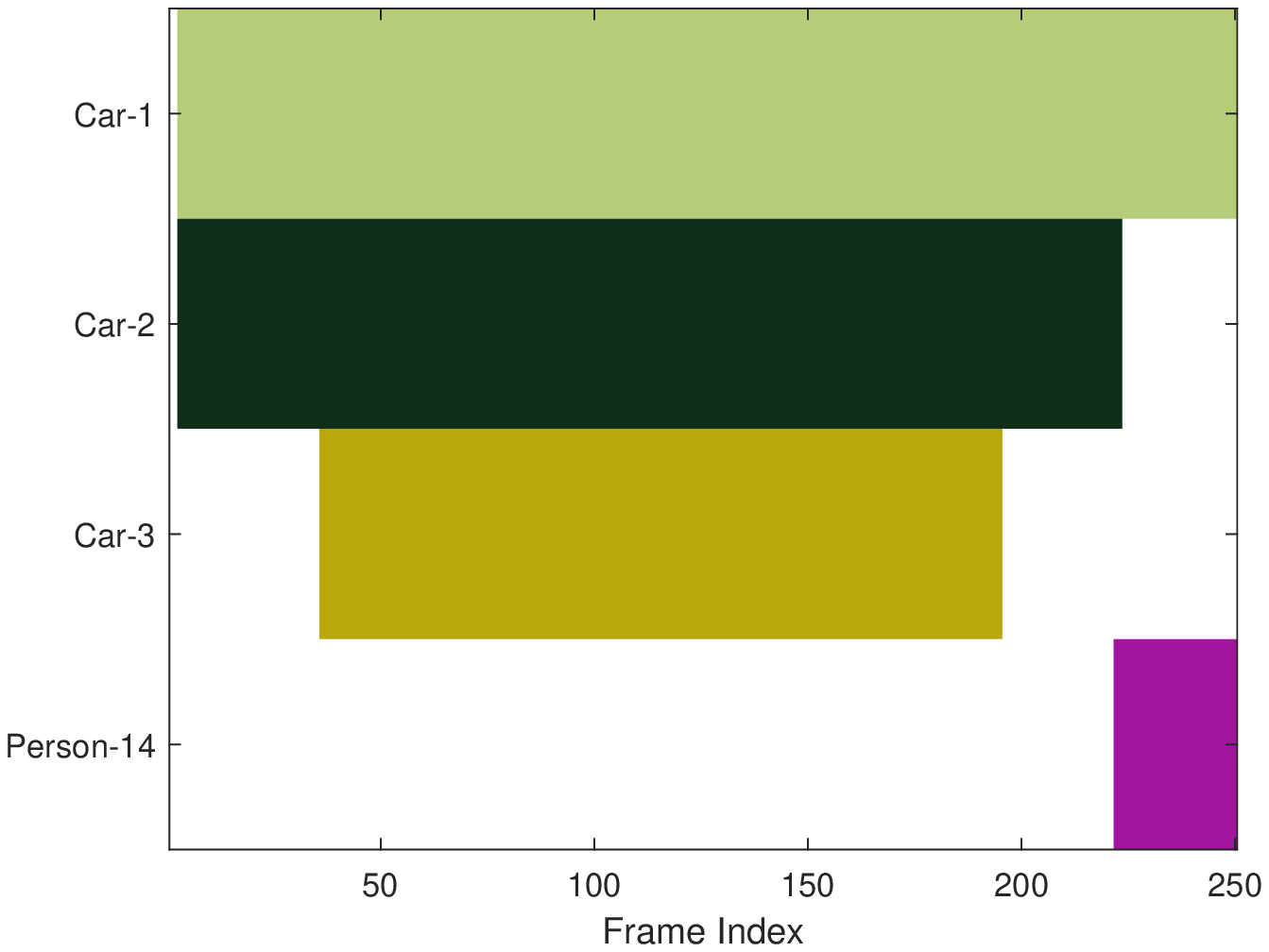}
  \caption{Filtered State Sequence}
%   \label{fig:GED_video_det_a}
\end{subfigure}%
  \caption{State sequences for the first video. Semantic extractor labels Car-2 as absent when it is in the scene for some time instances. Viterbi Algorithm is able to correct the errors in this particular example.}
       \label{fig:images_raw_seq_01}
\end{figure*}

\begin{figure*}[tb]
 \centering
 \begin{subfigure}[b]{.32\textwidth}
  \centering
  \includegraphics[width=1\linewidth]{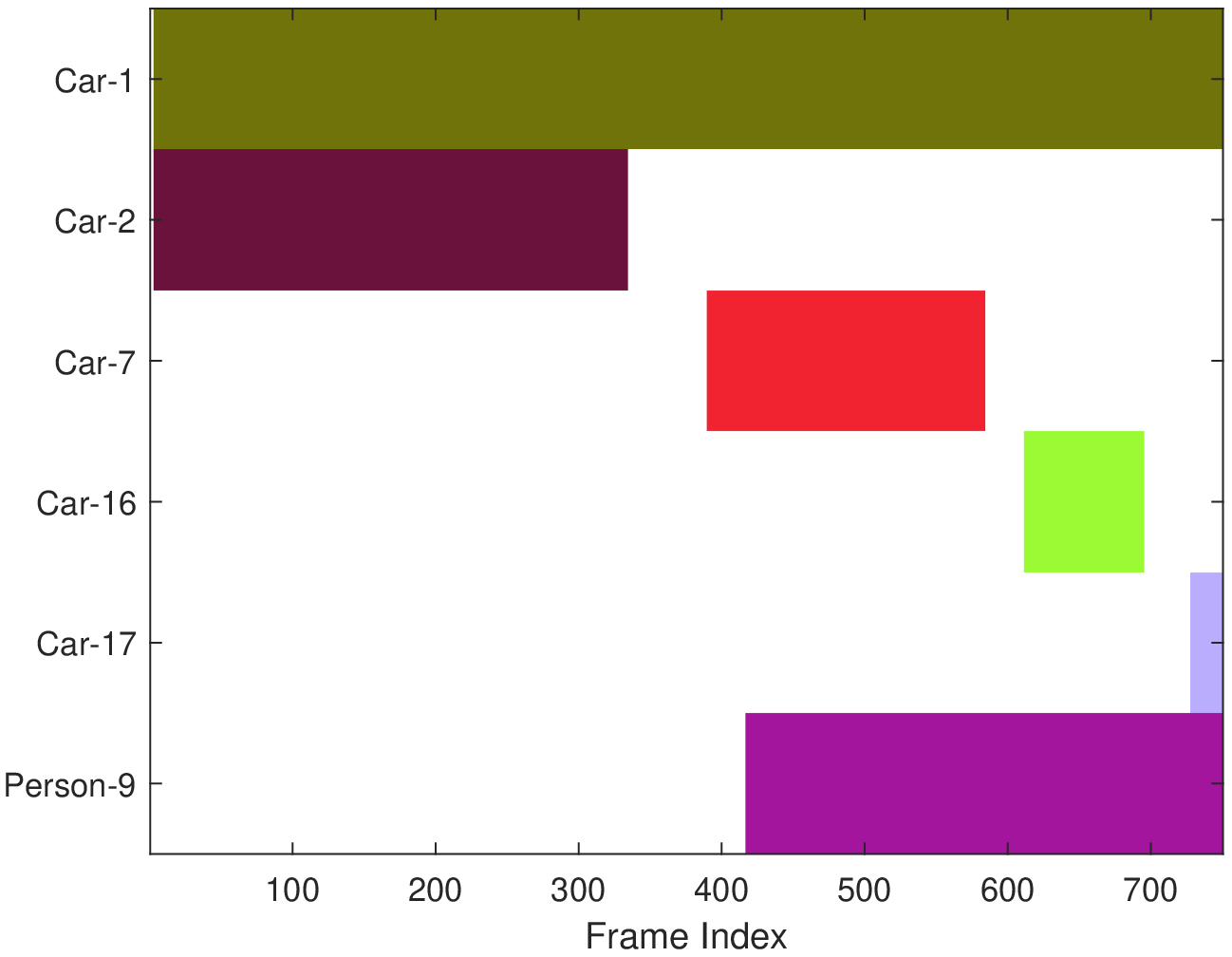}
  \caption{True State Sequence}
\end{subfigure}%
\begin{subfigure}[b]{.32\textwidth}
  \centering
  \includegraphics[width=1\linewidth]{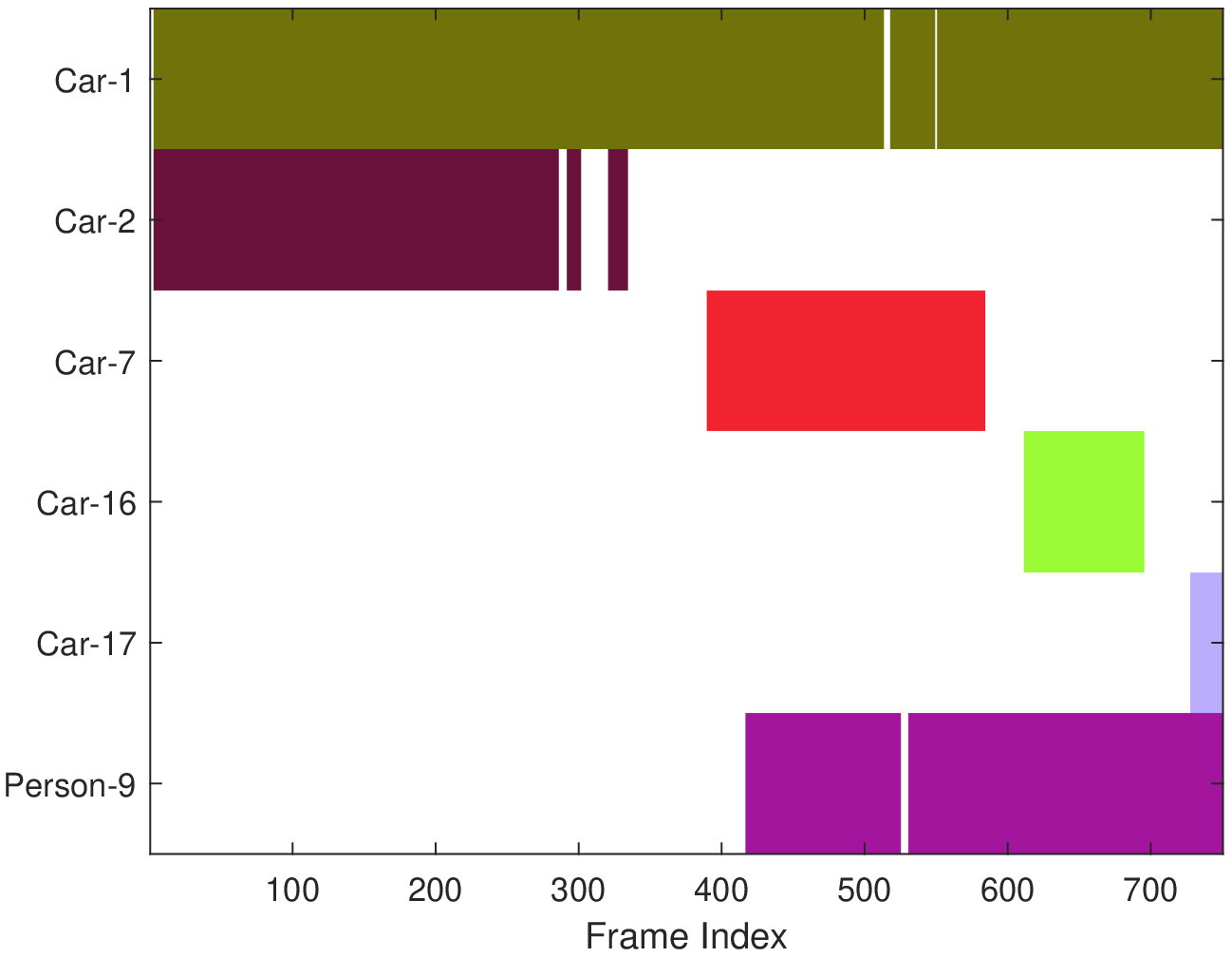}
  \caption{Extracted State Sequence}
\end{subfigure}%
\begin{subfigure}[b]{.32\textwidth}
  \centering
  \includegraphics[width=1\linewidth]{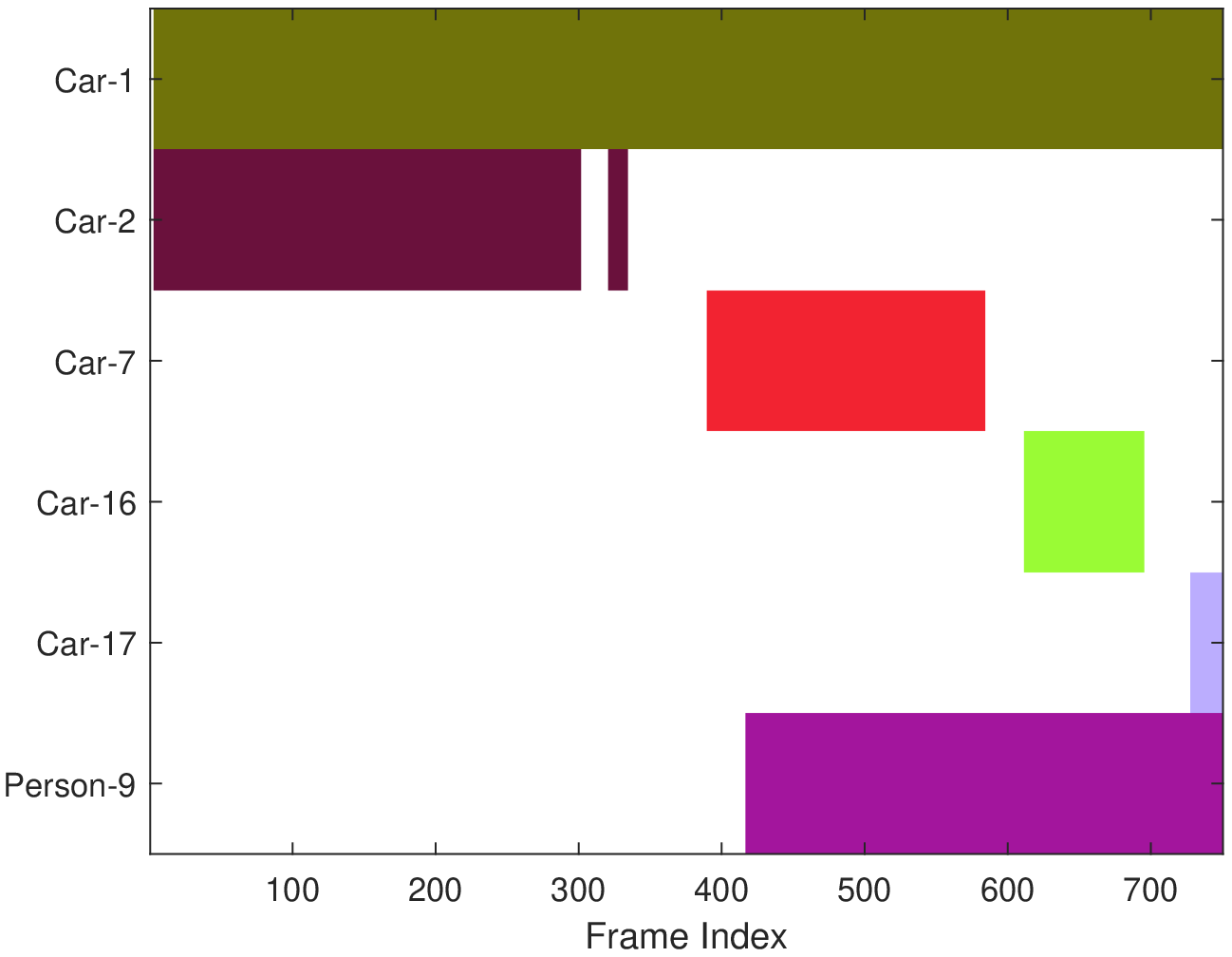}
  \caption{Filtered State Sequence}
\end{subfigure}%
    \caption{State sequences for the second video. Semantic extractor labels Car-1,Car-2 and Person-9 as absent incorrectly in some time instances. Viterbi Algorithm is shown to corrects these mistakes except for a short blank interval for Car-2.}
    \label{fig:images_raw_seq_02}
\end{figure*}

The true, observed, and estimated state sequences for the first video example are given in~Fig.~\ref{fig:images_raw_seq_01}. The true state sequence is defined manually and represents the state sequence that would be observed if the semantic extractor did not make any errors. 
The Viterbi Algorithm smooths out the errors introduced by the raw semantic extractor in the first video. As seen in Fig.~\ref{fig:images_raw_seq_01}, the semantic extractor erroneously decides that Car-2 is not in the scene in some instances. The Viterbi Algorithm corrects this mistake without introducing any further errors. For the second video illustrated in Fig.~\ref{fig:images_raw_seq_02}, the raw semantic extractor produces incorrect results for Car-1, Car-2, and Person-9 in some time instances. The proposed algorithm corrects these errors except for a short interval in Car-2's state sequence, where Car-2 is erroneously missing from the detections. 

\begin{figure*}[tb]
 \centering
  \begin{subfigure}[b]{.32\textwidth}
  \centering
  \includegraphics[width=1\linewidth]{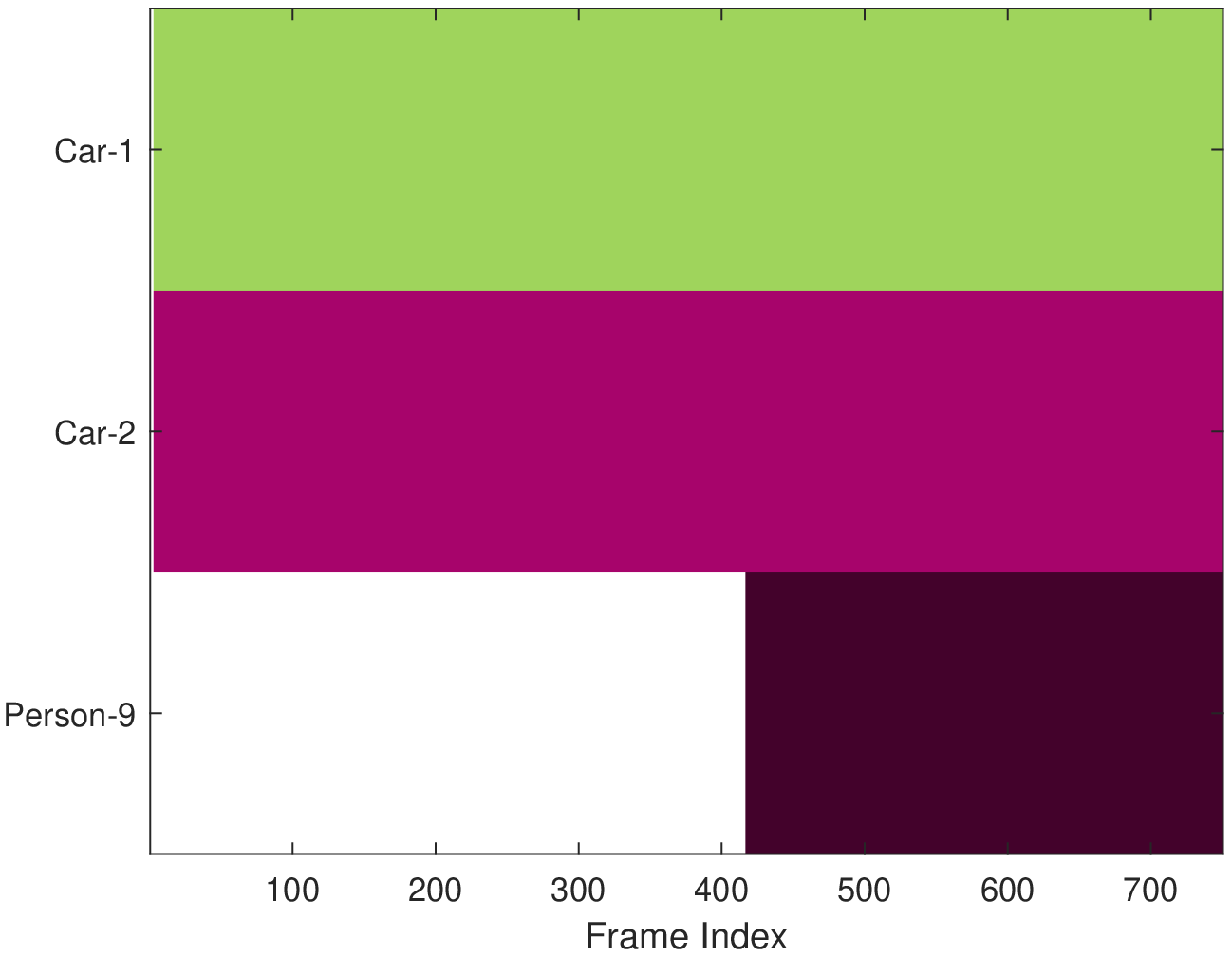}
  \caption{True State Sequence}
%   \label{fig:GED_video_det_a}
\end{subfigure}%
\begin{subfigure}[b]{.32\textwidth}
  \centering
  \includegraphics[width=1\linewidth]{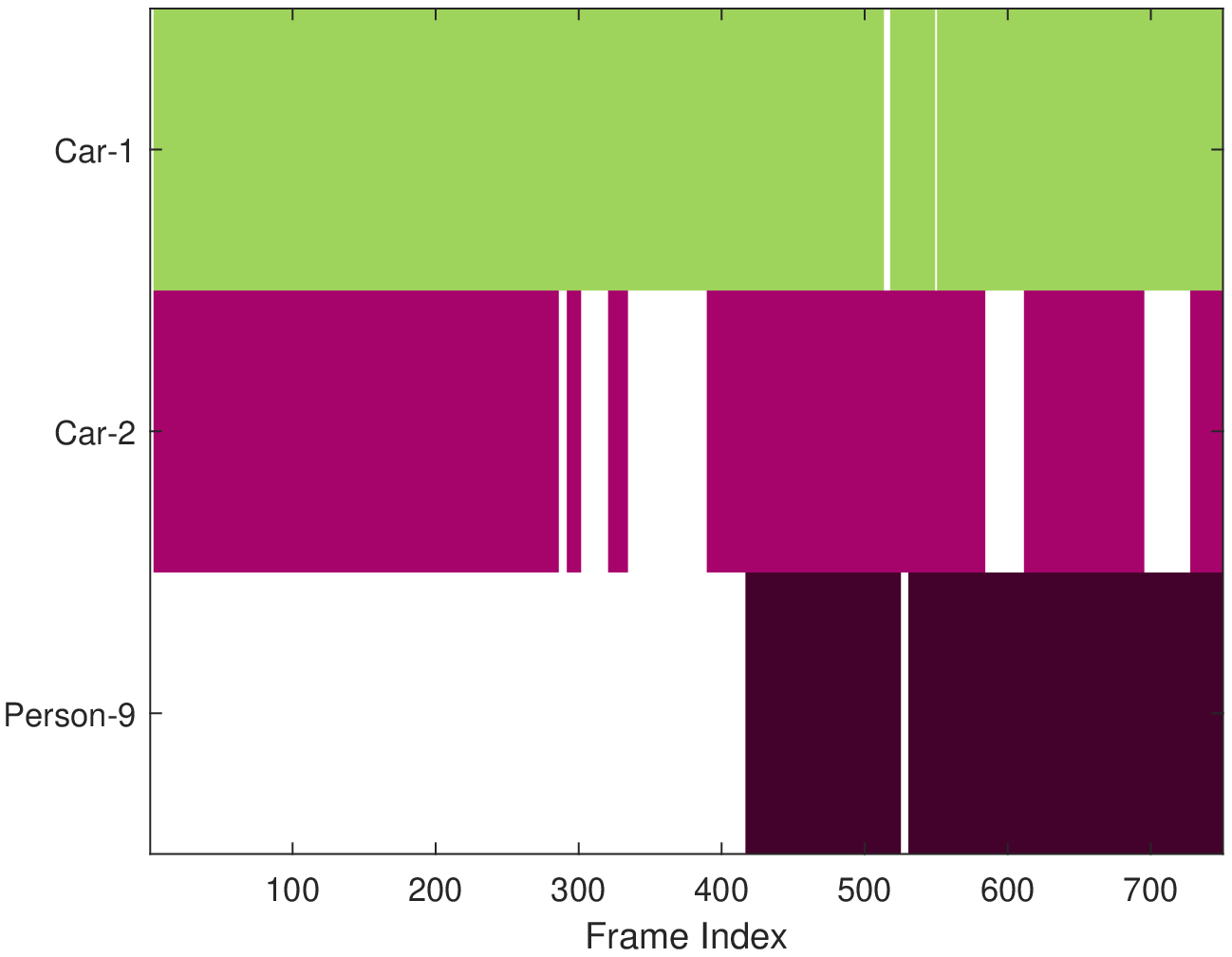}
  \caption{Extracted State Sequence}
%   \label{fig:GED_video_det_a}
\end{subfigure}%
\begin{subfigure}[b]{.32\textwidth}
  \centering
  \includegraphics[width=1\linewidth]{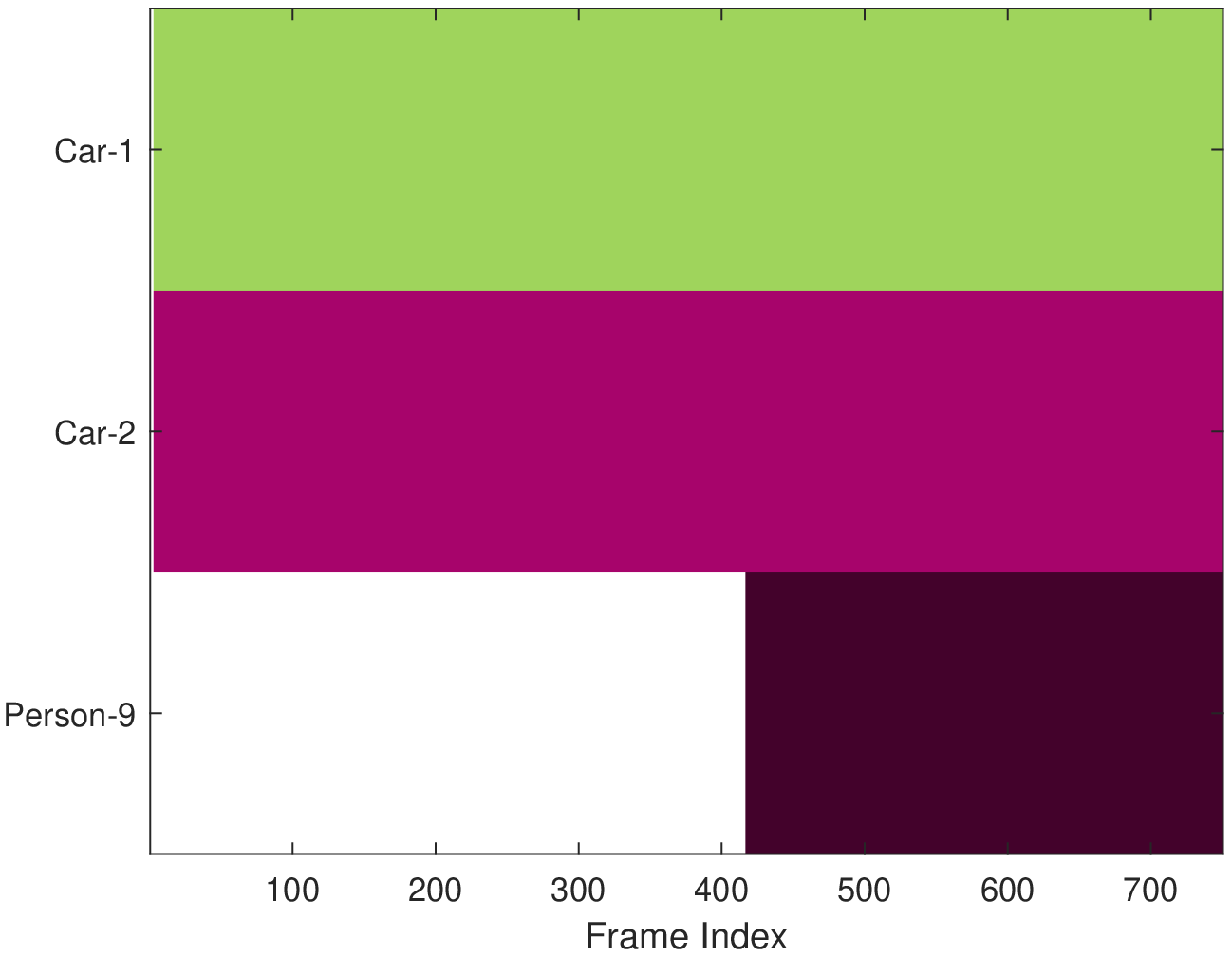}
  \caption{Filtered State Sequence}
%   \label{fig:GED_video_det_a}
\end{subfigure}%
    \caption{State sequences for the second video with reconciliation of the components, combined state sequence of Car-2, 7, 16, 17 in \ref{fig:images_raw_seq_02} is represented as Car-2. Semantic extractor labels Car-1,Car-2 and Person-9 as absent incorrectly in some time instances, Viterbi algorithm is shown to correct these mistakes. The model parameters are selected as $P_{S_0,S_0} = 0.80$, $P_{S_1,S_1} = 0.85$, $P_{S_0,O_0} = 0.70$, $P_{S_1,O_1} = 0.55$ for the car class and $P_{S_0,S_0} = 0.90$, $P_{S_1,S_1} = 0.90$, $P_{S_0,O_0} = 0.65$, $P_{S_1,O_1} = 0.55$ for the person class.
}
    \label{fig:images_concat_seq_01}
\end{figure*}

Note that in the second video, Car-2, Car-7, Car-16, and Car-17 are different instance identifications of the same underlying object (Car-2). The proposed HMM and Viterbi algorithm cannot reconcile these erroneous multiple identifications by itself. To produce a smooth and reliable output while reconciling among multiple identifications, we use the methods presented in this paper sequentially, as shown in Fig.~\ref{fig:semantic_extractor}. The \textit{Attribute Tracking} example given in Section~\ref{subsec:attr} in Figs.~\ref{fig:occlusions}--\ref{fig:Objectreconciliation2} specifically illustrates the reconciliation among the multiple identifications of Car-2, Car-7, Car-16 and Car-17. Using the reconciled output of the \textit{attribute tracking} module as the input of the \textit{graph signal tracking} module, the output of the filtered semantic output is shown in Fig.~\ref{fig:images_concat_seq_01}. As illustrated in Fig.~\ref{fig:images_concat_seq_01}, the Viterbi algorithm is shown to correct the mistakes for each component, producing a filtered state sequence that is identical to the true sequence. 

\section{Discussion on the Semantic Rate of Innovation}
\label{sec:innovation}
The proposed semantic extraction framework that has been defined and demonstrated so far is shown to generate reliable semantic graph signals with embedded numerical attributes. The proposed framework and its building blocks also enable the identification of the innovation events either at the graph level (using HMM and/or GED) or at the attribute level (using subspace tracking). These events can be used to schedule transmission and/or storage events, depending on the device and the application of interest. Once the innovation events can be detected given a semantic signal, underlying statistical properties of innovations can be modeled and estimated. These estimates on innovation statistics can, in turn, be used to estimate the data throughput in a communications network. 

Consider the semantic graph structure presented in Section~\ref{sec:SSPprimer}, with atomic bipartite graphs denoted by $D_i$ that evolve across time as shown in Fig.~\ref{fig:graph_example_Dt}. Let the rate of innovation at the graph level for $D_i$ be $I_{D_i}$, and the corresponding average message length be $d_{D_i}$. The rates of innovation and the average message lengths at the attribute level can be similarly defined as 
\begin{equation}
    I_{A_i} = [I^1_{A_i}, I^2_{A_i}, \ldots, I^L_{A_i}],
\end{equation}
\begin{equation}
    d_{A_i} = [d^1_{A_i}, d^2_{A_i}, \ldots, d^L_{A_i}],
\end{equation}
with $L$ being the total number of attribute layers in~\eqref{eq:theta_ti}. The total rate of transmitted or stored information for $N$ atomic graphs without any goal-oriented filtering can be written as
\begin{equation}
    R = \sum\limits_{i = 1}^{N} \left( I_{D_i} d_{D_i} + \sum\limits_{l = 1}^{L} I_{A_i} d_{A_i} \right).
    \label{eq:R}
\end{equation}
Note that the semantic structure (components, predicates, and attributes) must be defined to optimize the rate of transmission using~\eqref{eq:R}. Intuitively, the higher abstraction provided by the semantic structure should lead to lower rates of innovation for the graph and individual attributes compared to the rate of innovation of the raw input signal. 

Building on~\eqref{eq:R}, another significant advantage of the highly organized and hierarchical semantic structure is the processing of the graph signals in a goal-oriented manner. Given a dynamic goal $\mathcal{G}_t$ (either internally or externally defined) that defines interest over a subset of the extracted semantic information, the rate of information can be significantly reduced. If we denote the most general goal that is interested in every output of the semantic extractor as $\mathcal{G}_0$, with $\mathcal{G}_t$ being a proper subset of $\mathcal{G}_0$, application of the goal $\mathcal{G}_t$ at a class and attribute level will lead to a reduced number of graphs $\hat{N}<N$ with a reduced number of attribute layers $\hat{L}<L$. The corresponding goal-oriented transmission rate can be written as
\begin{equation}
    \hat = \sum\limits_{i = 1}^{\hat{N}} \left( I_{D_i} d_{D_i} + \sum\limits_{l = 1}^{\hat{L}} I_{A_i} d_{A_i} \right),
    \label{eq:Rhat}
\end{equation}
with $\hat{R}<R$ for $\mathcal{G}_t \subset \mathcal{G}_0$. 

Some of the most significant advantages of moving beyond technical communications toward goal-oriented semantic communications are the reduction of the rates of innovation and the inherent compression of the underlying signals. The modeling of the rates of innovation as well as the efficient compression of the semantic signals under dynamic goals will enable massive deployment for the next generation of sensor networks.

\section{Conclusions and Future Research Directions}
\label{sec:Conclusion}

The next generation of signal processing and communication systems will employ intelligent agents that can generate semantic information from their local environment. This work introduces a semantic extraction framework, where the extracted graph-based imperfect semantic signals can be improved for better fidelity, filtered for semantic source noise, while enabling the identification of significant innovations in the semantic signal. The aforementioned tasks are achieved by exploiting known semantic characteristics of the environment and statistical characteristics of the front-end semantic extractor. The proposed methods provide reliable semantic outputs and enable efficient ways of identifying semantic innovation, while filtering out unwanted semantic noise. Note that the proposed metrics and methods can also be used to schedule transmission and storage events in semantics-enabled sensor devices.

As the semantic signal processing and the semantic extraction frameworks proposed in this work and in the literature move closer to massive deployment in the next generation of sensor networks, continued research on the practical implications of the proposed methods is required. Specifically, semantic extraction techniques for different signal modalities should be developed and standardized for a shared semantic structure/language for next-generation devices. Given standardized semantic extractors, semantically-aware time integration metrics should be modeled and estimated for application specific scenarios. Model estimation is also critical for the efficient implementation of the HMM-based modeling and processing proposed in this work. 

The Markov model and its extension (HMM) also enable quantifying the semantic rate of innovation through the entropy rate of the underlying Markov chain and the development of efficient compression and transmission schemes of semantic information. The modeling of the state sequences can further be improved by replacing the Markov model with a semi-Markov model, which can relieve the restriction of HMM on waiting time distributions and increase the representational power of the model.

We finally note that, depending on the type of sensor/device and its computational capabilities, the proposed methods can be used collectively or independently. As the signal processing and communications paradigms move towards semantic signal processing and transmission, we believe the proposed semantic extraction framework will be an essential building block in developing the next generation of sensor devices and networks.

% % use section* for acknowledgment
% \section*{Acknowledgment}


\ifCLASSOPTIONcaptionsoff
  \newpage
\fi



%\bibliographystyle{IEEEtran}
%\begin{thebibliography}{1}

%\newpage
\bibliographystyle{bib/IEEEtran.bst}
\bibliography{bib/JSAC_SemanticExtraction_ref}
%\balance
%\end{thebibliography}


% \end{IEEEbiography}


\vspace{-1cm}

% % insert where needed to balance the two columns on the last page with
% %\newpage

%\balance



%\vfill


\end{document}